\documentclass[a4paper,12pt]{article}
\usepackage{exscale}
\usepackage{amssymb,amsmath,latexsym,theorem}
\def \slas{\kern -6.2pt /}
\def \sla{\kern -5.4pt /}
\def \sl{\kern -4.0pt /}
\def \Cslas{\kern -6.8pt /}
\def \Dslas{\kern -7.4pt /}
\def \slass{\kern -7.4pt /}
\def \OO{{\cal O}}
\def \QQ{{\cal Q}}
\def \ii{\mathrm{i}}
\def \d{\mathrm{d}}

\def \lcd{\tilde{\partial}}
\def \pd{\partial}

\def \lcx{\tilde{x}}
\def \tl#1{\overset{\kern 2pt\circ}{#1}}
\def \tll#1{\overset{\kern -1pt\circ}{#1}}
\def \TL#1{\overset{\kern -28pt \circ}{#1}}
\def \TLL#1{\overset{\kern -7pt \circ}{#1}}
\def \relstack#1#2{\mathrel{\mathop{#2}\limits_{#1}}}
\newcommand{\SC}{\scriptstyle}
\def \Tensor#1{\overset{\,\leftarrow\!\!\!\!\!\!\rightarrow}{#1}}
\def \LD#1{\overset{\,\leftarrow}{#1}}
\def \RD#1{\overset{\,\,\rightarrow}{#1}}
\def \LDL#1{\overset{\!\!\!\leftarrow}{#1}}
\def \RDL#1{\overset{\!\rightarrow}{#1}}

\renewcommand{\theequation}{\arabic{section}.\arabic{equation}}
\newcommand{\D}{\displaystyle}

\renewcommand{\[}{\left[}

\newcommand{\bq}{\begin{equation}}
\newcommand{\eq}{\end{equation}}
\newcommand{\bea}{\begin{eqnarray}}
\newcommand{\eea}{\end{eqnarray}}

\newcommand\ka{\kappa_1}
\newcommand\kb{\kappa_2}

\newcommand\xx{\tilde{x}}

\begin{document}
\title{\bf Twist decomposition of \\ nonlocal light-cone operators II: \\
	General tensors of 2nd rank\\
	\phantom{II. General tensors of 2nd rank}}

%\vspace*{1cm}

\author{\large{Bodo~Geyer}\thanks{e-mail: geyer@itp.uni-leipzig.de}
		~~and~
        \large{Markus~Lazar}\thanks{e-mail: lazar@itp.uni-leipzig.de} \\	
        \emph{}\\
        \emph{Center for Theoretical Studies
	and Institute~of~Theoretical~Physics,}\\
        \emph{Leipzig University, 
	Augustusplatz~10, D-04109~Leipzig, Germany}\\[2ex]
        %\emph{}\\
        %\emph{Universit\"at~Leipzig}\\
        %\emph{University of Leipzig, Institute~for~Theoretical~Physics}\\
        %\emph{Augustusplatz~10}\\
        %\emph{Augustusplatz~10, D-04109~Leipzig, Germany}\\[2ex]
   %%  \large{Dieter~Robaschik}\\    
   %%  \emph{Karl--Franzens--Universit\"at~Graz,
   %%	Institut~f\"ur~Theoretische~Physik}\\
        %\emph{}\\
        %\emph{}\\
   %%   \emph{Universit\"atsplatz~5, A-8010~Graz, Austria}
}
\date{\today}
\maketitle
\thispagestyle{empty}
\vspace{1cm}
%
%\vfill
%\newpage
%\thispagestyle{empty}
\begin{abstract}
\noindent
A group theoretical procedure, introduced earlier in \cite{GLR990}, 
to decompose bilocal light--ray 
operators into (harmonic) operators of definite twist is applied 
to the case of arbitrary 2nd rank tensors.  
As a generic example the bilocal gluon operator is considered which
gets contributions of twist--2 up to twist--6 from four different
symmetry classes characterized by corresponding Young tableaux;
also the twist decomposition of the related vector and scalar operators
is considered. In addition, we extend these results to various
trilocal light--ray operators, like the Shuryak-Vainshtein, the three--gluon
and the four--quark operators, which are required for the consideration
of higher--twist distribution amplitudes. The present results rely on
the knowledge of harmonic tensor polynomials of any order $n$ 
which have been determined up to the case of 2nd rank symmetric 
tensor for arbitrary space-time dimension.
\end{abstract}
\vspace{1cm}
%%\hrulefill\hspace{10cm}\\
%%{\footnotesize 
%%e-mail: geyer@itp.uni-leipzig.de\\
%%e-mail: lazar@itp.uni-leipzig.de\\}

\newpage
\section{Introduction}
Any careful quantum field theoretical analysis of the experimental data
for the different light--cone dominated QCD--processes (for recent
experiments and analyses, see~\cite{exp,HTEXP}) necessarily requires a rigorous 
twist decomposition of the various nonlocal light--ray operators 
appearing in the theoretical set up of these processes. 
The matrix elements of these special 
operators are related to the distribution 
amplitudes (or the hadronic wave functions) being necessary for the 
phenomenological description of the above mentioned processes. Moreover, 
their anomalous dimensions determine the $Q^2$--evolution of the 
distribution amplitudes.

The notion of twist has been originally introduced for {\em local} operators
by Gross and Treiman~\cite{Gro71} as a geometric quantity, 
twist $\tau$\,=\,(canonical) dimension $d$\,--\,spin $j$, which
is directly related to the irreducible representations of the Lorentz 
group.\footnote{
Obviously, if $d$ is taken as scale dimension of the operators it is not 
related to the generators of the Lorentz group but to those 
of the subgroup $SO(2h)\otimes {\Bbb R}_+$ of the conformal group in 
$2h$-dimensions. Thus a local operator with definite twist is a 
irreducible finite dimensional representation of the group 
$SO(2h)\otimes {\Bbb R}_+$.}
The twist decomposition of the local operators of some 
well-defined tensor structure corresponds directly to their 
decomposition into 
irreducible tensor representations of the Lorentz group. These
representations are uniquely (up to equivalence) determined by 
their symmetry classes, i.e.,~Young frames, and by the 
(anti)\-symmetrization defined by corresponding Young operators, 
i.e.,~specific Young tableaux. 

In the case of {\em nonlocal} tensor operators, being given as 
(infinite) towers of local tensor operators of growing rank, 
the definition of (geometric) twist is more subtle. For that
reason another notion of (dynamical) twist, being related to the 
LC--quantization in the infinite momentum frame \cite{KS70}, has been 
introduced \cite{Jaf92} by counting powers of $1/Q$ which, 
from a phenomenological point of view, is quite advantageous.
However, this different notion of twist has serious theoretical 
disadvantages: It is not Lorentz invariant and, therefore, not 
immediately related to geometric twist; furthermore, it is 
process--dependent because it is only applicable to matrix elements 
of the corresponding LC--operators and not to the operators itself.

In order to be able to disentangle the twist content of 
nonlocal LC--operators which are relevant for various hard QCD 
processes a systematic group theoretical study has been started in 
an earlier paper \cite{GLR990}, thereafter cited as I. There, 
bilocal quark operators up to (antisymmetric) second rank have 
been considered. Here, we continue this study 
for general tensor operators of 2nd rank. We carry out the 
complete twist decomposition of all relevant LC--operators, 
e.g.,~bilocal gluon operators, trilocal quark--gluon correlation operators 
(related to so-called Shuryak-Vainshtein operators) %\cite{Shu82} 
and four--fermion operator, as well as multilocal quark and gluon operators.
 Thereby, generalized harmonic polynomials being tensors
up to rank 2 are introduced and the interior derivative on the 
light--cone is used.

The paper is organized as follows. In Chapt.~2 we shortly review 
the general method introduced in Part I. In Chapt.~3 we 
give the complete twist decomposition of the various gluonic 
bilocal light--ray tensor operators of second rank being specified 
by different symmetry classes; their twist content ranges from 2 
up to 6. In Chapt.~4 we extend these results with minor 
modifications to the trilocal light--ray operators mentioned 
above. Any of these nonlocal operators consists of an infinite tower 
of local operators which are characterized by equal twist. 
Appendix A contains some material about the tensorial harmonic 
functions.

\section{General procedure}

Let us now shortly review the general procedure, introduced in Part I, 
 of how to decompose 
arbitrary bilocal LC--operators into harmonic operators of definite twist. 
Those nonlocal operators which are relevant for the virtual Compton scattering 
\cite{BGR99} are obtained by the nonlocal LC expansion 
\cite{AS78,SLAC} of the (renormalized)
time-ordered product of two electromagnetic hadronic currents:
\begin{equation}
\label{GLCE}
%\hspace{-.5cm}
{\cal R}
T\big(J^\mu(\hbox{$y\!+\!\large\frac{x}{2}$})
J^\nu(\hbox{$y\!-\!\large \frac{x}{2}$})S\big) 
\!\relstack{x^2\rightarrow 0}{\approx}\!
\int%_{-\frac{1}{2}}^{\frac{1}{2}}
\!\d^2 \underline\kappa\,
C^{\mu\nu}_\Gamma(x^2\!, \underline\kappa) 
{\cal R}
T\big(\OO^\Gamma\!(y\!+\!\ka \xx, y\!+\!\kb \xx)S\big) 
\!+\cdots,
\nonumber
\end{equation}
%Strictly speaking, the r.h.s. of eq.~(\ref{GLCE}) are operator valued 
%distributions~(see e.g. \cite{Mack77}).
where 
$\Gamma=\{1,\gamma_\mu,\sigma_{\mu\nu};
\gamma_5,\gamma_\mu\gamma_5,\sigma_{\mu\nu}\gamma_5\}$
indicates the tensor structure of the nonlocal quark operators 
and $\xx$ is a light-like vector related to $x$, cf. \cite{AS78,LEIP,GLR990}. 
The un\-renormalized nonlocal (flavour singlet) operators  
at $y=0$ are given by
\begin{equation}
\label{GLCEOP}
\OO^\Gamma(\ka \xx, \kb \xx) 
=
\overline\psi(\ka \xx) \Gamma 
U(\ka \xx, \kb \xx) \psi(\kb \xx)%\nonumber
\end{equation}
with the path ordered phase factor
\begin{equation}
\label{phase}
U(\ka \xx, \kb \xx) 
 = 
{\cal P} \exp\left\{\ii g
\int_{\ka}^{\kb} \d\tau \,\xx^\mu A_\mu (\tau \xx)
\right\},%\quad\tau=(\ka-\kb)t+\kb \nonumber
\end{equation}
ensuring gauge invariance. 
As is well-known, under renormalization these operators mix with 
appropriate chiral-even (or chiral-odd) bilocal gluonic tensor operators, 
eventually multiplied by appropriate tensors built up from $\xx$: 
\begin{align}
\label{Gtensor}
G_{\alpha\beta}(\ka\xx,\kb\xx)
&=
F_\alpha^{\ \rho}(\ka\xx)
U(\ka\xx,\kb\xx)F_{\beta\rho}(\kb\xx)\,,
\\
\label{Gtensor5}
\widetilde{G}_{\alpha\beta}(\ka\xx,\kb\xx)
&=
F_\alpha^{\ \rho}(\ka\xx)
U(\ka\xx,\kb\xx)\widetilde{F}_{\beta\rho}(\kb\xx)\, ,
\end{align}
where $F_{\alpha\beta}$ and 
$\widetilde{F}_{\alpha\beta}= 
\frac{1}{2}\epsilon_{\alpha\beta\mu\nu}F^{\mu\nu}$
are the gluon field strength and its dual, respectively.
Here, the phase factors are to be taken in the adjoint representation. 

In general, the twist decomposition of an arbitrary bilocal light--ray 
operator 
may be formulated for any space-time dimension $D=2h$. Let us denote 
such operators for arbitrary values of $x$ as follows:
\begin{equation}
\label{O_Gamma}
%\hspace{-.2cm}
\OO^\Gamma(\kappa_1  x,\kappa_2 x)
=
{\Phi}'(\kappa_1  x)\Gamma 
U(\kappa_1  x,\kappa_2 x)
\Phi(\kappa_2 x)
%\nonumber
\end{equation}
where, suppressing any indices indicating the group representations, $\Phi$ 
generically denotes the various local fields, e.g.,~scalars 
($d=h-1$), Dirac spinors ($d=h-\frac{1}{2}$) as well as gauge field strength 
($d=h$). $A_\mu$ is the gauge potential having dimension $d=h-1$. Furthermore, 
$\Gamma$ labels the tensor structure as well as additional quantum numbers, if 
necessary.\\
\medskip
Now, the twist decomposition of operators (\ref{O_Gamma}) consists of the 
following steps:\\
\noindent
(1)\quad{\em Taylor expansion} of the nonlocal operators for 
{\em arbitrary} values of $x$ at the point $y=0$ into an infinite 
series of local tensor operators having definite rank $n$ and 
canonical dimension $d$:
\begin{align}
\hspace{-.3cm}
\label{O_ent}
\OO^\Gamma(\kappa_1  x,\kappa_2 x)
=
\sum_{n=0}^{\infty}
\frac{1}{n!}
{x}^{\mu_1}\ldots {x}^{\mu_n}
\Big[\Phi'(y)\Gamma
\Tensor D_{\mu_1}\!(\kappa_1, \kappa_2)
\ldots 
\Tensor D_{\mu_n}\!(\kappa_1, \kappa_2)
\Phi(y)\Big]_{y=0} 
%\nonumber
\end{align}
with the generalized covariant derivatives
\begin{eqnarray}
\label{D_kappa}
\Tensor D_\mu(\kappa_1, \kappa_2)
&\equiv& 
\kappa_1 \LD D_\mu+
\kappa_2 \RD D_{\mu}\,,
%\nonumber
\\
\RD D_\mu=\RDL{\pd^y_\mu} +\ii g A_\mu(y),
&&
\LD D_\mu=\LDL{\pd^y_\mu} -\ii g A_\mu(y)\,. 
\nonumber
\end{eqnarray}

\noindent
(2)\quad{\em Decomposition} of local operators with respect to 
{\em irreducible tensor representations} of the Lorentz group $SO(2h-1,1)$ 
or, equivalently, the orthogonal group 
$SO(2h)$.\footnote{
The corresponding representations are related through 
analytic continuation, cf. \cite{BR,Vil,GLR990}}
These representations are built up by traceless tensors of rank $m$
whose symmetry class is determined
by some (normalized) Young operators ${\cal Y}_{[m]}=(f_{[m]}/m!){\cal QP}$,
where $[m] = (m_1, m_2, \ldots m_r)$
with $m_1 \geq m_2\geq \ldots \geq m_r$
and $\sum^r_{i=1}\, m_i = m$ denotes the corresponding Young pattern. 
$\cal P$ and $\cal Q$, as usual, denote symmetrization and antisymmetrization 
with respect to that pattern. The allowed Young patterns for $SO(2h)$, which 
because of the tracelessness are restricted by 
$\ell_1+\ell_2\le 2h$ ($\ell$: length of {\em columns} of $[m]$), 
are for a fixed value of $m$:\\
\begin{enumerate}
\item[(i)] \unitlength0.4cm
\begin{picture}(30,1)
\linethickness{0.15mm}
\multiput(1,0)(1,0){13}{\line(0,1){1}}
\put(1,1){\line(1,0){12}}
\put(1,0){\line(1,0){12}}
\put(15,0){$j=m,m-2,m-4,\ldots$}
\end{picture}

\item[(ii)]\unitlength0.4cm
\begin{picture}(5,1)
\linethickness{0.15mm}
\multiput(3,0)(1,0){10}{\line(0,1){1}}
\multiput(1,-1)(1,0){2}{\line(0,1){2}}
\put(1,1){\line(1,0){11}}
\put(1,0){\line(1,0){11}}
\put(1,-1){\line(1,0){1}}
\put(15,0){$j=m-1,m-2,m-3,\ldots$}
\end{picture}

\item[(iii)] \unitlength0.4cm
\begin{picture}(5,2)
\linethickness{0.15mm}
\multiput(3,0)(1,0){9}{\line(0,1){1}}
\multiput(1,-2)(1,0){2}{\line(0,1){3}}
\put(1,1){\line(1,0){10}}
\put(1,0){\line(1,0){10}}
\put(1,-1){\line(1,0){1}}
\put(1,-2){\line(1,0){1}}
\put(15,0){$j=m-2,m-3,m-4,\ldots$}
\end{picture}

\item[(iv)] \unitlength0.4cm
\begin{picture}(5,3)
\linethickness{0.15mm}
\multiput(4,0)(1,0){8}{\line(0,1){1}}
\multiput(1,-1)(1,0){3}{\line(0,1){2}}
\put(1,1){\line(1,0){10}}
\put(1,0){\line(1,0){10}}
\put(1,-1){\line(1,0){2}}
\put(15,0){$j=m-2,m-3,m-4,\ldots$}
\end{picture}
\end{enumerate}
\vspace*{5mm}
$\qquad\qquad\vdots$
\\
\\
(Here, we depicted only those patterns which appear for $h=2$ where 
$\ell_1+\ell_2\le 4$.)
Loosely speaking, the lower spins in the above series correspond to the
various trace terms which define, in general, reducible higher twist
operators. These higher twist operators again have to be 
decomposed according to the considered symmetry classes.

Now, two remarks are in order. At first, this decomposition of the 
local operators 
appearing in (\ref{O_ent}) into irreducible components is independent 
of the 
special values of $\ka, \kb$. Therefore, we may choose $\ka=0, \kb=\kappa$ 
and factor out 
$\kappa^n$ for every term in eq.~(\ref{O_ent}), thus simplifying 
the explicit computations below. At the end of the computations the general 
values of $\kappa_i$ may be re-installed.
Furthermore, if these irreducible local tensor operators are multiplied by
${x}^{\mu_1}\ldots {x}^{\mu_n}$, as it is necessary according to 
eq.~(\ref{O_ent}), then harmonic (tensorial) polynomials of order $n$ appear 
whose remaining tensor structure results from $\Gamma$. For a detailed 
exposition of these objects, see Appendix A.
 
\noindent
(3)\quad{\em Resummation} of the local operators 
$\OO^{\tau}_{\Gamma\,n}(x)$
belonging to the same symmetry class (for any $n$) and having equal 
twist $\tau$. That infinite tower of local 
operators creates a {\em nonlocal} operator of definite twist,
\begin{equation}
\label{O_nonlocal}
\OO^\tau_\Gamma(0,\kappa x)=\bigoplus_{n=0}^\infty 
\frac{\kappa^n}{n!}
\OO^{\tau}_{\Gamma\,n}(x)\,.
\nonumber
\end{equation}
Obviously, from a group theoretical point of view, this nonlocal 
operator is built up as direct sum of irreducible local tensor operators.
The nonlocal operators $\OO^\tau_\Gamma(0,\kappa x)$ are 
harmonic tensor functions. 
Now, putting together all contributions, i.e.,~including the higher twist
contributions resulting from the trace terms, we get the (infinite) 
twist decomposition of the nonlocal operator we started with:
\begin{equation}
\OO_\Gamma(0,x)
=\bigoplus_{\tau=\tau_{\rm min}}^\infty \OO^\tau_\Gamma(0, x)\,.
\nonumber
\end{equation}

%%which consists of a nonlocal operator -- whose local parts show the correct
%%symmetry behaviour, but being not traceless -- as well as terms of higher twist
%%which are necessary to consecutively subtract the traces. 
%%In general, these higher twist operators are reducible and have to be 
%%decomposed according to the considered symmetry class.

\noindent
(4)\quad Finally, the {\em projection onto the light--cone},
$x\rightarrow\tilde{x}$, leads to the required light--cone operator with 
well defined geometric twist. However, since the harmonic tensor
polynomials essentially depend on (infinite) sums of powers of $x^2$ and 
$\square$ as well as some specific differential operators in front
of it, in that limit only a finite number of terms survive:
\begin{equation}
\OO_\Gamma(\ka\xx,\kb\xx)=
\bigoplus_{\tau=\tau_{\rm min}}^{\tau_{\rm max}} 
\OO^\tau_\Gamma(\ka\xx,\kb\xx)\,.
\nonumber
\end{equation}
The resulting light--ray operators of definite twist, which have been 
rewritten for general arguments, are tensor functions on the light--cone 
which fulfil another kind of tracelessness conditions to be formulated 
with the help of the interior (on the cone) derivative. --- Let us remark 
that step~(3) and (4) can be interchanged without changing the result.

The advantages of our method are:
First, these nonlocal operators of definite twist are Lorentz covariant 
tensors (contrary to the phenomenological concept of twist).
Second, this twist decomposition is unique, process-- and model--independent.
Furthermore, the twist decomposition is independent from the dimension 
$2h$ of space-time.

%[Let us remark that coefficients like $g_{\alpha\beta}$, $x_\alpha$ 
%or $x_\beta$ do not belong to the higher twist operator itself. 
%Actually, the higher twist operators are the expressions behind these
%coefficients.]

%%%%%%%%%%%%%%%%%%%%%%%%%%%%%%%%%%%%%%%%%%%%
\newpage
\section{Twist decomposition of a general 2nd rank \\
tensor operator}
\label{tensor}
\setcounter{equation}{0}
%%%%%%%%%%%%%%%%%%%%%%%%%%%%%%%%%%%%%%%%%%%%
%
%
This Chapter is devoted to the twist decomposition of a general 
2nd rank tensor operator. For simplicity, we restrict ourselves 
to {\em four dimensional space-time} ($h=2$).
Furthermore, since the twist decomposition is independent from the 
chirality we demonstrate it only for the chiral-even gluon operator 
(\ref{Gtensor}). In addition,
we make the special choice $\ka = 0, \kb = \kappa$, thereby also 
simplifying the generalized
covariant derivatives, eq.~(\ref{D_kappa}), to the usual ones: 
\begin{align}
\label{Gtensor0}
G_{\alpha\beta}(0,\kappa x)
=&\,
F_\alpha^{\ \rho}(0)U(0,\kappa x)F_{\beta\rho}(\kappa x)%\\
=
\sum_{n=0}^\infty \frac{\kappa^n}{n!}x^{\mu_1} \ldots x^{\mu_n}
G_{\alpha\beta(\mu_1\ldots\mu_n)}\ ,
\\
%\end{equation}
%\begin{equation}
%\label{Gn_loc}
{\rm with}&\;\;\qquad G_{\alpha\beta(\mu_1\ldots\mu_n)}
=
F_\alpha^{\ \rho}(y) D^y_{(\mu_1}\ldots D^y_{\mu_n)}
F_{\beta\rho}(y)\big|_{y=0}\ ;
\nonumber
\end{align}
here the symbol $(\ldots)$ denotes symmetrization of the enclosed 
indices.

The local tensor operators 
$ G_{\alpha\beta(\mu_1\ldots\mu_n)} $  decompose according to
the Young patterns (i) to (iv) and possible higher ones
(i.e.,~for $m_1+m_2 = n+2, m_2 \geq 3$):
\begin{eqnarray}
\hspace{-1cm}
\label{G_loc}
G_{\alpha\beta(\mu_1\ldots\mu_n)}
\!\!\!&{=}&\!\!\!
G_{\alpha\beta(\mu_1\ldots\mu_n)}^{\mathrm{(i)}}
+
\alpha_{n+1}
%\hbox{\large$\frac{2n}{n+1}$}
G_{\alpha\beta(\mu_1\ldots\mu_n)}^{\mathrm{(ii)}}
\nonumber\\
\!\!\!&{}&\!\!\!
+\beta_n
G_{\alpha\beta(\mu_1\ldots\mu_n)}^{\mathrm{(iii)}}
+\gamma_n
G_{\alpha\beta(\mu_1\ldots\mu_n)}^{\mathrm{(iv)}}
+ \ldots\, ,
%%\nonumber
\end{eqnarray}
with the (nontrivial) normalization coefficients 
$f_{[m]}/m!$ of the Young operators given by
$\alpha_{n+1} = 2(n+1)/(n+2)$ for $[m] = (n+1,1)$,
$\beta_n = 3n/(n+2)$ for $[m] = (n,1,1)$
and
$\gamma_n = 4(n-1)/(n+1)$ for $[m] = (n,2)$.
The corresponding Clebsch-Gordan series in terms of 
representations $(j_1,j_2)$ of the Lorentz group is
\begin{eqnarray}
\hspace{-1cm}
\lefteqn{\hspace{-2cm}
\hbox{\large$\big(\frac{1}{2},\frac{1}{2}\big)$}
\otimes
\hbox{\large$\big(\frac{1}{2},\frac{1}{2}\big)$}
\otimes
\left(
\hbox{\large$\big(\frac{n}{2},\frac{n}{2}\big)$}
\oplus
\hbox{\large$\big(\frac{n-2}{2},\frac{n-2}{2}\big)$}
\oplus
\ldots\right)}
\nonumber\\
\!\!\!&=&\!\!\!
\hbox{\large$\big(\frac{n+2}{2},\frac{n+2}{2}\big)$}
\oplus
\left(\hbox{\large$\big(\frac{n+2}{2},\frac{n}{2}\big)$}
\oplus
\hbox{\large$\big(\frac{n}{2},\frac{n+2}{2}\big)$}\right)
\oplus
\hbox{\large$\big(\frac{n}{2},\frac{n}{2}\big)$}
\nonumber\\
\label{CGR1}
&&\oplus
\left(\hbox{\large$\big(\frac{n+2}{2},\frac{n-2}{2}\big)$}
\oplus
\hbox{\large$\big(\frac{n-2}{2},\frac{n+2}{2}\big)$}\right)
\\
&&\oplus
\left(\hbox{\large$\big(\frac{n}{2},\frac{n-2}{2}\big)$}
\oplus
\hbox{\large$\big(\frac{n-2}{2},\frac{n}{2}\big)$}\right)
\oplus
\hbox{\large$\big(\frac{n-2}{2},\frac{n-2}{2}\big)$}
%%\nonumber\\
%%&&\oplus
%%\left(\hbox{\large$\big(\frac{n}{2},\frac{n-4}{2}\big)$}
%%\oplus
%%\hbox{\large$\big(\frac{n-4}{2},\frac{n}{2}\big)$}\right)
\oplus\ldots\ ;
\nonumber
\end{eqnarray}
the corresponding tensor spaces will be denoted by ${\bf T} (j_1,j_2)$.
In the last line of eq.~(\ref{CGR1}) such representations are listed down
which correspond to higher twist contributions contained in the trace
terms of symmetry classes (i) -- (iv).
The canonical (or scale) dimension $d$ of the local operator 
(\ref{G_loc}) is $n+4$, 
and the spin of the various contributions in (\ref{CGR1}) 
ranges from $n+2$ up to $1$ or $0$ if $n$ is even or odd,
respectively;
therefore, the local operators with well-defined twist are 
irreducible tensors of the Lorentz group or, equivalently,
of the group 
$SL(2,{\Bbb C})\otimes {\Bbb R}_+$, having scale dimension $d$.

In general, according to the spin content and the rank of the corresponding 
local tensor operators, the nonlocal operator (\ref{Gtensor0}) for arbitrary
$x$ contains contributions of twist ranging from $\tau = 2$
until $\tau=\infty$. 
However, after projection onto the light--cone,
$x \rightarrow {\tilde x}$, this infinite series terminates
at least at $\tau = 6$ since higher order terms are
proportional to $x^2$. 

In order to be more explicit, as well as for later use, we 
introduce the (anti)sym\-metrization with respect to $\alpha$ and
$\beta$ and we define the following nonlocal operators:
\begin{eqnarray}
\label{GT}
G^\pm_{\alpha\beta}(0,\kappa x) 
&:=& 
\hbox{$\frac{1}{2}$} \left(G_{\alpha\beta}(0,\kappa x) \pm 
G_{\beta\alpha}(0,\kappa x)\right)\ ,
\\
\label{GV}
G^\pm_{\alpha }(0,\kappa x)
 &:=& x^\beta\ G^\pm_{\alpha\beta}(0,\kappa x)
 \equiv G^\pm_{\alpha\bullet}(0,\kappa x)\ ,
\\
\label{GS}
G (0,\kappa x)
&:=&  x^\alpha x^\beta\ G^+_{\alpha\beta}(0,\kappa x)\ .
\end{eqnarray}
In fact, the twist decomposition of these bilocal light--ray 
operators reads:\footnote{
We use the notation
$A_{(\alpha\beta)}\equiv\frac{1}{2}(A_{\alpha\beta}+A_{\beta\alpha})$
and 
$A_{[\alpha\beta]}\equiv\frac{1}{2}(A_{\alpha\beta}-A_{\beta\alpha})$.
}
\begin{align} 
\label{G^+_t}
G^+_{\alpha\beta}(0,\kappa\tilde{x})
=&\phantom{+}
 G^{\mathrm{tw2}}_{(\alpha\beta)}(0,\kappa\tilde{x})
+G^{\mathrm{tw3}}_{(\alpha\beta)}(0,\kappa\tilde{x}) 
+G^{\mathrm{tw4}}_{(\alpha\beta)}(0,\kappa\tilde{x})
\nonumber\\
&\!\!+\!G^{\mathrm{tw5}}_{(\alpha\beta)}(0,\kappa\tilde{x}) 
+G^{\mathrm{tw6}}_{(\alpha\beta)}(0,\kappa\tilde{x})\,,
\\
\label{G^-_t}
G^-_{\alpha\beta}(0,\kappa\tilde{x})
=&\phantom{+}
G^{\mathrm{tw3}}_{[\alpha\beta]}(0,\kappa\tilde{x}) 
+G^{\mathrm{tw4}}_{[\alpha\beta]}(0,\kappa\tilde{x}) 
+G^{\mathrm{tw5}}_{[\alpha\beta]}(0,\kappa\tilde{x})\,, 
\\
\label{G^+_vec}
G^+_{\alpha}(0,\kappa\tilde{x})
=&\phantom{+}
 G^{\mathrm{tw2}}_{(\alpha\bullet)}(0,\kappa\tilde{x})
+G^{\mathrm{tw3}}_{(\alpha\bullet)}(0,\kappa\tilde{x}) 
+G^{\mathrm{tw4}}_{(\alpha\bullet)}(0,\kappa\tilde{x}) \,,
\\
\label{G^-_vec}
G^-_{\alpha}(0,\kappa\tilde{x})
=&\phantom{+}
 G^{\mathrm{tw3}}_{[\alpha\bullet]}(0,\kappa\tilde{x}) 
+G^{\mathrm{tw4}}_{[\alpha\bullet]}(0,\kappa\tilde{x})\,,
\\
\label{G_sca}
G(0,\kappa\tilde{x})
=&\phantom{+}
 G^{\mathrm{tw2}}(0,\kappa\tilde{x})\,.
\end{align}
The various twist contributions individually decompose further 
according to the symmetry classes which may contribute.
The explicit expressions for generic tensor operators
are given in the following subsections, where
it will be shown that Young patterns (i) and (ii) as well as (iv) 
contribute to the symmetric tensor operators and related vector
and scalar operators, and Young patterns (ii) and (iii) contribute to
the antisymmetric tensor operators and related vector 
operators. %\footnote{
%Let us remark that for virtual Compton scattering only the symmetric gluon 
%vector operators mixes with the flavour singlet quark vector 
%operators (see, e.g.~\cite{BGR99}).}
Special tensor operators are considered in Chapt. 4.
%%%\medskip
%%%\noindent
\subsection{Tensor operators of symmetry class (i)}
%%%{\it 3.1.1~~
\subsubsection{Construction of nonlocal symmetric class-(i) operators 
of definite twist}
%%%\\
Let us start with the simplest case of the totally symmetric traceless 
tensors, and their contractions with $x$, which have twist $\tau =2$ and
are contained in the tensor space ${\bf T} (\frac{n+2}{2},\frac{n+2}{2})$. 
They have symmetry class (i) and are uniquely characterized by the 
following standard tableau (with normalizing factor 1):
\\
\\
%\vspace*{3mm}
\unitlength0.5cm
\begin{picture}(30,1)
\linethickness{0.075mm}
\put(1,0){\framebox(1,1){$\alpha$}}
\put(2,0){\framebox(1,1){$\beta$}}
\put(3,0){\framebox(1,1){$\mu_1$}}
\put(4,0){\framebox(3,1){$\ldots$}}
\put(7,0){\framebox(1,1){$\mu_n$}}
\put(9.5,0){$\stackrel{\wedge}{=}$}
\put(11,0){$\relstack{\alpha\beta\mu_1\ldots\mu_n}{\cal S}
F_\alpha^{\ \rho}(0) 
\left(D_{\mu_1}\ldots D_{\mu_n}F_{\beta\rho}\right)(0)
- \mathrm{trace~terms~,}$}
\end{picture}
\\
\\
The irreducible local twist--2 operator reads
\begin{equation}
\label{G2_loc}
G^{\mathrm{tw2(i)}}_{(\alpha\beta\mu_1\ldots\mu_n)}
=
F_{(\alpha}^{\, \rho}(0) D_{\mu_1}\ldots D_{\mu_n}F_{\beta)\rho}(0)
-\mathrm{trace~terms}\,.
\nonumber
\end{equation}
Let us postpone the determination of the trace terms and make the 
resummation to the corresponding nonlocal operator in advance. 
This is obtained, at first, by contracting with $x^{\mu_1}\ldots x^{\mu_n}$ 
and rewriting in the following form:
\begin{eqnarray}
\label{2betan}
\hspace{-.2cm}
G^{\mathrm{tw2(i)}}_{(\alpha\beta) n}(x)
=
x^{\mu_1}\ldots x^{\mu_n}\
G^{\mathrm{tw2(i)}}_{(\alpha\beta\mu_1\ldots\mu_n)}
%\nonumber\\
\label{O2_x}
=%&=&
\hbox{\large$\frac{1}{(n+2)(n+1)}$}\,
\pd_\alpha\pd_\beta \tl G_{n+2}(x), 
\nonumber
\end{eqnarray}
with the denotation
\begin{eqnarray}
\tl G_{n+2}(x) &=& G_{n+2}(x)-\mathrm{~trace~terms~,}
\nonumber\\
G_{n+2}(x) &\equiv& x^\mu x^\nu F_\mu^{\ \rho} (0) (xD)^n F_{\nu\rho}(0)\ .
\nonumber
\end{eqnarray}
Here, $\tl G_{n+2}(x)$ is a harmonic polynomial of order $n+2$, 
cf.~eq.~(\ref{T_harm_d}), Appendix~\ref{trace}:
\begin{equation}
\label{Proj_tw2}
\hspace{-.3cm}
\tl G_{n+2} (x)
=
H_{n+2}^{(4)}\!\left(x^2|\square\right) G_{n+2}(x)
\equiv
\sum_{k=0}^{[\frac{n+2}{2}]}\frac{(n+2-k)!}{(n+2)!k!}
\!\left(\frac{-x^2}{4}\right)^{\!\!k}\!\square^k
G_{n+2}(x)\,.
\nonumber
\end{equation}
Then, using $((n+2)(n+1))^{-1} = \int^1_0 \d\lambda 
(1-\lambda)\lambda^n$ and
the integral representation of Euler's beta function we obtain
the nonlocal {\em twist--2 tensor operator}:\footnote{
Here, and in the following, we omit the indication of the symmetry class
of the nonlocal twist--2 operators because only totally symmetric 
tensors contribute. However, the trace terms being of higher twist 
must be classified according to their symmetry type.}
\begin{equation}
\label{G_tw2_tens}
%\label{O2_beta}
G^{\mathrm{tw2}}_{(\alpha\beta)} (0,\kappa x)
=
\sum_{n=0}^{\infty}
\frac{\kappa^n}{n!}\ 
G^{\mathrm{tw2(i)}}_{(\alpha\beta)n}(x)
=
\pd_\alpha\pd_\beta
\int_{0}^{1} \d\lambda (1-\lambda)
\tl G(0,\kappa\lambda x) ,
\vspace{-.3cm}
\end{equation}
with
\begin{equation}
\label{proj_tw2}
\hspace{-.3cm}
\tl G(0,\kappa x)
=
G(0,\kappa x)
+\sum_{k=1}^{\infty}\int_0^1\!\!\d t\, t
\left(\frac{-x^2}{4}\right)^{\!\!k}\!\!
\frac{\square^k}{k!(k-1)!}
\left(\frac{1-t}{t}\right)^{\!\! k-1}
\!\!G(0,\kappa tx).
%\nonumber
\end{equation}
%
%Of course, the derivative $\pd_\beta$ could be taken outside the 
%parentheses in eq. (\ref{2betan}) (and the integral) 
%because the Young operators should be applied
%to traceless tensors. (For notational definiteness we remark
%that partial derivatives are everywhere with respect to $x$ only.)
%
%Note that $x^\beta O^{\mathrm{tw2}}_{\beta} (0,\kappa x)=
%\tl O(0,\kappa x)$ is obtained 
%from eq.~(\ref{O_tw2}) by partial integration 
%observing that for any function $f(\lambda x)$ 
%the equality $x^\mu\partial f/\partial x^\mu =
%\lambda \partial f/\partial \lambda$ holds. 
%
The operator $G^{\mathrm{tw2}}_{(\alpha\beta)} (0,\kappa x)$ 
satisfies the conditions of a harmonic tensor function:
\begin{equation}
\label{cond}
g^{\alpha\beta}G^{\mathrm{tw2}}_{(\alpha\beta)}(0,\kappa x)=0 ,
\quad
\square G^{\mathrm{tw2}}_{(\alpha\beta)}(0,\kappa x)=0 ,
\quad
\pd^\alpha G^{\mathrm{tw2}}_{(\alpha\beta)}(0,\kappa x)=0\, .
%=\pd^\beta G^{\mathrm{tw2}}_{\alpha\beta}(0,\kappa x)
\end{equation}
%%%\medskip
%%\noindent
%%%{\it 3.1.2~~
\subsubsection{Reduction to vector and scalar operators}
Now, we specify to the nonlocal vector and scalar operators. 
Multiplying eq.~(\ref{G_tw2_tens}) by $x^\beta$ and observing
the equality
$(x\pd_x) \tl G(0,\kappa\lambda x) = (\lambda \partial_\lambda +2) 
\tl G(0,\kappa\lambda x)$ gives the {\em twist--2 vector operator},
\begin{equation}
\label{G_tw2_v}
G^{\mathrm{tw2}}_{(\alpha\bullet)}(0,\kappa x)
=
\pd_\alpha\int_{0}^{1} \d\lambda\,\lambda\, \tl G(0,\kappa\lambda x),
\end{equation}
which satisfies the conditions
\begin{equation}
\label{cond2}
\square\ G^{\mathrm{tw2}}_{(\alpha\bullet)}(0,\kappa x)=0,
\quad
\pd^\alpha\ G^{\mathrm{tw2}}_{(\alpha\bullet)}(0,\kappa x)=0\,.
\end{equation}
Multiplying with $x^\alpha x^\beta$  gives the {\em twist--2 scalar 
operator},
\begin{equation}
\label{G_tw2_s}
G^{\mathrm{tw2}}(0,\kappa x) = \tl G(0,\kappa x)\,,
\end{equation}
which, by definition, satisfies the condition
\begin{equation}
\label{cond20}
\square G^{\mathrm{tw2}}(0,\kappa x)=0\,.
\end{equation}
The expression (\ref{proj_tw2}) for the scalar operator already has
been given by Balitsky and Braun \cite{BB88}.

Comparing eqs.~(\ref{G_tw2_tens}), (\ref{G_tw2_v}) and (\ref{G_tw2_s})
we may recognize that in the case of symmetry class (i) the tensor 
and vector operators are obtained from the scalar operator by very
simple operations. Furthermore, we observe how in the case of the
scalar operator the trace terms -- being proportional to $x^2$ --
are to be subtracted from $G(0,\kappa x)$ in order to make that
operator traceless. In the case of vector and tensor operators
such subtraction, because of the appearance of the derivatives,
is more complicated.
%%%\medskip
%%%\noindent
%%%{\it 3.1.3~~
\subsubsection{Projection onto the light--cone}
Let us now project onto the light--cone and, at the same time,
also extend to the case of general values $(\ka, \kb)$.
Because of the derivatives $\pd_\alpha$ and $\pd_\beta$,
appearing in eq.~(\ref{G_tw2_tens}), only the terms with $k=1,2$ 
in eq.~(\ref{proj_tw2}) contribute. The final expression for the
{\em symmetric twist--2 light--cone tensor operator} is given by
%\begin{align}
\begin{align}
\hspace{-.3cm}
\label{Gtw2_gir}
G^{\mathrm{tw2}}_{(\alpha\beta)}(\ka\xx,\kb\xx)
=
&\,
\pd_\alpha\pd_\beta\!\!\int_{0}^{1}\!\!\!\d\lambda (1-\lambda)
G(\kappa_1\lambda x,\kappa_2\lambda x)\big|_{x=\tilde{x}}\!
%\nonumber\\
-G_{(\alpha\beta)}^{\mathrm{>(i)}}(\kappa_1\lcx,\kappa_2\lcx),
\\
\hspace{-.3cm}
\label{G-high(i)}
G_{(\alpha\beta)}^{\mathrm{>(i)}}(\kappa_1\lcx,\kappa_2\lcx)
=&
\int_0^1\d\lambda\Big\{(1-\lambda+\lambda\ln\lambda)
\left(\hbox{\large$\frac{1}{2}$}
g_{\alpha\beta}+x_{(\alpha}\pd_{\beta)}\right)\square
\\
\hspace{-.3cm}
&+\hbox{\large$\frac{1}{4}$}\big(2(1-\lambda)+(1+\lambda)\ln\lambda\big)
x_\alpha x_\beta\square^2\Big\}
G(\kappa_1\lambda x,\kappa_2\lambda x)\big|_{x=\tilde{x}}.\nonumber
\end{align}
The operator $G_{(\alpha\beta)}^{\mathrm{>(i)}}(\kappa_1\xx,\kappa_2\xx)$
contains the {\em higher twist contributions} which have to be 
subtracted from the first term, eq.~(\ref{Gtw2_gir}),
in order to make the whole expression traceless. 
In order to disentangle the different terms of well-defined twist
we observe that the tensor operator (\ref{G-high(i)}) 
consists of scalar and vector parts. 
Here, an operator being contained in the trace terms is 
called a scalar and vector part, respectively, if the expression 
multiplied by $g_{\alpha\beta}$, $x_\alpha$ or $x_\beta$ 
is a scalar, like $\square G(\ka x,\kb x)$ or $\pd^\mu G_\mu(\ka x,\kb x)$, 
and a vector, like $\pd_\beta G(\ka x,\kb x)$ or 
$\pd_\beta\pd^\mu G_\mu(\ka x,\kb x)$, respectively.\footnote{
Strictly speaking, $x_\alpha$ and $x_\beta$ as well as $g_{\alpha\beta}$ 
which are necessary for the dimension and tensor structure of the whole
expression do not belong to the higher twist operator itself.}
The scalar operators, $\square G(\ka x,\kb x)|_{x=\tilde{x}}$ and 
$\square^2 G(\ka x,\kb x)|_{x=\tilde{x}}$,
occurring in eq.~(\ref{G-high(i)}) 
already have well-defined twist $\tau =4$ and $\tau =6$, respectively. 
To ensure that the vector operator 
$\pd_\beta \square G(\ka x,\kb x)|_{x=\tilde{x}}$ also obtains
well-defined twist, it is necessary to subtract its own trace terms 
which are of twist  $\tau=6$.

The higher twist operators contained in the trace terms of the 
twist--2 tensor operator are the following:
\begin{align}
%\label{M_tw2_b}
G^{\mathrm{>(i)}}_{(\alpha\beta)}
(\kappa_1\tilde{x},\kappa_2\tilde{x})
=&\;
G^{\mathrm{tw4(i)a}}_{(\alpha\beta)}(\kappa_1\tilde{x},\kappa_2\tilde{x})+
G^{\mathrm{tw4(i)b}}_{(\alpha\beta)}(\kappa_1\tilde{x},\kappa_2\tilde{x})
\nonumber\\
&+ G^{\mathrm{tw6(i)a}}_{(\alpha\beta)}(\kappa_1\tilde{x},\kappa_2\tilde{x})
+
G^{\mathrm{tw6(i)b}}_{(\alpha\beta)}(\kappa_1\tilde{x},\kappa_2\tilde{x})
\,,
\end{align}
with
\begin{align}
G_{(\alpha\beta)}^{\mathrm{tw4(i)a}}(\kappa_1\lcx,\kappa_2\lcx)
=&\,\hbox{\large$\frac{1}{2}$}g_{\alpha\beta}\square
\int_0^1\d\lambda(1-\lambda+\lambda\ln\lambda)
G(\kappa_1\lambda x,\kappa_2\lambda x)\big |_{x=\tilde{x}}\,,
\nonumber\\
G_{(\alpha\beta)}^{\mathrm{tw4(i)b}}(\kappa_1\lcx,\kappa_2\lcx)
=&\, x_{(\alpha}\pd_{\beta)}\square
\int_0^1\d\lambda(1-\lambda+\lambda\ln\lambda)
G(\kappa_1\lambda x,\kappa_2\lambda x)\big |_{x=\tilde{x}}
\nonumber\\
&-G_{(\alpha\beta)}^{\mathrm{tw6(i)b}}(\kappa_1\lcx,\kappa_2\lcx)\,,
\nonumber
\end{align}
and
\begin{align}
G_{(\alpha\beta)}^{\mathrm{tw6(i)a}}(\kappa_1\lcx,\kappa_2\lcx)
=&
\hbox{\large$\frac{1}{4}$}x_\alpha x_\beta\square^2\!
\int_0^1\!\!\!\d\lambda
\big( 2(1\!-\!\lambda)+(1\!+\!\lambda)\ln\lambda\big)
G(\kappa_1\lambda x,\kappa_2\lambda x)\big |_{x=\tilde{x}}\,,
\nonumber\\
G_{(\alpha\beta)}^{\mathrm{tw6(i)b}}(\kappa_1\lcx,\kappa_2\lcx)
=&\,
\hbox{\large$\frac{1}{4}$}x_\alpha x_\beta\square^2\!
\int_0^1\!\!\!\d\lambda
\Big(\! \hbox{\large$\frac{(1-\lambda)^2}{\lambda}$}-
\hbox{\large$\frac{1-\lambda^2}{2\lambda}$}-\lambda\ln\lambda\!\Big)
G(\kappa_1\lambda x,\kappa_2\lambda x)\big|_{x=\tilde{x}}\,.
\nonumber
\end{align}

Obviously, by inspection of the spin content of the symmetry type (i),
the local twist--4 and twist--6 operators carry the 
representation ${\bf T} (\frac{n}{2},\frac{n}{2})$ and 
${\bf T} (\frac{n-2}{2},\frac{n-2}{2})$, respectively. 

Finally,
completing the twist decomposition of the symmetric nonlocal tensor 
operator, we write down the symmetric twist--4 light--cone
tensor operator contained in the expression (\ref{G-high(i)}) 
\begin{align}
G_{(\alpha\beta)}^{\mathrm{tw4(i)}}(\kappa_1\lcx,\kappa_2\lcx)
&=G_{(\alpha\beta)}^{\mathrm{tw4(i)a}}(\kappa_1\lcx,\kappa_2\lcx)
+G_{(\alpha\beta)}^{\mathrm{tw4(i)b}}(\kappa_1\lcx,\kappa_2\lcx)
\nonumber\\
&=
\int_0^1\d\lambda\Big\{(1-\lambda+\lambda\ln\lambda)
\left(\hbox{\large$\frac{1}{2}$}
g_{\alpha\beta}+x_{(\alpha}\pd_{\beta)}\right)\square
\\
&-\hbox{\large$\frac{1}{4}$}
\Big(\hbox{\large$\frac{(1-\lambda)^2}{\lambda}$}
-\hbox{\large$\frac{1-\lambda^2}{2\lambda}$}-\lambda\ln\lambda\Big)
x_\alpha x_\beta\square^2\Big\}
\left.G(\kappa_1\lambda x,\kappa_2\lambda x)\right|_{x=\tilde{x}}\,,
\nonumber
\end{align}
whereas the symmetric twist--6 light--cone operator contained 
in (\ref{G-high(i)}) reads
\begin{align}
G_{(\alpha\beta)}^{\mathrm{tw6(i)}}(\kappa_1\lcx,\kappa_2\lcx)
&=G_{(\alpha\beta)}^{\mathrm{tw6(i)a}}(\kappa_1\lcx,\kappa_2\lcx)
+G_{(\alpha\beta)}^{\mathrm{tw6(i)b}}(\kappa_1\lcx,\kappa_2\lcx)\\
&=
\hbox{\large$\frac{1}{4}$}x_\alpha x_\beta\square^2
\int_0^1\d\lambda
\Big( \hbox{\large$\frac{1-\lambda^2}{2\lambda}$}+\ln\lambda\Big)
\left.G(\kappa_1\lambda x,\kappa_2\lambda x)\right|_{x=\tilde{x}}\nonumber\, .
\end{align}
%%\noindent
%%%{\it 3.1.4~~
\subsubsection{Vector and scalar light--ray operators}
%%\subsubsection{Vector and scalar operators of symmetry class (i)}
%%Let us now evaluate the twist--2 light--ray vector operator. 
Contracting eq.~(\ref{Gtw2_gir}) with $\lcx^\beta$, making use of formula
$(x\pd_x)f(\lambda x)=\lambda\pd_\lambda f(\lambda x)$ and performing the 
partial integrations we obtain the final version of the {\em twist--2 
light--cone vector operator},
\begin{equation}
\label{G_tw2_vec}
G^{\mathrm{tw2}}_{(\alpha\bullet)}(\kappa_1\lcx,\kappa_2\lcx)
=\int_0^1\d\lambda\, \lambda
\Big[\pd_\alpha+\hbox{\large$\frac{1}{2}$}(\ln\lambda)x_\alpha \square\Big]
G(\kappa_1\lambda x,\kappa_2\lambda x)\big |_{x=\tilde{x}}\,,
\end{equation}
which already has been used in \cite{BGR99},
and the twist--4 light--cone vector operator,
\begin{equation}
\label{G4i}
G^{\mathrm{tw4(i)}}_{(\alpha\bullet)}(\kappa_1\lcx,\kappa_2\lcx)=
-\hbox{\large$\frac{1}{2}$}x_\alpha \square
\int_0^1\d\lambda\, \lambda(\ln\lambda)
G(\kappa_1\lambda x,\kappa_2\lambda x)\big |_{x=\tilde{x}}\, .
\end{equation}

In order to obtain the {\em scalar twist--2 light--ray operator} 
we multiply (\ref{G_tw2_vec}) by $\lcx^\alpha$. Then, 
the twist--4 part vanishes and the remaining twist--2 operator
restores the scalar operator (compare eq.~(\ref{G_sca})), 
\begin{equation}
G^{\mathrm{tw2}}(\kappa_1\lcx,\kappa_2\lcx)
=
G(\kappa_1\lcx,\kappa_2\lcx)\,.
\end{equation}
%This is the expression cited in Eq.~(\ref{G_sca}). 
Let us point to the fact that the trace of the original gluon
tensor, $g^{\alpha\beta} G_{\alpha\beta}(\ka\xx,\kb\xx)$, is a
twist--4 scalar operator. It is contained in 
$G^{\rm tw4(i)a}_{(\alpha\beta)}(\ka\xx,\kb\xx)$
as well as similar expressions occurring below, cf.~also 
eq.~(\ref{G_trace}).
%%%\medskip
%%%\noindent
%%%{\it 3.1.5~~
\subsubsection{Condition of tracelessness on the cone}
Finally, it should be remarked, that the conditions (\ref{cond}),
(\ref{cond2}) and (\ref{cond20}),
if translated into the corresponding ones containing derivatives 
with respect to $\xx$, 
no longer hold for the light--cone operators. 
This is clear because, by projecting onto the light--cone, part 
of the original structure of the operators has been lost.
Nevertheless, the conditions of tracelessness of the 
light--cone operators may be formulated by using the 
{\em interior derivative} on the light--cone \cite{Bargmann,Dobrev77}
which has been used extensively for the construction of local
conformal operators \cite{Dobrev76,Dobrev82b}.
In four dimensions it is given by
\begin{equation}
\d_\alpha\equiv
\big(1+\lcx\lcd\big)\lcd_\alpha
-\hbox{\large$\frac{1}{2}$}
\lcx_\alpha\lcd^2\quad\text{with}\quad
\lcd_\alpha\equiv\frac{\pd}{\pd\lcx^\alpha}\,,
\nonumber
\end{equation}
and has the following properties
\begin{equation}
\d^2 = 0\, ,\quad
[\d_\alpha,\d_\beta]=0
\quad\text{and}\quad
\d_\alpha\lcx^2=\lcx^2\big(\d_\alpha+2\lcd_\alpha\big)\,.
\nonumber
\end{equation}
Then, the conditions of tracelessness simplify, namely, they read:
\begin{align}
g^{\alpha\beta} 
G^{\mathrm{tw2}}_{(\alpha\beta)}(\ka\xx,\kb\xx)=0\,,
\qquad
%\text{and}\qquad
\d^\alpha G^{\mathrm{tw2}}_{(\alpha\beta)}(\ka\xx,\kb\xx)=0\,,
\nonumber
\end{align}
as well as
\begin{align}
\d^\alpha G^{\mathrm{tw2}}_{(\alpha\bullet)}(\ka\xx,\kb\xx)=0\,.
\nonumber
\end{align}

Analogous conditions hold for the light--cone operators of
definite twist in the case of symmetry classes (ii) -- (iv)
obtained below. 

In passing we remark that the interior derivative
may be used in defining more directly totally symmetric
local light--cone operators \cite{GL99}. For example it holds
\begin{align}
G^{\mathrm{tw2(i)}}_{(\alpha\beta)n} (\xx)
&=
\hbox{\large$\frac{1}{(n+2)^2(n+1)^2}$}\,
\d_\alpha\d_\beta G_{n+2}(\xx)\,,
\nonumber\\
G^{\mathrm{tw2(i)}}_{(\alpha\bullet)n} (\xx)
&=
\hbox{\large$\frac{1}{(n+2)^2}$}\,
\d_\alpha G_{n+2}(\xx)\,. 
\nonumber
\end{align}
However, in case of symmetry types (ii) -- (iv)  
the corresponding expressions should be more involved.

\subsection{Tensor operators of symmetry class (ii)}
%%%{\it 3.2.1~~
\subsubsection{Construction of nonlocal (anti)symmetric class-(ii)
operators of definite twist: Young tableau A}
%%%\phantom{3.2.1~~} twist: Young tableau A}\\
Now we consider tensor operators, and their contractions with $x$,
having symmetry class (ii) and whose local twist--3 parts are contained
in ${\bf T}\hbox{$(\frac{n+2}{2},\frac{n}{2})
\oplus {\bf T}(\frac{n}{2},\frac{n+2}{2})$}$.  Contrary to the 
totally symmetric case we have different possibilities to put 
the tensor indices into the corresponding Young pattern. Without
presupposing any symmetry of indices $\alpha$ and
$\beta$ we should start with the following Young tableau:
\\
\\
\unitlength0.5cm
\begin{picture}(30,1)
\linethickness{0.075mm}
\put(1,-1){\framebox(1,1){$\alpha$}}
\put(1,0){\framebox(1,1){$\mu_1$}}
\put(2,0){\framebox(1,1){$\mu_2$}}
\put(3,0){\framebox(3,1){$\ldots$}}
\put(6,0){\framebox(1,1){$\mu_n$}}
\put(7,0){\framebox(1,1){$\beta$}}
\put(8.5,0){$\stackrel{\wedge}{=}$}
\put(9.5,0)
{${\hbox{\large$\frac{2(n+1)}{n+2}$}}\!
\relstack{\alpha\mu_1}{\cal A}\;
\relstack{\beta\mu_1\ldots\mu_n}{\cal S} \!\!\!
F_\alpha^{\ \rho}(0)D_{\mu_1}\ldots D_{\mu_n}F_{\beta\rho}(0)
\! - \!\mathrm{trace~terms,} $}
\end{picture}
\\ \\
with normalizing factor $\alpha_{n+1}$. (The tableau with 
$\alpha\leftrightarrow\beta$ will be considered thereafter.)
Denoting this
symmetry behaviour by (iiA) we may write the local twist-3 tensor
operator as follows:
\begin{align}
%\hspace{-1cm}
%\lefteqn{
G^{\mathrm{tw3(iiA)}}_{\alpha\beta\mu_1\ldots\mu_n}
%\label{M_tw2_l}
&=\;
\hbox{\large$\frac{1}{n+2}$}
\Big\{
F_\alpha^{\ \rho}(0)D_{(\mu_1}\ldots D_{\mu_n)}F_{\beta\rho}(0)
+
F_{\alpha}^{\ \rho}(0)D_{\beta} D_{(\mu_2}\ldots
D_{\mu_n}F_{\mu_1)\rho}(0)
\nonumber\\
& \qquad\qquad
+\!\sum\limits_{l=2}^{n}\!
F_\alpha^{\ \rho}(0)
D_{(\mu_1}\ldots D_{\mu_{l-1}}D_{|\beta|} D_{\mu_{l+1}}
\ldots D_{\mu_n } F_{\mu_l)\rho}(0)
\nonumber\\
& \qquad\qquad
- (\alpha \leftrightarrow \mu_1)
%\!\sum\limits_{\alpha,l=2}^{n}\!
%F_{\mu_1}^{\ \rho}(0)
%D_{(\alpha}D_{\mu_2} \ldots D_{\mu_{l-1}}
%D_{|\beta|} D_{\mu_{l+1}}
%\ldots D_{\mu_n } F_{\mu_l)\rho}(0)\!
\Big\}-\mathrm{trace~terms}\, .%\qquad
\nonumber
\end{align}
Proceeding in the same manner as in the last Subsection we 
multiply by $x^{\mu_1}\ldots x^{\mu_n}$ and obtain:
\begin{align}
%\label{Mx_tw2_l}
G^{\mathrm{tw3(iiA)}}_{\alpha\beta n}(x)
&=
\hbox{\large$\frac{1}{n(n+2)}$}
\Big(\delta_\alpha^\mu (x\pd)
-x^\mu\pd_\alpha\Big)
\pd_\beta \tl G_{\mu\bullet|n+1}(x) 
\nonumber
\end{align}
with
\begin{align}
\tl G_{\mu\bullet|n+1}(x)
&\equiv x^\nu \tl G_{\mu\nu|n}(x)
\nonumber\\
&= \Big\{\delta^\alpha_\mu - 
\hbox{\large$\frac{1}{(n+2)^2}$}
\Big[x_\mu\pd^\alpha (x\pd) -
\hbox{\large$\frac{1}{2}$}x^2\pd_\mu\pd^\alpha\Big]\Big\}
H^{(4)}_{n+1}(x^2|\square)G_{\alpha\bullet|n+1}(x) \,.
\nonumber
\end{align}
Here, $\tl G_{\mu\bullet|n+1}(x)$ is the harmonic vector polynomial
of order $n+1$
(see, Eq.~(\ref{Proj}), Appendix A)
 and the symmetry behaviour of class (ii) is obtained 
through the differential operator in front of it.
Now, using 
$((n+2)n)^{-1}=\int_0^1\d\lambda(1-\lambda^2)\lambda^n/(2\lambda)$
and Euler's beta function we sum up to obtain the following nonlocal 
twist--3 tensor operator:
\begin{align}
\hspace{-1cm}
\label{G_tw3_nl_1}
G^{\mathrm{tw3(iiA)}}_{\alpha\beta} (0,\kappa x)
%&=
% \int_{0}^{1} \d\lambda
%\Big(\lambda\delta_\alpha^\mu
%-\hbox{\large$\frac{1-\lambda^2}{2\lambda}$}
%x^\mu\pd_\alpha\Big)
%\pd_\beta \tl G_{\mu\cdot}(0,\kappa \lambda x)\\
&=
\int_{0}^{1} \d\lambda\hbox{\large$\frac{1-\lambda^2}{2\lambda}$}
\Big(\delta_\alpha^\mu(x\pd)-x^\mu\pd_\alpha\Big)
\pd_\beta \tl G_{\mu\bullet}(0,\kappa \lambda x)\,,
%\qquad
\end{align}
with the nonlocal traceless (vector) operator
\begin{align}
\tl G_{\alpha\bullet}(0,\kappa x)
=&\,
G_{\alpha\bullet}(0,\kappa x)
+\sum_{k=1}^\infty\int_0^1\!\d t
\left(\frac{-x^2}{4}\right)^{\!k}\!
\frac{\square^k}{k!(k-1)!}
\left(\frac{1-t}{t}\right)^{\!k-1}\!\!
G_{\alpha\bullet}(0,\kappa t x)
\nonumber\\
\label{M2_tl}
& 
- \big[ x_\alpha\pd^\mu(x\pd)
- \hbox{\large$\frac{1}{2}$} x^2\pd_\alpha\pd^\mu\big]
\\
& 
\times
\sum_{k=0}^\infty
\int_0^1\!\d\tau\tau \int_0^1\!\d t\,t
\left(\frac{-x^2}{4}\right)^{\!k}
\frac{\square^k}{k!k!}
\left(\frac{1-t}{t}\right)^{\!k}\!
G_{\mu\bullet}(0,\kappa\tau t x)\ .
\nonumber
\end{align}

The following decomposition into symmetric and antisymmetric 
part
is useful for the further calculations
\begin{equation}
\label{G3A}
G^{\mathrm{tw3(iiA)}}_{\alpha\beta} (0,\kappa x)=
G^{\mathrm{tw3(iiA)}}_{[\alpha\beta]} (0,\kappa x)+
G^{\mathrm{tw3(iiA)}}_{(\alpha\beta)} (0,\kappa x)\, ,
\end{equation}
with
\begin{align}
\label{G3as}
G^{\mathrm{tw3(iiA)}}_{[\alpha\beta]} (0,\kappa x)
&=
\int_0^1\d\lambda\, \lambda\delta^\mu_{[\alpha}\pd_{\beta]} 
\tl G_{\mu\bullet}(0,\kappa \lambda x)\,,
\\
\label{G3sym}
G^{\mathrm{tw3(iiA)}}_{(\alpha\beta)} (0,\kappa x)
&=
\int_0^1\d\lambda\, \hbox{\large$\frac{1}{\lambda}$}
\Big(\delta^\mu_{(\alpha}\pd_{\beta)}
-\hbox{\large$\frac{1-\lambda^2}{2}$}
\pd_\alpha\pd_\beta x^\mu\Big) 
\tl G_{\mu\bullet}(0,\kappa \lambda x)\\
&=
\int_{0}^{1} \d\lambda\hbox{\large$\frac{1-\lambda^2}{2\lambda}$}
\Big(\delta_{(\alpha}^\mu(x\pd)-x^\mu\pd_{(\alpha}\Big)
\pd_{\beta)} \tl G_{\mu\bullet}(0,\kappa \lambda x)\,.
\end{align}

Again, these operators are harmonic tensor functions.
The conditions of tracelessness for the tensor operator are
\begin{eqnarray}
\label{M2harm}
g^{\alpha\beta} G^{\mathrm{tw3(iiA)}}_{\alpha\beta}(0,\kappa x) = 0\,,
&\quad&
\square G^{\mathrm{tw3(iiA)}}_{\alpha\beta}(0,\kappa x) = 0\,,
\nonumber\\
\pd^\alpha G^{\mathrm{tw3(iiA)}}_{\alpha\beta}(0,\kappa x) =0\,,
&\quad&
\pd^\beta G^{\mathrm{tw3(iiA)}}_{\alpha\beta}(0,\kappa x) =0\,.
\end{eqnarray}
%%%\medskip
%%%\noindent
%%%{\it 3.2.2~~ 
\subsubsection{Reduction to vector operators}
The corresponding twist--3 vector operator is obtained 
from eq.~(\ref{G_tw3_nl_1}) by multiplication with $x^\beta$: 
\begin{equation}
\label{Mab2}
G^{\mathrm{tw3(iiA)}}_{\alpha\bullet}(0,\kappa x)
=
 \int_{0}^{1} \d\lambda\,\lambda
\Big(\delta_\alpha^\mu(x\pd)-x^\mu\pd_\alpha\Big)
\tl G_{\mu\bullet}(0,\kappa\lambda x).
\end{equation}
Obviously, a corresponding scalar operator does not exist.

This vector operator fulfils the following conditions of tracelessness
\begin{eqnarray}
\square G_{\alpha\bullet}^{\rm tw3(iiA)}(0,\kappa x) =0\,,
\qquad
\pd^\alpha G_{\alpha\bullet}^{\rm tw3(iiA)}(0,\kappa x)=0\,.
\end{eqnarray}
%%%\medskip
%%%\noindent
%%%{\it 3.2.3~~
\subsubsection{Projection onto the light--cone}
%%%Now, we project onto the light--cone. 
(a)~~The calculation of the {\em antisymmetric} tensor operator
$G^{\mathrm{tw3(iiA)}}_{[\alpha\beta]}(\ka\xx,\kb\xx)$ 
on the light--cone is similar to 
that of $M^{\mathrm{tw2}}_{[\alpha\beta]}(\ka\xx,\kb\xx)$
in Part I; for the details we refer to it. The resulting expression is:
\begin{eqnarray}
\label{M_tw2_ir}
\hspace{-.5cm}
G^{\mathrm{tw3(iiA)}}_{[\alpha\beta]} (\ka\xx,\kb\xx)
\!\!\!&=&\!\!\!\!
 \int_{0}^{1}\!\!\!\d\lambda\,\lambda
\delta^\mu_{[\alpha}\pd_{\beta]}
G_{\mu\bullet}(\ka\lambda\xx,\kb\lambda\xx)
\big|_{x=\tilde{x}}
\!-\!
G^{\mathrm{>(iiA)}}_{[\alpha\beta]} (\ka\xx,\kb\xx),
\\
\label{M_hi.tw_a}
\hspace{-.5cm}
G^{\mathrm{>(iiA)}}_{[\alpha\beta]} (\ka\xx,\kb\xx)
\!\!\!&=&\!\!\!
\hbox{\large$\frac{1}{2}$}
\!\! \int_{0}^{1}\!\!\d\lambda(1-\lambda)
\Big\{\big(2 x_{[\alpha}\pd_{\beta]}\pd^\mu
- 
x_{[\alpha}\delta_{\beta]}^\mu\square\big)
G_{\mu\bullet}(\ka\lambda x,\kb\lambda x) \big|_{x=\tilde{x}}
\nonumber\\
&& \!\!\!
-(1-\lambda+\lambda\ln\lambda)x_{[\alpha}\pd_{\beta]}\square 
G(\ka\lambda x,\kb\lambda x)\big|_{x=\tilde{x}}\Big\}\,.
\end{eqnarray}
$G^{\mathrm{>(iiA)}}_{[\alpha\beta]} (\ka\xx,\kb\xx)$ 
contains twist--4 and twist--5 contributions, but
twist--6 contributions do not appear due to $x_{[\alpha}x_{\beta]}=0$.
The higher twist operator (\ref{M_hi.tw_a}) contains two
%$G^{\mathrm{higher}}_{[\alpha\beta]} (0,\kappa\xx)$ 
vector operators multiplied by $x_\alpha$ and $x_\beta$, respectively.
For their twist decomposition one has to take into account Young pattern (i) 
as well as (ii). The procedure is analogous to the decomposition 
of the vector operator $O_\alpha(\ka\xx,\kb\xx)$ made
in Part~I. After a straightforward calculation we obtain
\begin{align}
%\label{M_tw2_b}
G^{\mathrm{>(iiA)}}_{[\alpha\beta]}
(\kappa_1\tilde{x},\kappa_2\tilde{x})
=&\,
G^{\mathrm{tw4(iiA)a}}_{[\alpha\beta]}(\kappa_1\tilde{x},\kappa_2\tilde{x})+
G^{\mathrm{tw4(iiA)b}}_{[\alpha\beta]}(\kappa_1\tilde{x},\kappa_2\tilde{x})
\nonumber\\
&+
G^{\mathrm{tw5(iiA)a}}_{[\alpha\beta]}(\kappa_1\tilde{x},\kappa_2\tilde{x})
%\nonumber
\end{align}
with
\begin{eqnarray}
\hspace{-.5cm}
\label{G_tw4_iiA}
G^{\mathrm{tw4(iiA)a}}_{[\alpha\beta]}
(\kappa_1\tilde{x},\kappa_2\tilde{x})
\!\!\!&=&\!\!\!
\hbox{\large$\frac{1}{2}$}
x_{[\alpha}\pd_{\beta]}\pd^{\mu} %x^\nu
\int_{0}^{1}\d\lambda\,
\hbox{\large$\frac{1-\lambda^2}{\lambda}$}
G_{\mu\bullet}(\kappa_1\lambda x,\kappa_2\lambda x)
\big|_{x=\xx}\,,
\nonumber\\
\hspace{-.5cm}
G^{\mathrm{tw4(iiA)b}}_{[\alpha\beta]}
(\kappa_1\tilde{x},\kappa_2\tilde{x})
\!\!\!&=&\!\!\!
- \hbox{\large$\frac{1}{2}$}x_{[\alpha}\pd_{\beta]}\square
\int_{0}^{1}\d\lambda
\Big(
\hbox{\large$\frac{1-\lambda^2}{2\lambda}$}+\lambda\ln\lambda
\Big)
G(\kappa_1\lambda x,\kappa_2\lambda x)
\big|_{x=\xx}\,,
\nonumber\\
\hspace{-.5cm}
G^{\mathrm{tw5(iiA)a}}_{[\alpha\beta]}
(\kappa_1\tilde{x},\kappa_2\tilde{x})
\!\!\!&=&\!\!\!
\label{M_tw5_ii}
- 
x_{[\alpha}\big(\delta_{\beta]}^{\mu}(x\pd)
-x^{\mu}\pd_{\beta]}\big)\square\!\! %x^\nu
\int_{0}^{1}\!\!\!\d\lambda 
\hbox{\large$\frac{(1\!-\!\lambda)^2}{4\lambda}$}
G_{\mu\bullet}(\kappa_1\lambda x,\kappa_2\lambda x)
\big|_{x=\xx}\,.
\nonumber
\end{eqnarray}
(b)~~Now, we determine the {\em symmetric} tensor operator
$G^{\mathrm{tw3(iiA)}}_{(\alpha\beta)}(\ka\xx,\kb\xx)$
on the light--cone. Putting
eq.~(\ref{M2_tl}) into (\ref{G3sym}), after some lengthy 
but straightforward calculation (taking into account  only the 
relevant terms of the $k$-summation and performing some
partial integrations), we get the following result:
\begin{align}
G^{\mathrm{tw3(iiA)}}_{(\alpha\beta)} (\ka\xx,\kb\xx)
=&
\int_0^1\d\lambda 
\hbox{\large$\frac{1-\lambda^2}{2\lambda}$}
\Big(\delta^\mu_{(\alpha}(x\pd)
-x^\mu\pd_{(\alpha}\Big)\pd_{\beta)} %x^\nu
G_{\mu\bullet} (\lambda x,\kb \lambda x)\big|_{x=\xx}
\nonumber\\
&-G^{\mathrm{>(iiA)}}_{(\alpha\beta)} (\ka\xx,\kb\xx)\, ,
\end{align}
with the higher twist contributions of the trace terms
\begin{align}
\label{G_high_(ii)_sy}
G^{\mathrm{>(iiA)}}_{(\alpha\beta)} (\ka\xx,&\kb\xx)
=
\int_{0}^{1}\!\!\d\lambda\Big\{\Big[
\hbox{\large$\frac{1-\lambda^2}{2\lambda}$}
g_{\alpha\beta}\pd^\mu
+\hbox{\large$\frac{1-\lambda}{2\lambda}$}
\delta^\mu_{(\alpha}x_{\beta)}\square
+(1-\lambda)x_{(\alpha}\pd_{\beta)}\pd^\mu\nonumber\\
&-\hbox{\large$\frac{(1-\lambda)^2}
{4\lambda}$}x_\alpha x_\beta\pd^\mu\square
\Big] G_{\mu\bullet} (\ka\lambda\xx,\kb\lambda\xx)
-\Big[\hbox{\large$\frac{1}{2}$}\Big(
\hbox{\large$\frac{1-\lambda^2}{2\lambda}$}
+\lambda\ln\lambda\Big)
g_{\alpha\beta}\square
\nonumber\\
&+\Big(\hbox{\large$\frac{1}{2}$}(1-\lambda)+
\hbox{\large$\frac{1-\lambda^2}{4\lambda}$}
+\lambda\ln\lambda\Big)
x_{(\alpha}\pd_{\beta)}\square\nonumber\\
&+\hbox{\large$\frac{1}{4}$}\Big(
\hbox{\large$\frac{1-\lambda^2}{2\lambda}$}
-\hbox{\large$\frac{(1-\lambda)^2}{\lambda}$}
+\lambda\ln\lambda\Big)
x_\alpha x_\beta\square^2\Big] G(\ka\lambda x,\kb\lambda x)
\Big\}\Big|_{x=\tilde{x}} .
\end{align}
It is obvious that eq.~(\ref{G_high_(ii)_sy}) contains scalar and vector
operators.
Again, using Young patterns (i) and (ii) and subtracting the trace terms, 
we can decompose the vector part appearing in eq.~(\ref{G_high_(ii)_sy}) 
into twist--4, twist--5 and twist--6 operators. This procedure is analogous
to the twist decomposition of the (vector) quark operator 
$O_\alpha(\ka\lcx,\kb\lcx)$ in Part I.
From the twist--4 and twist--5 vector part we have to subtract their trace terms 
being of twist--6  and add it to the other twist--6 scalar operator.
Let us recall that the scalar twist--4 and twist--6 operators are already 
traceless on the cone.
In that manner we get the following decomposition:
\begin{align}
%\label{M_tw2_b}
G^{\mathrm{>(iiA)}}_{(\alpha\beta)} (\ka\xx,\kb\xx)
&=\!
G^{\mathrm{tw4(iiA)a}}_{(\alpha\beta)} (\ka\xx,\kb\xx)\!+\!
G^{\mathrm{tw4(iiA)b}}_{(\alpha\beta)} (\ka\xx,\kb\xx)\!+\!
G^{\mathrm{tw4(iiA)c}}_{(\alpha\beta)} (\ka\xx,\kb\xx)
\nonumber\\
&\!\!\!\!\!\!\!\!\!+
G^{\mathrm{tw4(iiA)d1}}_{(\alpha\beta)} (\ka\xx,\kb\xx)+
G^{\mathrm{tw4(iiA)d2}}_{(\alpha\beta)} (\ka\xx,\kb\xx)+
G^{\mathrm{tw4(iiA)e}}_{(\alpha\beta)} (\ka\xx,\kb\xx)
\nonumber\\
&\!\!\!\!\!\!\!\!\!+
G^{\mathrm{tw5(iiA)d}}_{(\alpha\beta)} (\ka\xx,\kb\xx)+
G^{\mathrm{tw6(iiA)a}}_{(\alpha\beta)} (\ka\xx,\kb\xx)+
G^{\mathrm{tw6(iiA)b}}_{(\alpha\beta)} (\ka\xx,\kb\xx)
\nonumber\\
&\!\!\!\!\!\!\!\!\!+
G^{\mathrm{tw6(iiA)c}}_{(\alpha\beta)}(\ka\xx,\kb\xx)+
G^{\mathrm{tw6(iiA)d1}}_{(\alpha\beta)} (\ka\xx,\kb\xx)+
G^{\mathrm{tw6(iiA)d2}}_{(\alpha\beta)} (\ka\xx,\kb\xx)
\nonumber\\
&\!\!\!\!\!\!\!\!\!+
G^{\mathrm{tw6(iiA)d3}}_{(\alpha\beta)} (\ka\xx,\kb\xx)+
G^{\mathrm{tw6(iiA)d4}}_{(\alpha\beta)} (\ka\xx,\kb\xx)+
G^{\mathrm{tw6(iiA)e}}_{(\alpha\beta)} (\ka\xx,\kb\xx),
\end{align}
with  
\begin{align}
G_{(\alpha\beta)}^{\mathrm{tw4(iiA)a}}(\ka\xx,\kb\xx)
=&-\hbox{\large$\frac{1}{2}$} g_{\alpha\beta}\square
\int_0^1\d\lambda
\Big(%%\hbox{\large$\frac{5}{4}$}
\hbox{\large$\frac{1-\lambda^2}{2\lambda}$}
+\lambda\ln\lambda\Big)
G(\ka\lambda x,\kb\lambda x)\big|_{x=\tilde{x}}\,,
\nonumber\\
G_{(\alpha\beta)}^{\mathrm{tw4(iiA)b}}(\ka\xx,\kb\xx)
=&\,- \hbox{\large$\frac{1}{2}$} x_{(\alpha}\pd_{\beta)}\square
\int_0^1\d\lambda\Big(
(1-\lambda)+\hbox{\large$\frac{1-\lambda^2}{2\lambda}$}
\nonumber\\&
+2\lambda\ln\lambda\Big)
G(\ka\lambda x,\kb\lambda x)\big|_{x=\tilde{x}}
%%%\nonumber\\&
-G_{(\alpha\beta)}^{\mathrm{tw6(iiA)b}}(\ka\xx,\kb\xx)\,,
\nonumber\\
G_{(\alpha\beta)}^{\mathrm{tw4(iiA)d1}}(\ka\xx,\kb\xx)
=&\,- \hbox{\large$\frac{1}{2}$} x_{(\alpha}\pd_{\beta)}\square
\int_0^1\d\lambda\Big(
\hbox{\large$\frac{1-\lambda}{\lambda}$}
+\hbox{\large$\frac{\ln\lambda}{\lambda}$}\Big)
G(\ka\lambda x,\kb\lambda x)\big|_{x=\tilde{x}}
\nonumber\\
&-G_{(\alpha\beta)}^{\mathrm{tw6(iiA)d1}}(\ka\xx,\kb\xx)\,,
\nonumber
\end{align}
and
\begin{align}
G_{(\alpha\beta)}^{\mathrm{tw6(iiA)a}}(\ka\xx,\kb\xx)
=&
\hbox{\large$\frac{1}{4}$}x_\alpha x_\beta\square^2\!\!
\int_0^1\!\!\!\d\lambda
\Big( 
\hbox{\large$\frac{(1-\lambda)^2}{\lambda}$}
- \hbox{\large$\frac{1-\lambda^2}{2\lambda}$}
- \lambda\ln\lambda\Big)\!
G(\ka\lambda x,\kb\lambda x)\big|_{x=\tilde{x}}\,,
\nonumber\\
G_{(\alpha\beta)}^{\mathrm{tw6(iiA)b}}(\ka\xx,\kb\xx)
=&
-\hbox{\large$\frac{1}{4}$}x_\alpha x_\beta\square^2
\int_0^1\d\lambda
\Big( \hbox{\large$\frac{(1-\lambda)^2}{2\lambda}$}
-\hbox{\large$\frac{3}{2}$}
\hbox{\large$\frac{1-\lambda^2}{2\lambda}$}
\nonumber\\
&\qquad\qquad\qquad\qquad
-\lambda\ln\lambda
-\hbox{\large$\frac{\ln\lambda}{2\lambda}$}\Big)
G(\ka\lambda x,\kb\lambda x)\big|_{x=\tilde{x}}\,,
\nonumber\\
G_{(\alpha\beta)}^{\mathrm{tw6(iiA)d1}}(\ka\xx,\kb\xx)
=&
\hbox{\large$\frac{1}{4}$}x_\alpha x_\beta\square^2
\int_0^1\d\lambda
\Big( \hbox{\large$\frac{1-\lambda}{\lambda}$}
+\hbox{\large$\frac{\ln\lambda}{\lambda}$}
+\hbox{\large$\frac{\ln^2\lambda}{2\lambda}$}\Big)
G(\ka\lambda x,\kb\lambda x)\big|_{x=\tilde{x}}\,,
\nonumber\\
G_{(\alpha\beta)}^{\mathrm{tw6(iiA)d3}}(\ka\xx,\kb\xx)
=&
-\hbox{\large$\frac{1}{2}$}x_\alpha x_\beta\square^2
\int_0^1\d\lambda
\Big( \hbox{\large$\frac{1-\lambda}{\lambda}$}
+\hbox{\large$\frac{\ln\lambda}{\lambda}$}
+\hbox{\large$\frac{\ln^2\lambda}{2\lambda}$}\Big)
G(\ka\lambda x,\kb\lambda x)\big|_{x=\tilde{x}}\,,
\nonumber
\end{align}
being related to the scalar operator $G(\ka\xx,\kb\xx)$, 
as well as
\begin{align}
G_{(\alpha\beta)}^{\mathrm{tw4(iiA)c}}(\ka\xx,\kb\xx)
=&\hbox{\large$\frac{1}{2}$} g_{\alpha\beta}\pd^\mu 
\int_0^1\d\lambda
\hbox{\large$\frac{1-\lambda^2}{\lambda}$}
G_{\mu\bullet}(\ka\lambda x,\kb\lambda x)\big|_{x=\tilde{x}}\,,
\nonumber\\
G_{(\alpha\beta)}^{\mathrm{tw4(iiA)d2}}(\ka\xx,\kb\xx)
=&\,x_{(\alpha}\pd_{\beta)}\pd^\mu 
\int_0^1\d\lambda\Big(
\hbox{\large$\frac{1-\lambda}{\lambda}$}
+\hbox{\large$\frac{\ln\lambda}{\lambda}$}\Big)
G_{\mu\bullet}(\ka\lambda x,\kb\lambda x)\big|_{x=\tilde{x}}
\nonumber\\
&-G_{(\alpha\beta)}^{\mathrm{tw6(iiA)d2}}(\ka\xx,\kb\xx)\,,
\nonumber\\
%\nonumber\\
G_{(\alpha\beta)}^{\mathrm{tw4(iiA)e}}(\ka\xx,\kb\xx)
=&\,x_{(\alpha}\pd_{\beta)}\pd^\mu 
\int_0^1\d\lambda(1-\lambda)
G_{\mu\bullet}(\ka\lambda x,\kb\lambda x)\big|_{x=\tilde{x}}
\nonumber\\
&-G_{(\alpha\beta)}^{\mathrm{tw6(iiA)e}}(\ka\xx,\kb\xx)\,,
\nonumber
\end{align}
and
%%\intertext{and}
\begin{align}
G_{(\alpha\beta)}^{\mathrm{tw5(iiA)}}(\ka\xx,\kb\xx)
=&
-\hbox{\large$\frac{1}{2}$}
x_{(\alpha}\big(\delta^\mu_{\beta)}(x\pd)-x^\mu\pd_{\beta)}\big)
\square 
\nonumber\\
&\qquad\qquad
\times\int_0^1\d\lambda\Big(
\hbox{\large$\frac{1-\lambda}{\lambda}$}
+\hbox{\large$\frac{\ln\lambda}{\lambda}$}\Big)
G_{\mu\bullet}(\ka\lambda x,\kb\lambda x) \big |_{x=\tilde{x}}\,,
\nonumber\\
&-G_{(\alpha\beta)}^{\mathrm{tw6(iiA)d3}}(\ka\xx,\kb\xx)
-G_{(\alpha\beta)}^{\mathrm{tw6(iiA)d4}}(\ka\xx,\kb\xx)\,,
\nonumber
\end{align}
and
%%\intertext{and}
\begin{align}
G_{(\alpha\beta)}^{\mathrm{tw6(iiA)c}}(\ka\xx,\kb\xx)
\!=&
-\hbox{\large$\frac{1}{4}$}x_\alpha x_\beta\square\pd^\mu 
\int_0^1\d\lambda
\hbox{\large$\frac{(1-\lambda)^2}{\lambda}$}
G_{\mu\bullet}(\ka\lambda x,\kb\lambda x)\big|_{x=\tilde{x}}\,,
\nonumber\\
G_{(\alpha\beta)}^{\mathrm{tw6(iiA)d2}}(\ka\xx,\kb\xx)
\!=&
-\!\hbox{\large$\frac{1}{2}$}x_\alpha x_\beta\square\pd^\mu 
\!\!\int_0^1\!\!\!\d\lambda\Big(\!
\hbox{\large$\frac{1-\lambda}{\lambda}$}
\!+\!\hbox{\large$\frac{\ln\lambda}{\lambda}$}
\!+\!\hbox{\large$\frac{\ln^2\lambda}{2\lambda}$}\!\Big)\!
G_{\mu\bullet}(\ka\lambda x,\kb\lambda x)\big|_{x=\tilde{x}}\,,
\nonumber\\
G_{(\alpha\beta)}^{\mathrm{tw6(iiA)d4}}(\ka\xx,\kb\xx)
\!=&
-\!\hbox{\large$\frac{1}{2}$} x_\alpha x_\beta\square\pd^\mu 
\!\int_0^1\!\d\lambda\Big(
\hbox{\large$\frac{1-\lambda}{\lambda}$}
+\hbox{\large$\frac{\ln\lambda}{\lambda}$}
\Big)\!
G_{\mu\bullet}(\ka\lambda x,\kb\lambda x)\big|_{x=\tilde{x}}\,,
\nonumber\\
G_{(\alpha\beta)}^{\mathrm{tw6(iiA)e}}(\ka\xx,\kb\xx)
\!=&
\hbox{\large$\frac{1}{4}$}
x_\alpha x_\beta\square\pd^\mu  
\int_0^1\d\lambda
\hbox{\large$\frac{(1-\lambda)^2}{\lambda}$}
G_{\mu\bullet}(\ka\lambda x,\kb\lambda x)\big|_{x=\tilde{x}}\,,
\nonumber
\end{align}
being related to the vector operator 
$G_{\mu\bullet}(\ka\xx,\kb\xx)$.
%%%\medskip
%%%\noindent
%%%{\it 3.2.4~~
\subsubsection{Contributions of Young tableau B}
Let us now consider the other Young tableau which is obtained from the 
former one by exchanging $\alpha$ and $\beta$:
\\
\\
\unitlength0.5cm
\begin{picture}(30,1)
\linethickness{0.075mm}
\put(1,-1){\framebox(1,1){$\beta$}}
\put(1,0){\framebox(1,1){$\mu_1$}}
\put(2,0){\framebox(1,1){$\mu_2$}}
\put(3,0){\framebox(3,1){$\ldots$}}
\put(6,0){\framebox(1,1){$\mu_n$}}
\put(7,0){\framebox(1,1){$\alpha$}}
\put(8.5,0){$\stackrel{\wedge}{=}$}
\put(9.5,0)
{${\hbox{\large$\frac{2(n+1)}{n+2}$}}\!\!
\relstack{\beta\mu_1}{\cal A}\;
\relstack{\alpha\mu_1\ldots\mu_n}{\cal S} \!\!\!
F_\alpha^{\ \rho}(0)D_{\mu_1}\ldots D_{\mu_n}\!F_{\beta\rho}(0)
\! - \!\mathrm{trace~terms.} $}
\end{picture}
\\ \\
This symmetry behaviour will be denoted by (iiB). The corresponding 
nonlocal twist--3 operator which we obtain after analogous 
calculations is (with self-explaining denotations):
\begin{equation}
\hspace{-1cm}
\label{M_tw2_nl}
G^{\mathrm{tw3(iiB)}}_{\alpha\beta} (0,\kappa x)
=
\int_{0}^{1} \d\lambda\hbox{\large$\frac{1-\lambda^2}{2\lambda}$}
\Big(\delta_\beta^\mu(x\pd)-x^\mu\pd_\beta\Big)
\pd_\alpha \tl G_{\bullet\mu}(0,\kappa \lambda x)\,.
\end{equation}
Its decomposition into symmetric and antisymmetric part yields
\begin{equation}
\label{G3B}
G^{\mathrm{tw3(iiB)}}_{\alpha\beta} (0,\kappa x)=
G^{\mathrm{tw3(iiB)}}_{[\alpha\beta]} (0,\kappa x)+
G^{\mathrm{tw3(iiB)}}_{(\alpha\beta)} (0,\kappa x)\, ,
\end{equation}
with
\begin{align}
G^{\mathrm{tw3(iiB)}}_{[\alpha\beta]} (0,\kappa x)
&=
\int_0^1\d\lambda\, \lambda\delta^\mu_{[\beta}\pd_{\alpha]}%x^\nu
\tl G_{\bullet\mu} (0,\kappa \lambda x)\,,
\\
G^{\mathrm{tw3(iiB)}}_{(\alpha\beta)} (0,\kappa x)
&=
\int_{0}^{1} \d\lambda\hbox{\large$\frac{1-\lambda^2}{2\lambda}$}
\Big(\delta_{(\alpha}^\mu(x\pd)-x^\mu\pd_{(\alpha}\Big)
\pd_{\beta)} %x^\nu
\tl G_{\bullet\mu}(0,\kappa \lambda x)\,.
\end{align}
The projection onto the light--cone and the
calculation of the higher twist operators contained in the trace terms
is completely analogous to case (A) and should be omitted here.
%%%\medskip
%%%\noindent
%%%{\it 3.2.5~~
\subsubsection{Construction of the complete twist--3 tensor operators
on the light--cone}
In order to obtain the complete twist--3 operator it is necessary to add 
both twist--3 operators (\ref{G3A}) and (\ref{G3B}) resulting from 
the Young patterns (iiA) and (iiB).
Since no further Young pattern contributes to twist--3
operators we omit the index (ii).
After projection onto the light--cone the final result is
\begin{equation}
\label{G3full}
G^{\mathrm{tw3}}_{\alpha\beta} (\ka\xx,\kb\xx) =
G^{\mathrm{tw3}}_{[\alpha\beta]} (\ka\xx,\kb\xx) +
G^{\mathrm{tw3}}_{(\alpha\beta)} (\ka\xx,\kb\xx) \, ,
\end{equation}
with the {\em antisymmetric twist--3 light--cone tensor 
operator}\footnote
{
Here, we used the abbreviation $G^-_{\mu}\equiv x^\nu G_{[\mu\nu]}$ 
instead of $G_{[\mu\bullet]}$ 
in order not to come into conflict with the antisymmetrization
of the indices $\alpha$ and $\beta$.} 
\begin{align}
\label{G3_anti}
G^{\mathrm{tw3}}_{[\alpha\beta]} (\ka\xx,\kb\xx)
=&
2\!\int_0^1\!\!\d\lambda\, \lambda\pd_{[\alpha}
G^-_{\beta]} (\ka\lambda x,\kb\lambda x)\big|_{x=\xx} 
\nonumber\\
&-G^{\mathrm{tw4(ii)}}_{[\alpha\beta]} (\ka\xx,\kb\xx) 
-G^{\mathrm{tw5(ii)}}_{[\alpha\beta]} (\ka\xx,\kb\xx) \, ,
\end{align}
where
\begin{align}
\label{G_tw4_ii}
G^{\mathrm{tw4(ii)}}_{[\alpha\beta]}(\ka\xx,\kb\xx) 
&=
x_{[\alpha}\pd_{\beta]}\pd^{\mu}
\int_{0}^{1}\d\lambda\,
\hbox{\large$\frac{1-\lambda^2}{\lambda}$}\
G^-_{\mu} (\ka\lambda x,\kb\lambda x) 
\big|_{x=\xx}\,,\\
G^{\mathrm{tw5(ii)}}_{[\alpha\beta]}(\ka\xx,\kb\xx) 
&=
\label{G_tw5_ii}
-x_{[\alpha}\big(\delta_{\beta]}^{\mu}(x\pd)
-x^{\mu}\pd_{\beta]}\big)\square 
\int_{0}^{1}\d\lambda\
\hbox{\large$\frac{(1-\lambda)^2}{2\lambda}$}
G^-_{\mu}(\ka\lambda x,\kb\lambda x)
\big|_{x=\xx}.
\end{align}
The  local twist--4 operators are contained in the 
tensor space ${\bf T} (\frac{n}{2},\frac{n}{2})$
and the local twist--5 operators are contained in the 
tensor space ${\bf T} (\frac{n}{2},\frac{n-2}{2})\oplus
{\bf T} (\frac{n-2}{2},\frac{n}{2})$. 

In turn, let us remark that the bilocal light--ray operator 
$G^{\mathrm{tw3}}_{[\alpha\beta]} (\ka\xx,\kb\xx)$ 
is the same as the antisymmetric tensor operator which obtains
from the Young tableau
\unitlength0.35cm
\begin{picture}(8.5,2)
\linethickness{0.05mm}
\put(1,0){\framebox(1,1){$\SC\alpha$}}
\put(1,1){\framebox(1,1){$\SC\beta$}}
\put(2,1){\framebox(1,1){$\SC\mu_1$}}
\put(3,1){\framebox(1,1){$\SC\mu_2$}}
\put(4,1){\framebox(3,1){$\SC\ldots$}}
\put(7,1){\framebox(1,1){$\SC\mu_n$}}
\end{picture}\,.
That tableau has been used in Part I for the computation of 
$M^{\mathrm{tw2}}_{[\alpha\beta]} (\ka\xx,\kb\xx)$.
Furthermore, let us note that the local operators contained in
$M^{\mathrm{tw2}}_{[\alpha\beta]} (\ka\xx,\kb\xx)$ and in
$G^{\mathrm{tw3}}_{[\alpha\beta]} (\ka\xx,\kb\xx)$ are in 
complete agreement with the local operators on the light--cone 
determined by Dobrev and Ganchev~\cite{Dobrev82b} by means of the
interior derivative.

From the above point of view our result for the antisymmetric 
part of twist--3 operator (\ref{G3full}) could be obtained much easier.
However, the {\em symmetric twist--3 light--cone tensor operator}
 is much more involved. It is given 
by\footnote
{
Similarly, we use the notation $G^+_\mu$ instead of $G_{(\mu\bullet)}$.
}
\begin{align}
\label{G3_symm}
G^{\mathrm{tw3}}_{(\alpha\beta)} (\ka\xx,\kb\xx)
=&
\int_{0}^{1} \d\lambda\hbox{\large$\frac{1-\lambda^2}{2\lambda}$}
\Big(\delta_{(\alpha}^\mu(x\pd)-x^\mu\pd_{(\alpha}\Big)
\pd_{\beta)} 
G^+_{\mu} (\ka\lambda x,\kb\lambda x) \big|_{x=\xx}
\nonumber\\
&-G^{\mathrm{>(ii)}}_{(\alpha\beta)} (\ka\xx,\kb\xx)\, ,
\end{align}
where $G^{\mathrm{>(ii)}}_{(\alpha\beta)} (\ka\xx,\kb\xx)$ includes
all the trace terms having higher twist; these operators are given by
\begin{align}
%\hspace{-.5cm}
G_{(\alpha\beta)}^{\mathrm{tw4(ii)a}} (\ka\xx,\kb\xx)
=&
-g_{\alpha\beta}\square
\int_0^1\d\lambda
\Big(
\hbox{\large$\frac{1-\lambda^2}{2\lambda}$}
+\lambda\ln\lambda\Big)
G(\ka\lambda x,\kb \lambda x)\big|_{x=\tilde{x}}\,,
\nonumber\\
%\hspace{-.5cm}
G_{(\alpha\beta)}^{\mathrm{tw4(ii)b}} (\ka\xx,\kb\xx)
=&
-x_{(\alpha}\pd_{\beta)}\square\!\!
\int_0^1\!\!\!\d\lambda
\Big(\!(1-\lambda)\!+\!
\hbox{\large$\frac{1\!-\!\lambda^2}{2\lambda}$}
\!+\!2\lambda\!\ln\lambda\!
\Big)
G(\ka\lambda x,\kb \lambda x)\big|_{x=\tilde{x}}
\nonumber\\
%\hspace{-.5cm}
&
-G_{(\alpha\beta)}^{\mathrm{tw6(ii)b}} (\ka\xx,\kb\xx)\,,
\nonumber\\
%\hspace{-.5cm}
G_{(\alpha\beta)}^{\mathrm{tw4(ii)d1}} (\ka\xx,\kb\xx)
=&
-x_{(\alpha}\pd_{\beta)}\square\!\!
\int_0^1\!\!\!\d\lambda
\Big(\!
 \hbox{\large$\frac{1\!-\!\lambda}{\lambda}$}
+\hbox{\large$\frac{\ln\lambda}{\lambda}$}\Big)
G(\ka\lambda x,\kb \lambda x)\big|_{x=\tilde{x}}
\nonumber\\
%\hspace{-.5cm}
&
-G_{(\alpha\beta)}^{\mathrm{tw6(ii)d1}} (\ka\xx,\kb\xx)\,,
\nonumber\\
%\end{align}
\intertext{and}
%\begin{align}
%\hspace{-.5cm}
G_{(\alpha\beta)}^{\mathrm{tw6(ii)a}} (\ka\xx,\kb\xx)
=&
\hbox{\large$\frac{1}{2}$}x_\alpha x_\beta\square^2\!\!
\int_0^1\!\!\!\d\lambda
\Big( 
 \hbox{\large$\frac{(1-\lambda)^2}{\lambda}$}
- \hbox{\large$\frac{1-\lambda^2}{2\lambda}$}
- \lambda\ln\lambda
\Big)
G(\ka\lambda x,\kb\lambda x)\big|_{x=\tilde{x}}\,,
\nonumber\\
%\hspace{-.5cm}
G_{(\alpha\beta)}^{\mathrm{tw6(ii)b}} (\ka\xx,\kb\xx)
=&
-\hbox{\large$\frac{1}{2}$}x_\alpha x_\beta\square^2
\int_0^1\d\lambda
\Big( 
 \hbox{\large$\frac{(1-\lambda)^2}{2\lambda}$}
-\hbox{\large$\frac{3}{2}$}
 \hbox{\large$\frac{1-\lambda^2}{2\lambda}$}
\nonumber\\
%\hspace{-.5cm}
&\qquad\qquad\qquad\qquad
-\lambda\ln\lambda
-\hbox{\large$\frac{\ln\lambda}{2\lambda}$}
\Big)
G(\ka\lambda x,\kb \lambda x)\big|_{x=\tilde{x}}\,,
\nonumber\\
%\hspace{-.5cm}
G_{(\alpha\beta)}^{\mathrm{tw6(ii)d1}} (\ka\xx,\kb\xx)
=&
\hbox{\large$\frac{1}{2}$}x_\alpha x_\beta\square^2
\int_0^1\d\lambda
\Big( 
 \hbox{\large$\frac{1-\lambda}{\lambda}$}
+\hbox{\large$\frac{\ln\lambda}{\lambda}$}
+\hbox{\large$\frac{\ln^2\lambda}{2\lambda}$}
\Big)
G(\ka\lambda x,\kb \lambda x)\big|_{x=\tilde{x}}\,,
\nonumber\\
G_{(\alpha\beta)}^{\mathrm{tw6(ii)d3}}(\ka\xx,\kb\xx)
=&
-x_\alpha x_\beta\square^2
\int_0^1\d\lambda
\Big( \hbox{\large$\frac{1-\lambda}{\lambda}$}
+\hbox{\large$\frac{\ln\lambda}{\lambda}$}
+\hbox{\large$\frac{\ln^2\lambda}{2\lambda}$}\Big)
G(\ka\lambda x,\kb\lambda x)\big|_{x=\tilde{x}}\,,
\nonumber
\end{align}
%%%
being related to the scalar operator $G(\ka x, \kb x)$, 
as well as
\begin{align}
G_{(\alpha\beta)}^{\mathrm{tw4(ii)c}} (\ka\xx,\kb\xx)
=& g_{\alpha\beta}\pd^\mu 
\int_0^1\d\lambda
\hbox{\large$\frac{1-\lambda^2}{\lambda}$}
G^{\mathrm{+}}_{\mu} (\ka\lambda x,\kb \lambda x)
\big|_{x=\tilde{x}}\,,
\nonumber\\
G_{(\alpha\beta)}^{\mathrm{tw4(ii)d2}} (\ka\xx,\kb\xx)
=&\,2 x_{(\alpha}\pd_{\beta)}\pd^\mu 
\int_0^1\d\lambda\Big(
\hbox{\large$\frac{1-\lambda}{\lambda}$}
+\hbox{\large$\frac{\ln\lambda}{\lambda}$}\Big)
G^{\mathrm{+}}_{\mu} (\ka\lambda x,\kb \lambda x)
\big|_{x=\tilde{x}}
\nonumber\\
&-G_{(\alpha\beta)}^{\mathrm{tw6(ii)d2}} (\ka\xx,\kb\xx)\,,
\nonumber\\
G_{(\alpha\beta)}^{\mathrm{tw4(ii)e}} (\ka\xx,\kb\xx)
=&\,2 x_{(\alpha}\pd_{\beta)}\pd^\mu 
\int_0^1\d\lambda(1-\lambda)
G^{\mathrm{+}}_{\mu} (\ka\lambda x,\kb \lambda x)
\big|_{x=\tilde{x}}
\nonumber\\
&-G_{(\alpha\beta)}^{\mathrm{tw6(ii)e}} (\ka\xx,\kb\xx)\,,
\nonumber
\end{align}
and
%%\intertext{and}
\begin{align}
G_{(\alpha\beta)}^{\mathrm{tw5(ii)}} (\ka\xx,\kb\xx)
=&
-x_{(\alpha}\big(\delta^\mu_{\beta)}(x\pd)\!-\!x^\mu\pd_{\beta)}\big)
\square 
\nonumber\\
&\qquad
\times
\int_0^1\!\!\d\lambda\Big(
 \hbox{\large$\frac{1-\lambda}{\lambda}$}
+\hbox{\large$\frac{\ln\lambda}{\lambda}$}\Big)
G^+_{\mu} (\ka\lambda x,\kb \lambda x)
\big|_{x=\tilde{x}}\,,
\nonumber\\
&-G_{(\alpha\beta)}^{\mathrm{tw6(ii)d3}}(\ka\xx,\kb\xx)
-G_{(\alpha\beta)}^{\mathrm{tw6(ii)d4}}(\ka\xx,\kb\xx)\,,
\nonumber
\end{align}
and
%%\intertext{and}
\begin{align}
G_{(\alpha\beta)}^{\mathrm{tw6(ii)c}} (\ka\xx,\kb\xx)
=&
-\hbox{\large$\frac{1}{2}$}x_\alpha x_\beta\square\pd^\mu 
\int_0^1\d\lambda
\hbox{\large$\frac{(1-\lambda)^2}{\lambda}$}
G^{\mathrm{+}}_{\mu} (\ka\lambda x,\kb \lambda x)
\big|_{x=\tilde{x}}\,,
\nonumber\\
G_{(\alpha\beta)}^{\mathrm{tw6(ii)d2}} (\ka\xx,\kb\xx)
=&
-\!x_\alpha x_\beta\square\pd^\mu 
\!\!\int_0^1\!\!\!\d\lambda\Big(\!
 \hbox{\large$\frac{1-\lambda}{\lambda}$}
\!+\!\hbox{\large$\frac{\ln\lambda}{\lambda}$}
\!+\!\hbox{\large$\frac{\ln^2\!\lambda}{2\lambda}$}\!\Big)
G^{\mathrm{+}}_{\mu} (\ka\lambda x,\kb \lambda x)
\big|_{x=\tilde{x}}\,,
\nonumber\\
G_{(\alpha\beta)}^{\mathrm{tw6(ii)d4}} (\ka\xx,\kb\xx)
=&
-x_\alpha x_\beta\square\pd^\mu 
\!\int_0^1\!\!\!\d\lambda\Big(
 \hbox{\large$\frac{1-\lambda}{\lambda}$}
+\hbox{\large$\frac{\ln\lambda}{\lambda}$}\Big)
G^{\mathrm{+}}_{\mu} (\ka\lambda x,\kb \lambda x)
\big|_{x=\tilde{x}}\,,
\nonumber\\
G_{(\alpha\beta)}^{\mathrm{tw6(ii)e}} (\ka\xx,\kb\xx)
=&
\hbox{\large$\frac{1}{2}$}
x_\alpha x_\beta\square\pd^\mu\!\! 
\int_0^1\!\!\!\d\lambda
\hbox{\large$\frac{(1-\lambda)^2}{\lambda}$}
G^{\mathrm{+}}_{\mu} (\ka\lambda x,\kb \lambda x)
\big|_{x=\tilde{x}}\,,
\nonumber
\end{align}
being related to the (symmetric) vector operator
$G^+_{\mu} (\ka x,\kb x)$.
%%\subsubsection{Vector operator of symmetry class (ii)}
%%%\medskip
%%%\noindent
%%%{\it 3.2.6~~
\subsubsection{Determination of complete light--cone vector operators}
Now we are able to determine the twist--3 vector operator resulting
from the tensor operator (\ref{G3full}) by multiplying with $x^\beta$.
Because neither symmetry type (iii) nor (iv) may contribute to the
vector operator this will be the final expression.
The {\em  twist--3 light--cone vector operator} consists of
two parts, one originating from the symmetric  and the
other from the antisymmetric tensor operator, 
\begin{align}
\label{Gtw3-vec}
&G^{\mathrm{tw3}}_\alpha(\kappa_1\lcx,\kappa_2\lcx)
=
G^{\mathrm{tw3}}_{(\alpha\bullet)}(\kappa_1\lcx,\kappa_2\lcx)
+
G^{\mathrm{tw3}}_{[\alpha\bullet]}(\kappa_1\lcx,\kappa_2\lcx)\,,
%\\ &\qquad  =
%\int_0^1\d\lambda\, \lambda
%\Big[\delta_\alpha^\mu(x\pd)-x^\mu\pd_\alpha
%-x_\alpha\big(\pd^\mu+(\ln\lambda) \square x^\mu\big)\Big]
%G_\mu(\kappa_1\lambda x,\kappa_2\lambda x)
%\big|_{x=\tilde{x}}\,,\nonumber
\end{align}
where
\begin{align}
\hspace{-.3cm}
\label{G_tw3_vec_sy}
G^{\mathrm{tw3}}_{(\alpha\bullet)}(\kappa_1\lcx,\!\kappa_2\lcx)
&=\!\int_0^1\!\!\!\d\lambda \lambda
\Big[\delta_\alpha^\mu(x\pd)\!-\!x^\mu\pd_\alpha
\!-\!x_\alpha\big(\pd^\mu\!+\!\ln\lambda \square x^\mu\big)\!\Big]
G^+_\mu(\kappa_1\lambda x,\!\kappa_2\lambda x)
\big|_{x=\tilde{x}}\,,
\\
\hspace{-.3cm}
G^{\mathrm{tw3}}_{[\alpha\bullet]}(\kappa_1\lcx,\!\kappa_2\lcx)
\label{G_tw3_vec_asy}
&=\!\int_0^1\!\!\d\lambda\, \lambda
\Big[\delta_\alpha^\mu(x\pd)\!-\!x^\mu\pd_\alpha
\!-\!x_\alpha\pd^\mu\Big]
G^-_\mu(\kappa_1\lambda x,\kappa_2\lambda x)
\big|_{x=\tilde{x}}\,,
\end{align}
and the twist--4 vector operator which is contained in the 
trace terms of that twist--3 vector operator is given by
\begin{align}
\label{G4ii}
G^{\mathrm{tw4(ii)}}_\alpha(\kappa_1\lcx,\kappa_2\lcx)
&=
G^{\mathrm{tw4(ii)}}_{(\alpha\bullet)}(\kappa_1\lcx,\kappa_2\lcx)
+
G^{\mathrm{tw4}}_{[\alpha\bullet]}(\kappa_1\lcx,\kappa_2\lcx)\,,
%\\&=x_\alpha 
%\int_0^1\d\lambda\,\lambda\big[\pd^\mu+(\ln\lambda)\square x^\mu\big]
%G_\mu(\kappa_1\lambda x,\kappa_2\lambda x)\big|_{x=\tilde{x}}\,,\nonumber
\end{align}
where
\begin{align}
\label{G_tw4_vec_sy(ii)}
G^{\mathrm{tw4(ii)}}_{(\alpha\bullet)}(\kappa_1\lcx,\kappa_2\lcx)
&=x_\alpha 
\int_0^1\d\lambda\,\lambda\big[\pd^\mu+(\ln\lambda)\square x^\mu\big]
G^+_\mu(\kappa_1\lambda x,\kappa_2\lambda x)\big|_{x=\tilde{x}}\,,
\\
G^{\mathrm{tw4}}_{[\alpha\bullet]}(\kappa_1\lcx,\kappa_2\lcx)
\label{G_tw4_vec_asy}
&=x_\alpha 
\int_0^1\d\lambda\,\lambda\,\pd^\mu
G^-_\mu(\kappa_1\lambda x,\kappa_2\lambda x)\big|_{x=\tilde{x}}\,.
\end{align}
The antisymmetric part of the twist--4 vector operator is already
complete. The complete {\em twist--4 light--cone vector operator} 
is given by 
\begin{align}
\label{G_tw4_vec}
G^{\mathrm{tw4}}_\alpha(\kappa_1\lcx,\kappa_2\lcx)
&=
G^{\mathrm{tw4}}_{(\alpha\bullet)}(\kappa_1\lcx,\kappa_2\lcx)+
G^{\mathrm{tw4}}_{[\alpha\bullet]}(\kappa_1\lcx,\kappa_2\lcx)
\nonumber\\
&= x_\alpha 
\int_0^1\d\lambda\,\lambda\big[\pd^\mu+
\hbox{\large$\frac{\ln\lambda}{2}$}\square x^\mu\big]
G_\mu(\kappa_1\lambda x,\kappa_2\lambda x)\big|_{x=\tilde{x}}\,,
\end{align}
where the symmetric twist--4 vector operator obtains by adding together
expressions (\ref{G4i}) and (\ref{G4ii}),
\begin{align}
\label{G_tw4_vec_sy}
G^{\mathrm{tw4}}_{(\alpha\bullet)}(\kappa_1\lcx,\kappa_2\lcx)
&=
G^{\mathrm{tw4(i)}}_{(\alpha\bullet)}(\kappa_1\lcx,\kappa_2\lcx)+
G^{\mathrm{tw4(ii)}}_{(\alpha\bullet)}(\kappa_1\lcx,\kappa_2\lcx)
\\
&=x_\alpha 
\int_0^1\d\lambda\,\lambda\big[\pd^\mu+
\hbox{\large$\frac{\ln\lambda}{2}$}\square x^\mu\big]
G^+_\mu(\kappa_1\lambda x,\kappa_2\lambda x)\big|_{x=\tilde{x}}\,.\nonumber
%\\G^{\mathrm{tw4}}_{[\alpha\bullet]}(\kappa_1\lcx,\kappa_2\lcx)
%&=G^{\mathrm{tw4(ii)}}_{[\alpha\bullet]}(\kappa_1\lcx,\kappa_2\lcx)\,.
\end{align}
Furthermore, putting together expressions 
(\ref{G_tw2_vec}), (\ref{G_tw3_vec_sy}), (\ref{G_tw4_vec_sy}) 
and (\ref{G_tw3_vec_asy}), (\ref{G_tw4_vec_asy}), respectively,
we obtain for the complete decomposition of the {\em light--cone vector operator} 
the following results anticipated by eqs.~(\ref{G^+_vec}) and (\ref{G^-_vec}),
respectively:
\begin{align}
G_{(\alpha\bullet)}(\kappa_1\lcx,\kappa_2\lcx)
&=
G^{\mathrm{tw2}}_{(\alpha\bullet)}(\kappa_1\lcx,\kappa_2\lcx)+
G^{\mathrm{tw3}}_{(\alpha\bullet)}(\kappa_1\lcx,\kappa_2\lcx)+
G^{\mathrm{tw4}}_{(\alpha\bullet)}(\kappa_1\lcx,\kappa_2\lcx)\\
G_{[\alpha\bullet]}(\kappa_1\lcx,\kappa_2\lcx)
&=G^{\mathrm{tw3}}_{[\alpha\bullet]}(\kappa_1\lcx,\kappa_2\lcx)+
G^{\mathrm{tw4(ii)}}_{[\alpha\bullet]}(\kappa_1\lcx,\kappa_2\lcx)
\end{align}

%\newpage
%%%%%%%%%%%%%%%%%%%%%%%%%%%%%%%%
\subsection{Tensor operators of symmetry class (iii)}
%%%{\it 3.3.1~~
\subsubsection{Construction of the nonlocal antisymmetric class-(iii)
operators of definite twist}
%%%\phantom{3.3.1~~} twist}
Now we consider the twist--4 and twist--5 contributions originating
from the (unique) Young tableau related to the symmetry class (iii)
(with the normalizing factor $\beta_n$):\\
%\pagebreak
\\
\unitlength0.5cm
\begin{picture}(30,1)
\linethickness{0.075mm}
\put(1,-2){\framebox(1,1){$\alpha$}}
\put(1,-1){\framebox(1,1){$\beta$}}
\put(1,0){\framebox(1,1){$\mu_1$}}
\put(2,0){\framebox(1,1){$\mu_2$}}
\put(3,0){\framebox(1,1){$\mu_3$}}
\put(4,0){\framebox(3,1){$\ldots$}}
\put(7,0){\framebox(1,1){$\mu_n$}}
\put(8.5,0){$\stackrel{\wedge}{=}$}
\put(9.5,0)
{$\hbox{\large$\frac{3n}{n+2}$}\!
\relstack{\alpha\beta\mu_1}{\cal A}\,
\relstack{\mu_1\ldots\mu_n}{\cal S}\!\!
F_\alpha^{\ \rho}(0)D_{\mu_1}\ldots
D_{\mu_n}F_{\beta\rho}(0)\! -\mathrm{trace~terms\,.} $}
\end{picture}
\\ \\ \\
The corresponding traceless local operator having twist $\tau = 4$ 
and being contained in the tensor space
${\bf T}\!\left(\frac{n}{2},\frac{n}{2}\right)$ is
given by:
\begin{eqnarray}
\label{M3}
G^{\mathrm{tw4(iii)}}_{\alpha\beta\mu_1\ldots\mu_n}
%&=&\frac{3n}{n+2}
%\relstack{\alpha\beta\mu_1}{\cal A}\,
%\relstack{\mu_1\ldots\mu_n}{\cal S}
%\bar{\psi}(0)\sigma_{\alpha\beta} D_{\mu_1}\ldots
%D_{\mu_n} \psi (0) -\mathrm{trace~terms} \nonumber \\
\!\!\!&=&\!\!\!
\hbox{\large$\frac{n}{n+2}$}
\Big(
F_{[\alpha|}^{\ \ \rho}(0)D_{(\mu_1}\ldots D_{\mu_n)} F_{|\beta]\rho}(0)- 
F_{[\alpha|}^{\ \ \rho}(0)D_{(\beta}\D_{\mu_2}
\ldots D_{\mu_n)} F_{|\mu_1]\rho}(0)
\nonumber\\
\label{M_tw3_l}
& & \qquad
-F_{[\mu_1|}^{\ \ \ \rho}(0) D_{(\alpha}D_{\mu_2}
\ldots D_{\mu_n) } F_{|\beta]\rho}(0)
\Big)
-\mathrm{trace~terms}.
\nonumber
\end{eqnarray}
Now, contracting this expression  
with $x^{\mu_1}\ldots x^{\mu_n}$ we obtain:
\begin{eqnarray}
\label{Mx_tw3_l}
G^{\mathrm{tw4(iii)}}_{[\alpha\beta] n}(x)
&=&
x^{\mu_1}\ldots x^{\mu_n}
G^{\mathrm{tw4}}_{[\alpha\beta]\mu_1\ldots\mu_n}  
\nonumber\\
&=&
\hbox{\large$\frac{1}{n+2}$}
\left(%\hbox{\large$\frac{1}{2}$}
(x\pd)\delta^\nu_{[\beta}
- 2 x^\nu\pd_{[\beta}\right)\delta_{\alpha]}^\mu
\tl G_{[\mu\nu] n}(x)\,.
     %-\mathrm{trace~terms}\Big)\,,
\nonumber
\end{eqnarray}
with the (antisymmetric) tensorial harmonic polynomials of order $n$
(see %Appendix \ref{trace}, 
eq.~(\ref{Proj5})):
\begin{align}
%\lefteqn{
\tl G_{[\alpha\beta] n}(x)
=&
\bigg\{\delta_\alpha^\mu\delta_\beta^\nu
+\hbox{\large$\frac{1}{(n+1)n}$}
\left(
2 x_{[\alpha}\delta_{\beta]}^{[\mu}\pd^{\nu]}(x\pd)
-
%%\hbox{\large$\frac{1}{2}$} 
x^2\pd_{[\alpha}\delta_{\beta]}^{[\mu}\pd^{\nu]}\right)\nonumber\\
&
-\hbox{\large$\frac{2}{(n+2)(n+1)n}$}\,
x_{[\alpha}\pd_{\beta]}x^{[\mu}\pd^{\nu]}\bigg\}
%}
H^{(4)}_n \!\big(x^2|\square\big)
G_{[\mu\nu] n}(x).
\nonumber
\end{align}

Resumming these local terms gives the nonlocal twist--4 operator 
as follows
\begin{eqnarray}
\label{M_tw3_nl_ir}
G^{\mathrm{tw4(iii)}}_{[\alpha\beta]}(0,\kappa x)
=
\int_0^1\d \lambda\,\lambda
\delta^\mu_{[\alpha}\left(
(x\pd)\delta^\nu_{\beta]}
- 2 x^\nu\pd_{\beta]}\right)
\tl G_{[\mu\nu]}(0,\kappa\lambda x)\, ,
\end{eqnarray}
where, using the integral representation of 
the additional factor $1/n$ (the remaining factors of the 
denominator are taken together with $1/n!$ to get $1/(n+2)!$) 
and of the beta function, we introduced:
\begin{eqnarray}
\label{G0anti}
%\lefteqn{
\hspace{-.6cm}
&&\tl G_{[\alpha\beta]}(0,\kappa x)
=
G_{[\alpha\beta]}(0,\kappa x)
+\sum_{k=1}^\infty\int_0^1\!
\frac{\d t}{t}
\!\left(\!\frac{-x^2}{4}\right)^{\!k}\!\!
\frac{\square^k}{k!(k-1)!}
\left(\!\frac{1-t}{t}\right)^{\!k-1}\!\!
G_{[\alpha\beta]}(0,\kappa t x)
%}\qquad
\nonumber\\
\hspace{-.6cm}
&&\quad
+\int_0^1\!\frac{\d\tau}{\tau}\bigg\{\!\!
\left(
2x_{[\alpha}\delta_{\beta]}^{[\mu}\pd^{\nu]}(x\pd)
- x^2\pd_{[\alpha}
\delta_{\beta]}^{[\mu}\pd^{\nu]}\right)
\!\sum_{k=0}^\infty
\int_0^1\!\!\d t
\left(\!\frac{-x^2}{4}\right)^{\!\!k}\!\!
\frac{\square^k}{k!k!}\!
\left(\!\frac{1-t}{t}\right)^{\!\!k}
\nonumber\\
\hspace{-.8cm}
&&\quad
-\,2x_{[\alpha}\pd_{\beta]}x^{[\mu}\pd^{\nu]}
\sum_{k=0}^\infty\!
\int_0^1\!\!\d t\,t\!
\left(\!\frac{-x^2}{4}\right)^{\!\!k}\!\!
\frac{\square^k}{(k+1)!k!}
\left(\!\frac{1-t}{t}\right)^{\!\!k+1}
\!\bigg\}
G_{[\mu\nu]}(0,\kappa\tau t x).\qquad\quad
\end{eqnarray}

Then, by construction, eq.~(\ref{M_tw3_nl_ir}) defines 
a nonlocal operator of twist--4. As is easily seen by partial 
integration it fulfils the following relation
\begin{equation}
\label{M_tw3}
G^{\rm tw4(iii)}_{[\alpha\beta]}(0,\kappa x)
=\tl G_{[\alpha\beta]}(0,\kappa x)
- G^{\rm tw3}_{[\alpha\beta]}(0,\kappa x)\, 
\end{equation}
and, because of eq.~(\ref{M2harm}) and the properties of 
(\ref{G0anti}), it is a harmonic tensor operator: 
\begin{equation}
\label{M0harm}
\square 
G^{\rm tw4(iii)}_{[\alpha\beta]}(0,\kappa x)=0\ ,
\qquad
\pd^\alpha G^{\rm tw4(iii)}_{[\alpha\beta]}(0,\kappa x)
=0\,.
%=\pd^\beta G^{\rm tw4(iii)}_{[\alpha\beta]}(0,\kappa x)\,.
\end{equation}
%%%\noindent
%%%{\it 3.3.2~~
\subsubsection{Projection onto the light--cone}
Let us now project onto the light--cone. After the same 
calculations as has been carried out for the operator
$M^{\rm tw3}_{[\alpha\beta]}(0,\kappa\lcx)$ in Part I %\cite{GLR990}
we obtain 
\begin{align}
\label{G_tw4_iii}
G^{\rm tw4(iii)}_{[\alpha\beta]}(\ka\xx,\kb\xx)
=&
\int_{0}^{1}\d\lambda\, \lambda
\delta^\mu_{[\alpha}\left( (x\pd)\delta_{\beta]}^\nu
- 2 x^\nu\pd_{\beta]}\right)
G_{[\mu\nu]}(\ka\lambda x,\kb\lambda x)\big|_{x=\xx}
\\
& 
- G^{\rm tw5(iii)}_{[\alpha\beta]}(\ka\xx,\kb\xx),
\nonumber
\end{align}
where the twist--5 part is determined by the trace
terms, namely
\begin{align}
\label{G_tw5_iii}
%\lefteqn{
%\hspace{-.5cm}
G^{\rm tw5(iii)}_{[\alpha\beta]}(\ka\xx,\kb\xx)
= 
& - \int_{0}^{1}\d\lambda 
\hbox{\large$\frac{1-\lambda^2}{\lambda}$}\Big\{
x_{[\alpha}\big(\delta_{\beta]}^{[\mu}(x\pd)
-x^{[\mu}\pd_{\beta]}\big)\pd^{\nu]}
\nonumber\\
&\qquad\qquad 
- x_{[\alpha}\delta_{\beta]}^{[\mu}x^{\nu]}\square\Big\}
G_{[\mu\nu]}(\ka\lambda x,\kb\lambda x)
\big|_{x=\xx}\,. 
\end{align}
Obviously, there exist no vector (and scalar) operators of
symmetry type (iii).
%%%\medskip
%%%\noindent
%%%{\it 3.3.3~~
\subsubsection{Determination of the complete antisymmetric
light--cone tensor operator}
This finishes the computation of twist contributions of the 
antisymmetric light--cone tensor operator. The complete 
{\em antisymmetric twist--4 light--cone tensor operator} obtains from
eq.~(\ref{G_tw4_iii}) and the twist--4 trace terms of the 
twist--3 operator, eq.~(\ref{G_tw4_ii}):
\begin{eqnarray}
\label{G_tw4_as}
\lefteqn{G^{\rm tw4}_{[\alpha\beta]}(\ka\xx,\kb\xx)
=
G^{\rm tw4(ii)}_{[\alpha\beta]}(\ka\xx,\kb\xx)
+
G^{\rm tw4(iii)}_{[\alpha\beta]}(\ka\xx,\kb\xx)}
\nonumber\\
&&=\!
\int_{0}^{1}\!\!\d\lambda\Big\{
\lambda\left(%\hbox{\large$\frac{1}{2}$}
(x\pd)\delta_{[\beta}^\nu
- 2 x^\nu\pd_{[\beta}\right)\delta_{\alpha]}^{\mu}
-\hbox{\large$\frac{1-\lambda^2}{\lambda}$}\Big(
x_{[\alpha}\pd_{\beta]}x^{[\mu}\pd^{\nu]}
\\
& &\quad
+ x_{[\alpha}\big(\delta_{\beta]}^{[\mu}(x\pd)
- x^{[\mu}\pd_{\beta]}\big)\pd^{\nu]}
- x_{[\alpha}\delta_{\beta]}^{[\mu}x^{\nu]}\square\Big)
\Big\}G_{[\mu\nu]}(\kappa_1\lambda x,\kappa_2\lambda x)
\big|_{x=\xx}.
\nonumber
\end{eqnarray}
Furthermore, the complete {\em  antisymmetric twist--5 
light--cone tensor operator} obtains from 
eqs.~(\ref{G_tw5_ii}) and (\ref{G_tw5_iii}) as follows:
\begin{eqnarray}
\label{G_tw5_as}
 \lefteqn{
G^{\rm tw5}_{[\alpha\beta]}(\ka\xx,\kb\xx)
=
G^{\rm tw5(ii)}_{[\alpha\beta]}(\ka\xx,\kb\xx)
+
G^{\rm tw5(iii)}_{[\alpha\beta]}(\ka\xx,\kb\xx)}
\\
&&=\!
 \int_{0}^{1}\d\lambda 
\hbox{\large$\frac{1-\lambda}{\lambda}$}\Big\{
x_{[\alpha}\delta_{\beta]}^{[\mu}x^{\nu]}\square
-2x_{[\alpha}\big(\delta_{\beta]}^{[\mu}(x\pd)
-x^{[\mu}\pd_{\beta]}\big)\pd^{\nu]}
\Big\}G_{[\mu\nu]}(\kappa_1\lambda x,\kappa_2\lambda x)
\big|_{x=\xx}.\nonumber
\end{eqnarray}
Together with the twist-3 part, eq.~(\ref{G3_anti}), we finally 
obtain the complete decomposition of the {\em antisymmetric 
light--cone tensor operator} (compare eq.~(\ref{G^-_t})):
\begin{equation}
G_{[\alpha\beta]}(\ka\xx,\kb\xx)=
G^{\rm tw3}_{[\alpha\beta]}(\ka\xx,\kb\xx)+
G^{\rm tw4}_{[\alpha\beta]}(\ka\xx,\kb\xx)+
G^{\rm tw5}_{[\alpha\beta]}(\ka\xx,\kb\xx)\, .
\end{equation}

\subsection{Tensor operators of symmetry class (iv)}
%%%{\it 3.4.1~~
\subsubsection{Construction of the nonlocal symmetric class-(iv) 
operator of definite twist}
Now let us investigate the symmetric tensor operators 
of the symmetry class (iv) which are given by the following Young tableau
(with normalizing factor $\gamma_{n}$):
\\
\\
\unitlength0.5cm
\begin{picture}(30,1)
\linethickness{0.075mm}
\put(1,-1){\framebox(1,1){$\alpha$}}
\put(2,-1){\framebox(1,1){$\beta$}}
\put(1,0){\framebox(1,1){$\mu_1$}}
\put(2,0){\framebox(1,1){$\mu_2$}}
\put(3,0){\framebox(1,1){$\mu_3$}}
\put(4,0){\framebox(3,1){$\ldots$}}
\put(7,0){\framebox(1,1){$\mu_n$}}
\put(8.5,0){$\stackrel{\wedge}{=}$}
\put(9.5,0)
{${\hbox{\large$\frac{4(n-1)}{n+1}$}}\!\!
\relstack{\alpha\mu_1}{\cal A}\,
\relstack{\beta\mu_2}{\cal A}\,
\relstack{\alpha\beta}{\cal S}\,
\relstack{\mu_1\ldots\mu_n}{\cal S} \!\!\!
F_\alpha^{\ \rho}(0)D_{\mu_1}\ldots D_{\mu_n}F_{\beta\rho}(0)
\! - \!\mathrm{traces\,.} $}
\end{picture}
\\ \\
These local tensor operators have twist--4 and transform according to
%the (reducible w.r.~to $SO(1,3)$) 
the representation
${\bf T}\hbox{$(\frac{n+2}{2},\frac{n-2}{2})\oplus
{\bf T}(\frac{n-2}{2},\frac{n+2}{2})$}$. 
They are given by:
\begin{align}
\hspace{-1cm}
%\lefteqn{
G^{\mathrm{tw4(iv)}}_{\alpha\beta\mu_1\ldots\mu_n}
%%\label{M_tw4_1}
=&\;
\hbox{\large$\frac{1}{(n+1)n}$}
\Big(G_{(\alpha\beta)(\mu_1\ldots\mu_n)}
-G_{(\alpha\mu_2)(\mu_1\beta\mu_3\ldots\mu_n)}
%\nonumber\\
-G_{(\mu_1\beta)(\alpha\mu_2\ldots\mu_n)}
\nonumber\\
\hspace{-1cm}
&\qquad\qquad
+G_{(\mu_1\mu_2)(\alpha\beta\mu_3\ldots\mu_n)}\Big)
%F_{|\mu_2)\rho}(0)D_{\mu_n)}F_{|\mu_2)\rho}(0)\Big)
%\nonumber\\
-\mathrm{trace~terms}\,,
\nonumber
\end{align}
with $G_{(\alpha\beta)(\mu_1\ldots\mu_n)}\equiv
F_{(\alpha|}^{\ \ \rho}(0)D_{(\mu_1}\ldots D_{\mu_n)} F_{|\beta)\rho}(0)$.
Multiplying with $x^{\mu_1}\ldots x^{\mu_n}$ we obtain:
\begin{align}
%\label{Gx_tw4_l}
G^{\mathrm{tw4(iv)}}_{(\alpha\beta) n}(x)
&=x^{\mu_1}\ldots x^{\mu_n}G^{\mathrm{tw4(iv)}}_{\alpha\beta\mu_1\ldots\mu_n}
\nonumber\\
&=\hbox{\large$\frac{1}{(n+1)n}$}
\Big(\delta_\alpha^\mu\delta_\beta^\nu x^\rho (x\pd)\pd_\rho
%-x^\mu\delta_\beta^\nu (x\pd)\pd_\alpha
%\nonumber\\
%&\qquad\qquad
-2x^\mu(x\pd) \delta_{(\alpha}^\nu \pd_{\beta)}
+x^\mu x^\nu\pd_\alpha\pd_\beta\Big)
\tl G_{(\mu\nu) n}(x)\, .
\nonumber
\end{align}
Here, $\tl G_{(\mu\nu) n}(x)$ is the harmonic symmetric tensor
polynomial of order $n$ (cf. Eq.~(\ref{Proj6}), Appendix~\ref{trace}):
\begin{align}
%\label{symTenpol}
\tl G_{(\alpha\beta) n}(x)
=&\;\Big\{ 
\delta_\alpha^\mu\delta_\beta^\nu
+\hbox{\large$\frac{n}{(n+2)^2}$} g_{\alpha\beta} x^{(\nu}\pd^{\mu)}
\nonumber\\
&-\hbox{\large$\frac{1}{(n+2)(n+1)}$} 
\Big(2n x_{(\alpha}\delta_{\beta)}^{(\nu}\pd^{\mu)}
-x^2\delta_{(\alpha}^{(\nu}\pd_{\beta)}\pd^{\nu)}\Big)
\nonumber\\
&-\hbox{\large$\frac{1}{(n+2)^2(n+1)}$}
\Big(2n x_{(\alpha}\pd_{\beta)}x^{(\nu}\pd^{\mu)}
-x^2 \pd_{\alpha}\pd_{\beta} x^{(\nu}\pd^{\mu)}\Big
)\nonumber\\
&+\hbox{\large$\frac{(n+3)n}{(n+2)^2(n+1)^2}$} 
\Big(x_{\alpha}x_{\beta} 
-\hbox{\large$\frac{1}{2}$}x^2 g_{\alpha\beta}\Big)\pd^{\mu}\pd^{\nu}
\nonumber\\
&-\hbox{\large$\frac{1}{(n+2)^2(n+1)^2}$}
\Big(2 x^2 \pd_{(\alpha}x_{\beta)} +\hbox{\large$\frac{1}{4}$}
x^4 \pd_{\alpha}\pd_{\beta}\Big) \pd^{\mu}\pd^{\nu}
\Big\}
%%%%H^{(4)}_n\left(x^2|\square\right) 
\breve{G}_{(\mu\nu) n}(x)\,.
\nonumber
\end{align}
Resumming all local operators of symmetry class (iv) 
and using the integral representation 
$((n+1)n)^{-1}=\int_0^1\d\lambda \lambda^n(1-\lambda)/\lambda$ 
gives
\begin{align}
\label{G_tw4_1}
G^{\mathrm{tw4(iv)}}_{(\alpha\beta)}(0,\kappa x)
&=
\Big(\delta_\alpha^\mu\delta_\beta^\nu x^\rho (x\pd)\pd_\rho
-2 x^\mu (x\pd)\delta_{(\beta}^\nu\pd_{\alpha)}
+x^\mu x^\nu\pd_\alpha\pd_\beta\Big)
\nonumber\\
%%\label{G_tw4_1}
&\qquad
\times
\int_0^1\d\lambda\hbox{\large$\frac{1-\lambda}{\lambda}$}
\,\tl G_{(\mu\nu)}(0,\kappa\lambda x)\, ,
\end{align}
where $\tl G_{(\mu\nu)}(0,\kappa x)$ is given by
\begin{eqnarray}
\lefteqn{\tl G_{(\alpha\beta)}(0,\kappa x)
=
\breve{G}_{(\alpha\beta)}(0,\kappa x)
+\sum_{k=1}^\infty\int_0^1\!
\frac{\d t}{t}
\!\left(\!\frac{-x^2}{4}\right)^{\!\!k}\!\!
\frac{\square^k}{k!(k-1)!}\!
\left(\!\frac{1-t}{t}\right)^{\!\!k-1}\!\!\!
\breve{G}_{(\alpha\beta)}(0,\kappa t x)}
\nonumber\\
%%& &+\int_0^1\!\frac{\d\tau}{\tau}\bigg\{
%%\left(
&&\!\!-\big[2x_{(\alpha}\delta_{\beta)}^{(\mu}\pd^{\nu)}(x\pd)
- x^2\pd_{(\alpha}\delta_{\beta)}^{(\mu}\pd^{\nu)}\big]
\nonumber\\
&&\qquad
\times
\!\sum_{k=0}^\infty
\int_0^1\!\!\d t\, t
\left(\!\frac{-x^2}{4}\right)^{\!\!k}\!\!
\frac{\square^k}{k!(k+1)!}
\left(\!\frac{1-t}{t}\right)^{\!\!k+1}\!\!
\breve{G}_{(\mu\nu)}(0,\kappa t x)
\nonumber\\
& &\!\!-\bigg\{\big[2x_{(\alpha}\pd_{\beta)}x^{(\mu}\pd^{\nu)}(x\pd)
-g_{\alpha\beta}x^{(\mu}\pd^{\nu)}(x\pd)\big(x\pd+1\big)
-x^2\pd_\alpha\pd_\beta x^{(\mu}\pd^{\nu)}\big]
\!\!\int_0^1\!\!\d \lambda\,\lambda\nonumber\\
& &\!\!-\big[\big(x_\alpha x_\beta\!-\!\hbox{\large$\frac{x^2}{2}$}%x^2 
g_{\alpha\beta}\big)
\pd^{\mu}\pd^{\nu}(x\pd)\big(x\pd\!+\!3\big)
%%\nonumber\\& &
\!-\!\big(2x^2\pd_{(\alpha}x_{\beta)}\!+\!\hbox{\large$\frac{x^4}{4}$} 
\pd_\alpha\pd_\beta\big)
\pd^\mu\pd^\nu\big]
\!\!\int_0^1\!\!\!\d \lambda(\!1\!-\!\lambda)\!\bigg\}
\nonumber\\
&&\qquad
\times
\sum_{k=0}^\infty\!
\int_0^1\!\!\d t\,t\!
\left(\!\frac{-x^2}{4}\right)^{\!\!k}\!\!
\frac{\square^k}{(k+1)!k!}
\left(\!\frac{1-t}{t}\right)^{\!\!k+1}
\breve{G}_{(\mu\nu)}(0,\kappa\lambda t x).\qquad\quad
\end{eqnarray}
Of course, eq.~(\ref{G_tw4_1}) we may be rewritten as follows
\begin{align}
\hspace{-.3cm}
G^{\rm tw4(iv)}_{(\alpha\beta)}(0,\kappa x)
\label{Gx_tw_2}
=&\;\tl G_{(\alpha\beta)}(0,\kappa x)
-2\!\!\int_0^1\!\!\!\d\lambda\hbox{\large$\frac{1}{\lambda}$}
\pd_{(\alpha}
\TLL {G^+_{\beta)}}(0,\kappa\lambda x)
%\nonumber\\&
+\!\!\int_0^1\!\!\!\d\lambda \hbox{\large$\frac{1-\lambda}{\lambda}$}
\pd_\alpha \pd_{\beta}
\tl G(0,\kappa\lambda x),
\nonumber
\end{align}
which will be used in the further considerations.
Contrary to any of the formerly obtained symmetric twist--4 tensor 
operators, this operator is the `true tensor part', i.e.,~not being
proportional to $g_{\alpha\beta}$ or $x_\alpha$ and $x_\beta$. 
Obviously, this twist--4 operator satisfies the conditions of tracelessness
\begin{equation}
\square G^{\rm tw4(iv)}_{(\alpha\beta)}(0,\kappa x)=0,\quad
\pd^\alpha G^{\rm tw4(iv)}_{(\alpha\beta)}(0,\kappa x)=0, \quad
%\pd^\beta G^{\rm tw4(iv)}_{(\alpha\beta)}(0,\kappa x),\quad
g^{\alpha\beta} G^{\rm tw4(iv)}_{(\alpha\beta)}(0,\kappa x).
\end{equation}
%%%\noindent
%%%{\it 3.4.2~~
\subsubsection{Projection onto the light--cone}
The symmetric twist--4 tensor operator of symmetry type (iv) 
on the light--cone is obtained as follows
\begin{align}
\label{G_tw4_iv}
G^{\mathrm{tw4(iv)}}_{(\alpha\beta)} (\ka\xx,\kb\xx)
=&
\Big(\delta_\alpha^\mu\delta_\beta^\nu x^\rho (x\pd)\pd_\rho
-2x^\mu (x\pd)\delta_{(\beta}^\nu\pd_{\alpha)}
+x^\mu x^\nu\pd_\alpha\pd_\beta\Big) 
\\
&\times
\int_0^1\d\lambda\hbox{\large$\frac{1-\lambda}{\lambda}$}
 G_{(\mu\nu)}(\ka\lambda x,\kb\lambda x)
\big|_{x=\xx}%\nonumber\\
-G^{\mathrm{>(iv)}}_{(\alpha\beta)} (\ka\xx,\kb\xx)\, ,
\nonumber
\end{align}
where the trace terms of higher twist are given by
\begin{eqnarray}
\lefteqn{
\label{G_high_(iv)_sy}
G^{\mathrm{>(iv)}}_{(\alpha\beta)} (\ka\xx,\kb\xx)
=
%\hbox{\large$\frac{1}{2}$}
\int_{0}^{1}\!\!\d\lambda\Big\{
\Big(2\lambda x_{(\alpha}\delta_{\beta)}^\nu\pd^\mu
-(1-\lambda)x_\alpha x_\beta\pd^\mu\pd^\nu\Big) 
G_{(\mu\nu)}(\ka\lambda x,\kb\lambda x)
}\nonumber\\
&&-\Big(
\hbox{\large$\frac{1}{\lambda}$}g_{\alpha\beta}\pd^\mu
+\hbox{\large$\frac{1-\lambda^2}{\lambda}$}x_{(\alpha}\delta^\mu_{\beta)}\square
-\hbox{\large$\frac{(1-\lambda)^2}{2\lambda}$}x_\alpha x_\beta\square\pd^\mu
\Big) G_{\mu}^{+}(\ka\lambda x,\kb\lambda x) \nonumber\\
&&+\Big(\hbox{\large$\frac{1-\lambda}{2\lambda}$}g_{\alpha\beta}\square
+\hbox{\large$\frac{(1-\lambda)^2}{2\lambda}$}
x_{(\alpha}\pd_{\beta)}\square
%%\nonumber\\&
-\hbox{\large$\frac{1}{2}$}\Big(
\hbox{\large$\frac{1-\lambda^2}{2\lambda}$}+\ln\lambda\Big)
x_\alpha x_\beta\square^2\Big)G(\ka\lambda x,\kb\lambda x)\nonumber\\
&&+
%%\hbox{\large$\frac{1}{2}$}
\Big(
\hbox{\large$\frac{1}{2\lambda}$}g_{\alpha\beta}(x\pd)
-\lambda x_{(\alpha}\pd_{\beta)}
-\hbox{\large$\frac{\lambda\ln\lambda}{2}$}
x_\alpha x_\beta\square\Big) G^\rho_{\ \rho}(\ka\lambda x,\kb\lambda x)
\Big\}\Big|_{x=\tilde{x}}.%%\nonumber
\end{eqnarray}
The explicit twist decomposition of this operator 
%eq.~(\ref{G_high_(iv)_sy})
by means of the Young tableaux (i) and (ii) gives
\begin{align}
%\label{M_tw2_b}
G^{\mathrm{>(iv)}}_{(\alpha\beta)}
(\ka\xx,\kb\xx)
&=
G^{\mathrm{tw4(iv)a}}_{(\alpha\beta)}(\ka\xx,\kb\xx)+
G^{\mathrm{tw4(iv)b}}_{(\alpha\beta)}(\ka\xx,\kb\xx)+
G^{\mathrm{tw4(iv)c}}_{(\alpha\beta)}(\ka\xx,\kb\xx)
\nonumber\\
&+
G^{\mathrm{tw4(iv)d1}}_{(\alpha\beta)}(\ka\xx,\kb\xx)+
G^{\mathrm{tw4(iv)d2}}_{(\alpha\beta)}(\ka\xx,\kb\xx)+
G^{\mathrm{tw4(iv)e1}}_{(\alpha\beta)}(\ka\xx,\kb\xx)
\nonumber\\
&+
G^{\mathrm{tw4(iv)e2}}_{(\alpha\beta)}(\ka\xx,\kb\xx)+
G^{\mathrm{tw4(iv)f}}_{(\alpha\beta)}(\ka\xx,\kb\xx)+
G^{\mathrm{tw4(iv)h}}_{(\alpha\beta)}(\ka\xx,\kb\xx)
\nonumber\\
&+
G^{\mathrm{tw5(iv)d}}_{(\alpha\beta)}(\ka\xx,\kb\xx)+
G^{\mathrm{tw5(iv)e}}_{(\alpha\beta)}(\ka\xx,\kb\xx)
\nonumber\\
&+
G^{\mathrm{tw6(iv)a}}_{(\alpha\beta)}(\ka\xx,\kb\xx)+
G^{\mathrm{tw6(iv)b}}_{(\alpha\beta)}(\ka\xx,\kb\xx)+
G^{\mathrm{tw6(iv)c}}_{(\alpha\beta)}(\ka\xx,\kb\xx)
\nonumber\\
&+
G^{\mathrm{tw6(iv)d1}}_{(\alpha\beta)}(\ka\xx,\kb\xx)+
G^{\mathrm{tw6(iv)d2}}_{(\alpha\beta)}(\ka\xx,\kb\xx)+
G^{\mathrm{tw6(iv)d3}}_{(\alpha\beta)}(\ka\xx,\kb\xx)
\nonumber\\
&+
G^{\mathrm{tw6(iv)d4}}_{(\alpha\beta)}(\ka\xx,\kb\xx)+
G^{\mathrm{tw6(iv)e1}}_{(\alpha\beta)}(\ka\xx,\kb\xx)+
G^{\mathrm{tw6(iv)e2}}_{(\alpha\beta)}(\ka\xx,\kb\xx)
\nonumber\\
&+G^{\mathrm{tw6(iv)e3}}_{(\alpha\beta)}(\ka\xx,\kb\xx)+
G^{\mathrm{tw6(iv)e4}}_{(\alpha\beta)}(\ka\xx,\kb\xx)+
G^{\mathrm{tw6(iv)f}}_{(\alpha\beta)}(\ka\xx,\kb\xx)
\nonumber\\
&+
G^{\mathrm{tw6(iv)h}}_{(\alpha\beta)}(\ka\xx,\kb\xx)+
G^{\mathrm{tw6(iv)k}}_{(\alpha\beta)}(\ka\xx,\kb\xx)\,,
\end{align}
with the following operators of well-defined twist
\begin{align}
G_{(\alpha\beta)}^{\mathrm{tw4(iv)a}}(\ka\xx,\kb\xx)
=& g_{\alpha\beta}\square
\int_0^1\d\lambda
%%\Big(\hbox{\large$\frac{5}{4}$}
\hbox{\large$\frac{1-\lambda}{2\lambda}$}
%%+\lambda\ln\lambda\Big)
G(\ka\lambda x,\kb\lambda x)\big|_{x=\tilde{x}}\,,
\nonumber\\
G_{(\alpha\beta)}^{\mathrm{tw4(iv)b}}(\ka\xx,\kb\xx)
=&\,  x_{(\alpha}\pd_{\beta)}\square
\int_0^1\d\lambda
\hbox{\large$\frac{(1-\lambda)^2}{2\lambda}$}
G(\ka\lambda x,\kb\lambda x)\big|_{x=\tilde{x}}
\nonumber\\&
-G_{(\alpha\beta)}^{\mathrm{tw6(iv)b}}(\ka\xx,\kb\xx)\,,
\nonumber\\
G_{(\alpha\beta)}^{\mathrm{tw4(iv)d1}}(\ka\xx,\kb\xx)
=&\,  x_{(\alpha}\pd_{\beta)}\square
\int_0^1\d\lambda\Big(
\hbox{\large$\frac{1-\lambda^2}{2\lambda}$}
+\hbox{\large$\frac{\ln\lambda}{\lambda}$}\Big)
G(\ka\lambda x,\kb\lambda x)\big|_{x=\tilde{x}}
\nonumber\\&
-G_{(\alpha\beta)}^{\mathrm{tw6(iv)d1}}(\ka\xx,\kb\xx)\,,
\nonumber\\
%\end{align}
\intertext{and}
%\begin{align}
G_{(\alpha\beta)}^{\mathrm{tw6(iv)a}}(\ka\xx,\kb\xx)
=&\,
-\hbox{\large$\frac{1}{2}$}x_\alpha x_\beta\square^2\!\!
\int_0^1\!\!\!\d\lambda
\Big( \hbox{\large$\frac{1-\lambda^2}{2\lambda}$}+\ln\lambda\Big)
G(\ka\lambda x,\kb\lambda x)\big|_{x=\tilde{x}}\,,
\nonumber\\
G_{(\alpha\beta)}^{\mathrm{tw6(iv)b}}(\ka\xx,\kb\xx)
=&
-\hbox{\large$\frac{1}{2}$}x_\alpha x_\beta\square^2\!\!
\int_0^1\!\!\!\d\lambda
\Big( \hbox{\large$\frac{1-\lambda}{\lambda}$}
-\hbox{\large$\frac{1-\lambda^2}{4\lambda}$}
+\hbox{\large$\frac{\ln\lambda}{2\lambda}$}\Big)
G(\ka\lambda x,\kb\lambda x)\big|_{x=\tilde{x}}\,,
\nonumber\\
G_{(\alpha\beta)}^{\mathrm{tw6(iv)d1}}(\ka\xx,\kb\xx)
=&\,
-\hbox{\large$\frac{1}{4}$}x_\alpha x_\beta\square^2\!\!
\int_0^1\!\!\!\d\lambda
\Big( \hbox{\large$\frac{1-\lambda^2}{2\lambda}$}
+\hbox{\large$\frac{\ln\lambda}{\lambda}$}
+\hbox{\large$\frac{\ln^2\lambda}{\lambda}$}\Big)
G(\ka\lambda x,\kb\lambda x)\big|_{x=\tilde{x}}\,,
\nonumber\\
G_{(\alpha\beta)}^{\mathrm{tw6(iv)d3}}(\ka\xx,\kb\xx)
=&\,
\hbox{\large$\frac{1}{2}$}x_\alpha x_\beta\square^2\!\!
\int_0^1\!\!\!\d\lambda
\Big( \hbox{\large$\frac{1-\lambda^2}{2\lambda}$}
+\hbox{\large$\frac{\ln\lambda}{\lambda}$}
+\hbox{\large$\frac{\ln^2\lambda}{\lambda}$}\Big)
G(\ka\lambda x,\kb\lambda x)\big|_{x=\tilde{x}}\,,
\nonumber
\end{align}
being related to the scalar operator,
as well as 
\begin{align}
G_{(\alpha\beta)}^{\mathrm{tw4(iv)c}}(\ka\xx,\kb\xx)
=& - g_{\alpha\beta}\pd^\mu 
\int_0^1\d\lambda
\hbox{\large$\frac{1}{\lambda}$}
G^{+}_{\mu}(\ka\lambda x,\kb\lambda x)\big|_{x=\tilde{x}}\,,
\nonumber\\
G_{(\alpha\beta)}^{\mathrm{tw4(iv)d2}}(\ka\xx,\kb\xx)
=&\,-2 x_{(\alpha}\pd_{\beta)}\pd^\mu 
\int_0^1\d\lambda\Big(
\hbox{\large$\frac{1-\lambda^2}{2\lambda}$}
+\hbox{\large$\frac{\ln\lambda}{\lambda}$}\Big)
G^{\mathrm{+}}_{\mu}(\ka\lambda x,\kb\lambda x)\big|_{x=\tilde{x}}
\nonumber
\\
&-G_{(\alpha\beta)}^{\mathrm{tw6(iv)d2}}(\ka\xx,\kb\xx)\,,
\nonumber\\
G_{(\alpha\beta)}^{\mathrm{tw4(iv)e1}}(\ka\xx,\kb\xx)
=&\, x_{(\alpha}\pd_{\beta)}\pd^\mu 
\int_0^1\d\lambda \hbox{\large$\frac{1-\lambda^2}{\lambda}$}
G^{\mathrm{+}}_{\mu}(\ka\lambda x,\kb\lambda x)\big|_{x=\tilde{x}}
\nonumber\\&
-G_{(\alpha\beta)}^{\mathrm{tw6(iv)e1}}(\ka\xx,\kb\xx)\,,
\nonumber
\end{align}
and
%%\intertext{and}
\begin{align}
G_{(\alpha\beta)}^{\mathrm{tw5(iv)d}}(\ka\xx,\kb\xx)
=&\,
x_{(\alpha}\big(\delta^\mu_{\beta)}(x\pd)-x^\mu\pd_{\beta)}\big)
\square 
\nonumber\\
&\qquad
\times\int_0^1\!\!\d\lambda\Big(
\hbox{\large$\frac{1-\lambda^2}{2\lambda}$}
+\hbox{\large$\frac{\ln\lambda}{\lambda}$}\Big)
G^{\mathrm{+}}_{\mu}(\ka\lambda x,\kb\lambda x)\big|_{x=\tilde{x}}\,,
\nonumber\\
&-G_{(\alpha\beta)}^{\mathrm{tw6(iv)d3}}(\ka\xx,\kb\xx)
-G_{(\alpha\beta)}^{\mathrm{tw6(iv)d4}}(\ka\xx,\kb\xx)\,,
\end{align}
and
%%\intertext{and}%\nonumber\\
\begin{align}
G_{(\alpha\beta)}^{\mathrm{tw6(iv)c}}(\ka\xx,\kb\xx)
=&
\hbox{\large$\frac{1}{2}$}x_\alpha x_\beta\square\pd^\mu 
\int_0^1\d\lambda
\hbox{\large$\frac{(1-\lambda)^2}{\lambda}$}
G^{\mathrm{+}}_{\mu}(\ka\lambda x,\kb\lambda x)\big|_{x=\tilde{x}}\,,
\nonumber\\
G_{(\alpha\beta)}^{\mathrm{tw6(iv)d2}}(\ka\xx,\kb\xx)
=&\,\hbox{\large$\frac{1}{2}$}
x_\alpha x_\beta\square\pd^\mu \!\!
\int_0^1\!\!\!\d\lambda\Big(
\hbox{\large$\frac{1-\lambda^2}{2\lambda}$}
+\hbox{\large$\frac{\ln\lambda}{\lambda}$}
+\hbox{\large$\frac{\ln^2\lambda}{\lambda}$}\Big)
G^{\mathrm{+}}_{\mu}(\ka\lambda x,\kb\lambda x)\big|_{x=\tilde{x}}\,,
\nonumber\\
G_{(\alpha\beta)}^{\mathrm{tw6(iv)d4}}(\ka\xx,\kb\xx)
=&\,
x_\alpha x_\beta\square\pd^\mu \!
\int_0^1\!\d\lambda\Big(
\hbox{\large$\frac{1-\lambda^2}{2\lambda}$}
+\hbox{\large$\frac{\ln\lambda}{\lambda}$}\Big)
G^{\mathrm{+}}_{\mu}(\ka\lambda x,\kb\lambda x)\big|_{x=\tilde{x}}\,,
\nonumber\\
G_{(\alpha\beta)}^{\mathrm{tw6(iv)e1}}(\ka\xx,\kb\xx)
=&\,
-\hbox{\large$\frac{1}{2}$}
x_\alpha x_\beta\square\pd^\mu 
\int_0^1\d\lambda\Big(
\hbox{\large$\frac{1-\lambda^2}{2\lambda}$}
+\hbox{\large$\frac{\ln\lambda}{\lambda}$}\Big)
G^{\mathrm{+}}_{\mu}(\ka\lambda x,\kb\lambda x)\big|_{x=\tilde{x}}\,,
\nonumber\\
G_{(\alpha\beta)}^{\mathrm{tw6(iv)e3}}(\ka\xx,\kb\xx)
=&\,
x_\alpha x_\beta\square\pd^\mu 
\int_0^1\d\lambda\Big(
\hbox{\large$\frac{1-\lambda^2}{2\lambda}$}
+\hbox{\large$\frac{\ln\lambda}{\lambda}$}\Big)
G^{\mathrm{+}}_{\mu}(\ka\lambda x,\kb\lambda x)\big|_{x=\tilde{x}}\,,
\nonumber
\end{align}
being related to the symmetric vector operator,
%and
\begin{align}
G_{(\alpha\beta)}^{\mathrm{tw5(iv)e}}(\ka\xx,\kb\xx)
=&\,
x_{(\alpha}\big(\delta^\nu_{\beta)}(x\pd)-x^\nu\pd_{\beta)}\big)
\pd^\mu 
\int_0^1\d\lambda\hbox{\large$\frac{1-\lambda^2}{\lambda}$}
\left.G_{(\mu\nu)}(\ka\lambda x,\kb\lambda x)\right|_{x=\tilde{x}}\,,
\nonumber\\
&-G_{(\alpha\beta)}^{\mathrm{tw6(iv)e3}}(\ka\xx,\kb\xx)
G_{(\alpha\beta)}^{\mathrm{tw6(iv)e4}}(\ka\xx,\kb\xx)
\nonumber\\
G_{(\alpha\beta)}^{\mathrm{tw6(iv)e4}}(\ka\xx,\kb\xx)
=& x_\alpha x_\beta\pd^\mu\pd^\nu
\int_0^1\d\lambda \hbox{\large$\frac{1-\lambda^2}{\lambda}$}
G_{(\mu\nu)}(\ka\lambda x,\kb\lambda x)\big|_{x=\tilde{x}}\,,
\nonumber\\
G_{(\alpha\beta)}^{\mathrm{tw6(iv)f}}(\ka\xx,\kb\xx)
=&\,- x_\alpha x_\beta\pd^\mu\pd^\nu
\int_0^1\d\lambda(1-\lambda)
G_{(\mu\nu)}(\ka\lambda x,\kb\lambda x)\big|_{x=\tilde{x}}\,,
\nonumber
\end{align}
being related to the symmetric tensor operator and, furthermore,
\begin{align}
G_{(\alpha\beta)}^{\mathrm{tw4(iv)f}}(\ka\xx,\kb\xx)
=&\, g_{\alpha\beta}(x\pd)\int_0^1\d\lambda
\hbox{\large$\frac{1}{2\lambda}$}
G^\rho_{\ \rho}(\ka\lambda x,\kb\lambda x)\big|_{x=\tilde{x}}\,,
\nonumber\\
G_{(\alpha\beta)}^{\mathrm{tw4(iv)h}}(\ka\xx,\kb\xx)
=& -x_{(\alpha}\pd_{\beta)}\!\!\int_0^1\!\d\lambda\,\lambda
G^\rho_{\ \rho}(\ka\lambda x,\kb\lambda x)\big|_{x=\tilde{x}}\!
-G_{(\alpha\beta)}^{\mathrm{tw6(iv)h}}(\ka\xx,\kb\xx)\,,
\nonumber\\
G_{(\alpha\beta)}^{\mathrm{tw4(iv)e2}}(\ka\xx,\kb\xx)
=& -x_{(\alpha}\pd_{\beta)}
\!\!\int_0^1\!\!\!\d\lambda
\hbox{\large$\frac{1-\lambda^2}{\lambda}$}
G^\rho_{\ \rho}(\ka\lambda x,\kb\lambda x)\big|_{x=\tilde{x}}\!
-G_{(\alpha\beta)}^{\mathrm{tw6(iv)e2}}(\ka\xx,\kb\xx)\,,
\nonumber\\
\intertext{and}
G_{(\alpha\beta)}^{\mathrm{tw6(iv)h}}(\ka\xx,\kb\xx)
=&\,-\hbox{\large$\frac{1}{4}$}x_\alpha x_\beta\square
\int_0^1\d\lambda \hbox{\large$\frac{1-\lambda^2}{\lambda}$}
G^\rho_{\ \rho}(\ka\lambda x,\kb\lambda x)\big|_{x=\tilde{x}}\,,
\nonumber\\
G_{(\alpha\beta)}^{\mathrm{tw6(iv)e2}}(\ka\xx,\kb\xx)
=&\,
\hbox{\large$\frac{1}{2}$}x_\alpha x_\beta\square
\int_0^1\d\lambda
\Big( \hbox{\large$\frac{1-\lambda^2}{2\lambda}$}
+\hbox{\large$\frac{\ln\lambda}{\lambda}$}\Big)
G^\rho_{\ \rho}(\ka\lambda x,\kb\lambda x)\big|_{x=\tilde{x}}\,,
\nonumber\\
G_{(\alpha\beta)}^{\mathrm{tw6(iv)k}}(\ka\xx,\kb\xx)
=&\,-\hbox{\large$\frac{1}{2}$}x_\alpha x_\beta\square
\int_0^1\d\lambda\,\lambda(\ln\lambda)
G^\rho_{\ \rho}(\ka\lambda x,\kb\lambda x)\big|_{x=\tilde{x}}\,,
\nonumber
\end{align}
being related to the trace of the symmetric tensor operator.

Here, it should be remarked that, contrary to the former cases (i)
-- (iii), the trace terms (\ref{G_high_(iv)_sy}) contain operators
having the same twist, $\tau = 4$, as the primary operator of symmetry 
class (iv). However, these operators being multiplied by
either $g_{\alpha\beta}$ or $x_\alpha$ and $x_\beta$ are scalar or
vector operators corresponding to symmetry type (i). Whereas
the local operators resulting from the twist--4 tensor operator of 
symmetry class (iv) are contained in the tensor space
${\bf T}\hbox{$(\frac{n+2}{2},\frac{n-2}{2})\oplus
{\bf T}(\frac{n-2}{2},\frac{n+2}{2})$}$, 
the local operators of the twist--4 scalar and vector operators 
are totally symmetric traceless tensors and, therefore, contained
in the tensor space ${\bf T}\hbox{$(\frac{n}{2},\frac{n}{2})$}$.
%%%\medskip
%%%\noindent
%%%{\it 3.4.3~~
\subsubsection{Determination of the complete symmetric 
light--cone tensor operators}
Now, having finished the decomposition of a general 2nd rank tensor
operator, let us sum up the remaining symmetric tensor operators of
twist greater than 3, which appear in
the trace terms of the symmetric twist operators with Young symmetry (i), (ii)
as well as (iv), to  complete twist--4, twist--5 and twist--6 operators. 

The `scalar part' of the twist--4 light--ray tensor operator is given by
\begin{align}
\label{G_tw4sc_i}
\lefteqn{\hspace{-2cm}
G_{(\alpha\beta)}^{\mathrm{tw4,s}}(\ka\xx,\kb\xx)=-g_{\alpha\beta}
\bigg\{
\hbox{\large$\frac{1}{2}$}\square\int_0^1\d\lambda\,\lambda(\ln\lambda)
G(\ka\lambda x,\kb\lambda x)
}\\
&-(x\pd)\int_0^1\d\lambda\hbox{\large$\frac{1}{2\lambda}$}
G^\rho_{\ \rho}(\ka\lambda x,\kb\lambda x)%\nonumber\\&
+\pd^\mu\int_0^1\d\lambda\,\lambda
G^+_{\mu}(\ka\lambda x,\kb\lambda x)
\bigg\}\bigg|_{x=\tilde{x}}\,,
\nonumber
\end{align}
and the `vector part' of the twist--4 light--ray tensor operator reads
\begin{eqnarray}
\label{G_tw4ve_i}
\lefteqn{
G_{(\alpha\beta)}^{\mathrm{tw4,v}}(\ka\xx,\kb\xx)
=-\lcx_{(\alpha}\times\nonumber}\\
&&\!\!\!\bigg\{\!\!
\int_0^1\!\!\!\d\lambda\Big\{
\Big(\hbox{\large$\frac{1-\lambda^2}{2\lambda}$}+\lambda\ln\lambda\Big)
\pd_{\beta)}+\hbox{\large$\frac{1}{4}$}x_{\beta)}
\Big(\hbox{\large$\frac{1-\lambda^2}{\lambda}$}+\lambda\ln\lambda
+\hbox{\large$\frac{\ln\lambda}{\lambda}$}\Big)\square\Big\}\square
G(\ka\lambda x,\kb\lambda x)\nonumber\\
&&\!\!\!+\int_0^1\!\!\hbox{\large$\frac{\d\lambda}{\lambda}$}
\Big\{\pd_{\beta)}+\hbox{\large$\frac{1}{2}$}x_{\beta)}
(\ln\lambda)\square\Big\}
G^\rho_{\ \rho}(\ka\lambda x,\kb\lambda x)\\
&&\!\!\!-\int_0^1\!\!\hbox{\large$\frac{\d\lambda}{\lambda}$}
\Big\{2\big(1-\lambda^2\big)\pd_{\beta)}
+\hbox{\large$\frac{1}{2}$}x_{\beta)}
\big(1-\lambda^2+2\ln\lambda\big)\square\Big\}\pd^\mu
G^+_{\mu}(\ka\lambda x,\kb\lambda x)
\bigg\}\bigg|_{x=\tilde{x}}.
\nonumber
\end{eqnarray}
Furthermore, the contributions to the complete `vector part' of the 
twist--5 light--ray tensor operator are 
constructed by means of the Young tableau (ii), and their 
local operators are antisymmetric\footnote{
Note that antisymmetry is not in $\alpha$ and $\beta$ but results
from the differential operator!} traceless tensors contained
in the space ${\bf T}\hbox{$(\frac{n}{2},\frac{n-2}{2})\oplus
{\bf T}(\frac{n-2}{2},\frac{n}{2})$}$. 
This complete twist--5 tensor operator is given by
\begin{align}
\label{G_tw5_ii_sy}
\lefteqn{\hspace{-.5cm}
G_{(\alpha\beta)}^{\mathrm{tw5,v}}(\ka\xx,\kb\xx)
=-\lcx_{(\alpha}\times}\nonumber\\
&\bigg\{
\!\!\int_0^1\!\!\d\lambda\Big\{\!
\hbox{\large$\frac{(1-\lambda)^2}{2\lambda}$}\,
\big(\delta^\nu_{\beta)}(x\pd)\!-\!x^\nu\pd_{\beta)}\!-\!
x_{\beta)}\pd^\nu\big)
\square
-\!\Big(\!\hbox{\large$\frac{1-\lambda}{\lambda}$}
-\hbox{\large$\frac{1-\lambda^2}{4\lambda}$}
\!+\!\hbox{\large$\frac{1}{2\lambda}$}\ln\lambda\!\Big)
x_{\beta)}\square^2 x^\nu\!\Big\}\nonumber\\
&\hspace{8cm}\times G^+_\nu(\ka\lambda x,\kb\lambda x)\nonumber\\
&-\int_0^1\d\lambda\Big\{
\hbox{\large$\frac{1-\lambda^2}{\lambda}$}\,
\big(\delta^\nu_{\beta)}(x\pd)-x^\nu\pd_{\beta)}-x_{\beta)}\pd^\nu\big)\pd^\mu 
-\Big(\hbox{\large$\frac{1-\lambda^2}{2\lambda}$}
+\hbox{\large$\frac{\ln\lambda}{\lambda}$}\Big)
x_{\beta)}\square\pd^\mu x^\nu\Big\}\nonumber\\
&\hspace{8cm}\times G_{(\mu\nu)}(\ka\lambda x,\kb\lambda x)
\bigg\}\bigg|_{x=\tilde{x}}.
\end{align}
The `scalar part' of the twist--6 light--ray tensor operator has 
Young symmetry (i). 
Thus, the local twist--6 operators are totally symmetric traceless tensors 
lying in the space ${\bf T}\hbox{$(\frac{n-2}{2},\frac{n-2}{2})$}$. 
This nonlocal twist--6 operator is given by
\begin{align}
\label{G_tw6_i}
G_{(\alpha\beta)}^{\mathrm{tw6,s}}(\ka\xx,\kb\xx)
=-\lcx_\alpha\lcx_\beta&\bigg\{
\hbox{\large$\frac{1}{2}$}\square^2
\int_0^1\d\lambda
\Big(\hbox{\large$\frac{1-\lambda}{\lambda}$}
+\hbox{\large$\frac{1+\lambda}{2\lambda}$}\ln\lambda\Big)
G(\ka\lambda x,\kb\lambda x)\nonumber\\
&-\square\int_0^1\d\lambda\,
\hbox{\large$\frac{1-\lambda^2}{2\lambda}$}
G^\rho_{\ \rho}(\ka\lambda x,\kb\lambda x)\nonumber\\
&+\square\int_0^1\d\lambda\,
\hbox{\large$\frac{(1-\lambda)^2}{2\lambda}$}
\pd^\mu G^+_\mu(\ka\lambda x,\kb\lambda x)\nonumber\\
&-\pd^\mu\pd^\nu\int_0^1\d\lambda\,\hbox{\large$\frac{1-\lambda}{\lambda}$}
G_{(\mu\nu)}(\ka\lambda x,\kb\lambda x)
\bigg\}\bigg|_{x=\tilde{x}}.
\end{align}

Now, we may pick up all the contributions to the symmetric
(traceless) tensor operator.
Together with the twist--2 part, eq.~(\ref{Gtw2_gir}), the twist--3 
part, eq.~(\ref{G3_symm}), and the twist--4 part, eq.~(\ref{G_tw4_iv}), 
we finally obtain the following complete decomposition
of the {\em symmetric light--cone tensor operator} (compare
eq.~(\ref{G^+_t})):
\begin{align}
\label{complete_sym}
G_{(\alpha\beta)}(\ka\xx,\kb\xx)
&=
G^{\rm tw2(i)}_{(\alpha\beta)}(\ka\xx,\kb\xx)+
G^{\rm tw3(ii)}_{(\alpha\beta)}(\ka\xx,\kb\xx)+
G^{\rm tw4(iv)}_{(\alpha\beta)}(\ka\xx,\kb\xx)\quad\nonumber\\
&+
G^{\rm tw4,s}_{(\alpha\beta)}(\ka\xx,\kb\xx)+
G^{\rm tw4,v}_{(\alpha\beta)}(\ka\xx,\kb\xx)\quad\nonumber\\
&+
G^{\rm tw5,v}_{(\alpha\beta)}(\ka\xx,\kb\xx)+
G^{\rm tw6,s}_{(\alpha\beta)}(\ka\xx,\kb\xx)\,.%\nonumber
\end{align}

Let us remark that by construction the terms of the first line are
traceless, whereas the traces of the third line vanish because of
antisymmetry and $\xx^2 = 0$ on the light--cone, respectively.
However, the traces of the second line do not vanish but restore
the trace of the tensor operator,
\begin{eqnarray}
g^{\alpha\beta} G_{\alpha\beta} (\ka\xx,\kb\xx)
=
g^{\alpha\beta} G^{\rm tw4,s}_{(\alpha\beta)} (\ka\xx,\kb\xx)
+
g^{\alpha\beta} G^{\rm tw4,v}_{(\alpha\beta)} ((\ka\xx,\kb\xx)).
\end{eqnarray}
This may be proven explicitly by taking the trace of 
eqs.~(\ref{G_tw4sc_i}) and (\ref{G_tw4ve_i}), using 
$(x\pd) \cong \lambda \pd_\lambda$,
performing partial integrations and observing the equality
\begin{eqnarray}
\label{G_trace}
\big(g^{\alpha\beta} - 2 \pd^\alpha x^\beta 
+ (1/2)\square x^\alpha x^\beta \big)
G_{(\alpha\beta)} (\ka\lambda x,\kb\lambda x)\big|_{\lambda =0} = 0\,.
%\nonumber
\end{eqnarray}

Herewith, the twist decomposition of a generic bilocal 2nd rank
light--ray tensor operator is completed. In the next Chapter we 
show how this general formalism may be applied also to tri-- and
multi--local light--ray operators.

%%%%%%%%%%%%%%%%%%%%%%%%%%%%%%%%%%%%%%%%
%%%%
%%\newpage

\section{Extension to trilocal tensor
operators}
\setcounter{equation}{0}
The procedure reviewed in Chapt.~2 for bilocal operators may be 
extended to trilocal operators, too. These operators occur in higher 
twist contributions 
to light--cone dominated processes and, additionally, as counterparts in the 
renormalization of such higher twist operators which have been obtained
in Chapt.~3. In the following we consider various trilocal operators of 
minimal twist--3 and twist--4. 
The local versions of such higher twist operators have been considered
already in the early days of QCD in systematic studies of deep inelastic
scattering \cite{N,G} and on behalf of giving some parton interpretation
for the distribution amplitudes
\cite{P,EFP}. Various studies of local twist--4 operators determined
their anomalous dimension matrices and the behaviour of their structure 
functions \cite{twist_4}. The relevance of local twist--3 operators for
the structure function $g_2$ of polarized deep inelastic scattering is
also well-known \cite{twist_3,Shu82}. If written nonlocal all these 
operators are necessarily multilocal. However, nonlocal higher twist 
operators have been extensively used only in the study of the structure 
function $g_2$ \cite{BB88,Gey96b} and of the photon and 
vector meson wave functions \cite{BBK89,BBKT99}.

Here, our main interest is not to give an exhaustive study of the various
scattering processes but to present the twist decomposition of some 
characteristic multilocal operators.

\subsection{General trilocal
tensor operators: 
     Quark--antiquark--gluon, four--quark and four--gluon operators}

This Subsection is
devoted to the consideration of trilocal operators 
which share the same twist decomposition as the gluon operators 
$G_{\alpha\beta}$ (or their (anti)symmetric parts). We call them 
unconstrained because their truncation with $x^\alpha x^\beta$ does
not vanish identically. For compact notations, we will introduce 
the following abbreviations:
\begin{align}
\Gamma^i 
= \{ 1, \gamma_5\}, \quad
\Gamma^i_\alpha 
= \{\gamma_\alpha, \gamma_5\gamma_\alpha\}, \quad
\Gamma^i_{\alpha\beta} 
= \{\sigma_{\alpha\beta}, \gamma_5\sigma_{\alpha\beta}\}, \quad
F^i_{\mu\nu} = \{F_{\mu\nu}, \widetilde F_{\mu\nu}\}\,.
\nonumber
\end{align}
%%%{\em 4.1.1 
\subsubsection{Quark--antiquark--gluon operators}
First, we consider the following quark--antiquark--gluon 
operators,\footnote{
For simplicity we restrict to the nonsinglet case and suppress flavour 
matrices.}
\begin{align}
V_{\alpha\beta}^{ij}(\ka x,\tau x,\kb  x)
&=
\overline{\psi}(\ka x)U(\ka x,\tau x)\Gamma_\alpha^{i\,\rho} 
F^j_{\beta\rho}(\tau x)U(\tau x,\kb x)\psi(\kb x)\,,
\\
\intertext{and}
W^{ij}_{[\alpha\beta]}(\ka x,\tau x,\kb  x)
&=
\overline{\psi}(\ka x) U(\ka x,\tau x) \Gamma^i 
F^j_{\alpha\beta}(\tau x) U(\tau x,\kb x)\psi(\kb x)\,,
\end{align}
whose minimal twist is $\tau_{\rm min} = 3$ and $4$, respectively; 
they have the same symmetry as the gluon tensor operators 
$G_{\alpha\beta}$ and $G_{[\alpha\beta]}$, respectively. 

In order to be able to apply the general procedure of Chapt.~2 
as well as the results of Chapt.~3 we have to verify the local structure 
of these trilocal operators.
Let us study, for example, the first of the above quark--gluon
operators,
$V^{11}_{\alpha\beta}$, in some detail 
and Taylor expand its three local fields around
$y=0$:
\begin{align}
\lefteqn{\hspace{-.9cm}
V^{11}_{\alpha\beta}(\ka x,\tau x,\kb x)=
}\nonumber\\
&=\
\overline{\psi}(\ka x)U(\ka x,0)U(0,\tau x)\sigma_\alpha^{\ \rho}
F_{\beta\rho}(\tau x)U(\tau x,0)U(0,\kb x)\psi(\kb x)
\nonumber\\
&=
\!\sum_{n_1,m,n_2 =0}^\infty\!
\hbox{\large$\frac{\ka^{n_1}\tau^m\kb^{n_2}}{n_1!m!n_2!}$}
\Big[\overline\psi(y)\!\left
(\!\LD D_y x\right)^{\!\!n_1}\!\Big]
\sigma_\alpha^{\ \rho}
\Big[\!\left(x D_z\right)^m\! F_{\beta\rho}(z)\Big]\!
\Big[\!\!\left(x \RD D_y \right)^{\!\!n_2}\!\!\!
\psi(y)\Big]\Big|_{y=z=0}
\nonumber\\
&=
\sum_{N=0}^\infty
\hbox{\large$\frac{1}{N!}$}\, \overline\psi(0)\!
\left(
\sum_{n=0}^N \!\sum_{\ell=0}^n
\!
\hbox{\large$\binom{N}{n}  \binom{n}{\ell}$}
\!\left(\!\ka \!\LD D x \right)^{\!\!n-\ell}\!
\!\!\!\sigma_\alpha^{\ \rho}
\Big[\!\left(\tau x D \right)^{N-n}\!
F_{\beta\rho}(0) \!\Big]\!\!
\left(\!\kb x\! \RD D \right)^{\!\!\ell}\!\right)\!
\psi(0)\,,
\nonumber
\end{align}
where the left and right derivatives are given by
eqs.~(\ref{D_kappa}), 
and
\begin{equation}
D_\mu F_{\alpha\beta} 
= \pd_\mu F_{\alpha\beta} + \ii g [A_\mu, F_{\alpha\beta}]\,.
\end{equation}
This leads to the
following expansion into local tensor operators
\begin{equation}
\label{Vij}
V^{11}_{\alpha\beta}(\ka x,\tau x,\kb  x)
=
\sum_{N=0}^\infty \frac{1}{N!} x^{\mu_1}\ldots x^{\mu_N}
V^{11}_{\alpha\beta\mu_1\ldots\mu_N}(\ka,\tau,\kb)\,,
\end{equation}
 where
\begin{equation}
\label{V_local3}
V^{11}_{\alpha\beta\mu_1\ldots\mu_N}(\ka,\tau,\kb)
\equiv 
\overline\psi(0) \sigma_\alpha^{\ \rho}\
{\bf D}_{[\beta\rho]\mu_1\ldots\mu_N}(\ka,\tau,\kb) \psi(0)\,,
%\nonumber
\end{equation}
and
the field strength--dependent generalized covariant derivative 
of $N$th order is given by
\begin{align}
\label{D_F}
%\lefteqn{
&{\bf D}_{[\beta\rho]\mu_1\ldots\mu_N}(\ka,\tau,\kb)=
%}
\\
&\quad= N!\sum_{n=0}^N \sum_{\ell=0}^n
\hbox{\large$\frac{\ka^{n-\ell}}{(n-\ell)!}$}
\LD D_{\mu_1}\! \ldots \!\LD D_{\mu_{n-\ell}}
\hbox{\large$\frac{\tau^{N-n}}{(N-n)!}$}
\Big[D_{\mu_{n+1}}\!\ldots\!D_{\mu_{N}} 
F_{\beta\rho}(0)\Big]
\hbox{\large$\frac{\kb^\ell}{\ell!}$}
\RD D_{\mu_{n-\ell+1}}\!\ldots \!\RD D_{\mu_n}.
\nonumber
\end{align}
Obviously, if the field
strength were not present this operation 
would reduce to the product 
$\Tensor D_{\mu_1}(\ka,\kb)\ldots \Tensor D_{\mu_N}(\ka,\kb)$
of generalized derivatives
(\ref{D_kappa}) introduced in Chapt.~2.
Furthermore, for $\tau =0$ we are left with a more
simple expression 
which, however, cannot be reduced to the $N$th power of some 
extended
derivative because $F$ is equipped with some matrix structure. 
Anyway, the local tensor
operators (\ref{V_local3}) of rank $N+2$
with canonical dimension $d=N+5$, which are given
as a sum of
 $\frac{1}{2} (N+1)(N+2)$ terms, have to be decomposed 
according to their
geometric twist. In principle, this has to be done 
term-by-term. But, because of the linearity of
that procedure we are 
not required to do this for any term explicitly.

Due to the same general local structure of $V^{ij}_{\alpha\beta}$ and 
$W^{ij}_{\alpha\beta}$ -- which is governed by Taylor expansions 
completely analogous to eq.~(\ref{Vij}) -- their decomposition into 
terms of definite twist leads to the same expressions as we obtained 
in Chapt. 3 but
with $G_{\alpha\beta}$ and $G_{[\alpha\beta]}$ exchanged by 
$V^{ij}_{\alpha\beta}$ and $W^{ij}_{\alpha\beta}$, respectively. 
The only difference consists in a shift of any twist by one unit, 
$\tau \rightarrow \tau + 1$, in the various inputs of Table 1 and 2
below.
 
In contrast to the operators $V$ and $W$ the quark--antiquark--gluon
operators
\begin{align}
\label{OO}
\OO^{ij}_{\alpha\beta}(\ka x,\tau x,\kb x)
&= x^\rho\,\overline{\psi}(\ka x)U(\ka x,\tau x)\Gamma_\alpha^i 
F^j_{\beta\rho}(\tau x)U(\tau x,\kb x)\psi(\kb x)
%\nonumber
\\ 
\intertext{and}
\label{OO'}
{\OO'}^{ij}_{\alpha\beta}(\ka x,\tau x,\kb x)
&=x^\rho x^\sigma \overline{\psi}(\ka x)U(\ka x,\tau x)\Gamma^i_{\alpha\rho} 
F^j_{\beta\sigma}(\tau x)U(\tau x,\kb x)\psi(\kb x)
%\nonumber
\end{align}
are special, namely, despite of their arbitrary symmetry with respect to
$\alpha$ and $\beta$ they vanish identically if multiplied by $x^\beta$
and $x^\alpha$ or $x^\beta$, respectively. Therefore, they show some
peculiarities which will be treated in Subsection 4.2.
%%%\medskip
%%%\noindent
%%%{\em 4.1.2 
\subsubsection{Four--quark and four--gluon operators}
Let us now consider trilocal operators which are built up from 
four quark or four gluon fields. We denote them by 
${}^{I}\!\QQ^{ij}_{\alpha\beta}$ with 
$I=0$ and $I=2$, respectively (here, $I$ counts the
number of external $x$'s of the operators):
\begin{align}
\lefteqn{\hspace{-3.6cm}
{}^{I=0}\!\QQ^{ij}_{\alpha\beta}(\ka x,\tau x,\kb x)
= }\\
&=
\big(\,\overline\psi(\ka x)\Gamma^i_\alpha U(\ka x,\tau x)\psi(\tau x)\big)
\big(\,\overline\psi(\tau x) U(\tau x,\kb x)\Gamma^j_\beta\psi(\kb x)\big)
\nonumber
\end{align}
as well
as
\begin{align}
\lefteqn{\hspace{-2cm}
{}^{I=2}\!\QQ^{ij}_{\alpha\beta}(\ka x,\tau x,\kb x)=
x^\rho x^\sigma\,
{}^{I=2}\!\QQ^{ij}_{\alpha\rho\sigma\beta}(\ka x,\tau x,\kb x)\,,}
\\
&= x^\rho x^\sigma
\Big(F_{\alpha\mu}(\ka x)U(\ka x,\tau x)F_\rho^{i\mu}(\tau
x)\Big)
\Big(F_\sigma^{j\nu}(\tau x)U(\tau x,\kb x)F_{\beta\nu}(\kb x)\Big)\,.
\nonumber
\end{align}

Both sets of operators have minimal twist $\tau_{\rm min} =4$. 
Their local versions have been already considered by various authors, 
see e.g.~\cite{JS82}. The explicit form of the local operators
will be given below. Because
there are no restrictions concerning the free indices $\alpha$ and $\beta$
the Young patterns (i) -- (iv) are involved. The twist decomposition 
may be performed along the lines of Chap.~3 and, as in the
foregoing cases, the outcome will be almost the same as for 
$G_{\alpha\beta}$. Again, the difference is that now the value of 
twist raises by two units, $\tau \rightarrow \tau + 2$, relative to the 
gluon operators shown in Table 1 and 2 of the Conclusions. 
In addition, because of the change in the external $x$--factors 
(not being accompanied by $\lambda$)
the measures of the $\lambda$--integrations 
have to be changed according to (see also Subsection 4.2)
\begin{equation}
\d\lambda
\longrightarrow  \d\lambda\,\lambda^{I}\,.
\qquad 
\end{equation}

Open, up to now, is the structure of the local operators. Let us study 
in detail the first of the four--quark operators,
${}^{I=0}\! \QQ^{11}_{\alpha\beta}$, whose 
Taylor expansion may be written in two equivalent ways:
\begin{align}
\lefteqn{\hspace{-.4cm}
{}^{I=0}\!
\QQ^{11}_{\alpha\beta}(\ka x,\tau x,\kb x)=
}\nonumber\\
&=\
\left(\,\overline\psi(\ka x)\gamma_\alpha U(\ka x,0)
U(0,\tau x)\psi(\tau x)\right)
\left(\overline\psi(\tau x) U(\tau x,0)
U(0,\kb x)\gamma_\beta\psi(\kb x)\right) 
\nonumber\\
&= \!\sum_{n_1,m,n_2
=0}^\infty\! \!\!
\hbox{\large$\frac{\ka^{n_1}\tau^m\kb^{n_2}}{n_1!m!n_2!}$}
\Big[\overline\psi(y)\!\left(\!\LD D_y x\right)^{\!\!n_1}\!\Big]\!
\gamma_\alpha 
\!\Big[\!\left(xD_z\right)^m\! \left(\psi(z)\overline\psi(z)\right)\Big]\!
\gamma_\beta
\!\Big[\!\!\left(x \RD D_y \right)^{\!\!n_2}\!\!\! \psi(y)\Big]\Big|_{y=z=0}
\nonumber\\
&=
\!\sum_{n_1=0}^\infty\!
\hbox{\large$\frac{1}{n_1!}$}
\Big(\overline\psi(y)\!\left(x\Tensor D_y (\ka,\tau)\right)^{\!\!n_1}
\!\!\!\gamma_\alpha \psi(y)\!\Big)\Big|_{y=0}
\!\sum_{n_2=0}^\infty\! 
\hbox{\large$\frac{1}{n_2!}$}
\Big(\overline\psi(z)\!\left(x\Tensor D_z (\tau,\kb)\right)^{\!\!n_2}
\!\!\!\gamma_\beta \psi(z)\!\Big)\Big|_{z=0},
\nonumber
\end{align}
where the two possibilities also use different generalized covariant 
derivatives. The left, right and left-right derivatives are given
by eqs.~(\ref{D_kappa}), and, treating $\psi(0)\overline\psi(0)$ as a 
matrix in the group algebra, another form of the derivative obtains:
\begin{eqnarray}
D_\mu \left(\psi(0)\overline\psi(0)\right) 
&= 
\pd_\mu \left(\psi(0)\overline\psi(0)\right) 
+ \ii g 
[A_\mu, \left(\psi(0)\overline\psi(0)\right)]
\nonumber\\
&=
\left(\RD D_\mu
\psi(0)\right)\overline\psi(0) 
+
\psi(0)\left(\overline\psi(0)\LD D_\mu\right)
\,.
\nonumber
\end{eqnarray}
This leads to the following expansion into local
tensor operators 
\begin{equation}
%\lefteqn{
{}^{I=0}\! \QQ^{ij}_{\alpha\beta}(\ka x,\tau x,\kb  x)%=}\\
=
\sum_{N=0}^\infty \frac{1}{N!} x^{\mu_1}\ldots x^{\mu_N} \,
{}^{I=0}\!
\QQ^{ij}_{\alpha\beta\mu_1\ldots\mu_N}(\ka,\tau,\kb)\,,
\end{equation}
{\rm where}
\begin{equation}
\label{O_loc4}
{}^{I=0}\!
\QQ^{ij}_{\alpha\beta\mu_1\ldots\mu_N}(\ka,\tau,\kb)
\equiv 
\overline\psi(0)\gamma_\alpha
{\bf D}_{\mu_1\ldots\mu_N}(\ka,\tau,\kb) 
\gamma_\beta\psi(0)\,,
%\nonumber
\end{equation}
and the quark field--dependent generalized covariant derivative 
of $N$th
order is given by
\begin{align}
\label{D_psi}
\lefteqn{\hspace{-2cm}
{\bf D}_{\mu_1\ldots\mu_N}(\ka,\tau,\kb)
=
\sum_{n=0}^N
\hbox{\large$\binom{N}{n}$}
\Tensor D_{\mu_1}\ldots \Tensor D_{\mu_{n}}
\psi(0)\times%\Big)\Big(
\overline\psi(0)
\Tensor D_{\mu_{n+1}}\ldots \Tensor D_{\mu_N}
\qquad\qquad\qquad}
\\
&\quad=
\sum_{n=0}^N N!
\sum_{\ell=0}^n 
\LD D_{\mu_1}\ldots\LD D_{\mu_{n-\ell}}
\hbox{\large$\frac{\ka^{n-\ell}\tau^\ell}{(n-\ell)!\ell!}$}
\RD D_{\mu_{n-\ell+1}}\ldots\RD D_{\mu_{n}} \psi(0)
\nonumber\\
&\quad\qquad\quad
\times
\sum_k^{N-n} 
\overline\psi(0) \LD D_{\mu_{n+1}}\ldots\LD D_{\mu_{n+k}}
\hbox{\large$\frac{\tau^{k}\kb^{N-n-k}}{k![N-n-k)!}$}
\RD D_{\mu_{n+k+1}}\ldots\RD D_{\mu_{N}}\,.
\nonumber
\end{align}
Analogous generalized covariant derivatives occur for the four--gluon
operators. The only difference will be that instead of the quark fields
corresponding gluon field strengths appear in (\ref{D_psi}), and
that the derivatives are to be taken in the adjoint representation.
Despite having a complicated structure this does not matter in the
twist decomposition of the trilinear operators 
${}^{I}\! \QQ^{ij}_{\alpha\beta}(\ka x,\tau x,\kb  x)$.

Let us finish this Subsection with a short remark concerning so-called 
{\em four-particle} operators which also have been considered in the 
literature for local operators. These operators have the general structure,
cf. \cite{P,EFP,BBS99}
\begin{equation}
(\overline\psi U \psi)(\overline\psi U\psi), \quad
(\overline\psi U
F U F U \psi) \quad {\rm or}
%(\overline\psi U F U \widetilde F U \psi), 
\quad
(F U F U F U F), %\quad \rm{etc.},
\nonumber
\end{equation}
with any field being located at a different point
$\kappa_i x, i = 1,\ldots,4$.
The expansion into local tensors is obtained in the same way as above,
leading to a general expression of the following form:
\begin{eqnarray}
{}^{I}\! {\cal  R}_{\alpha\beta}(\kappa_i x)
=\sum_{N=0}^\infty \frac{1}{N!} x^{\mu_1}\ldots x^{\mu_N}
x^\rho \ldots x^\sigma\,
{}^{I}\!{\cal R}_{\alpha\beta\rho\ldots\sigma\mu_1\ldots\mu_N}(\kappa_i)\,, 
\end{eqnarray}
where the terms with equal value of $N$ are given as well-defined sums
of local tensor operators. Any of these operators decompose in the
same manner into irreducible representations of the Lorentz group.
The latter result is immediately related to the fact that all indices
$\mu_1\ldots\mu_N$ are to be symmetrized, i.e.,~lying in the first row
of any relevant Young tableau! Different symmetry classifications solely 
depend on the distribution of the remaining indices $\alpha\beta
\rho\ldots\sigma$ -- the others being somehow truncated -- 
to the various Young tableaux. As long as only
up to two free indices $\alpha\beta$ are relevant the twist
decomposition takes place according to the results of Chapt.~3,
eventually modified by the power $\lambda^I$ which is related to the
number of external $x$'s in ${}^{I}\!{\cal R}$. In addition, the twist 
of the various components may be shifted by some (equal) amount. 

\subsection{Constrained trilocal operators:
Shuryak-Vainshtein-- and three--gluon operators}

Now we consider the special quark--antiquark--gluon operator
$\OO^{ij}_{\alpha\beta}$, eq.~(\ref{OO}), and related three--gluon 
operators both having minimal twist--3; they will be denoted by
${}^{I}\! \OO^{ij}_{\alpha\beta}(\ka x,\tau x,\kb x)$ with $I=0$ and
$I=1$, respectively. The quark--antiquark--gluon operators are given by
\begin{align}
{}^{I=0}\!
\OO^{ij}_{\alpha\beta}(\ka x,\!\tau x,\!\kb x)
&= x^\rho \,{}^{I=0}\!
\OO^{ij}_{\alpha\beta\rho}(\ka x,\!\tau x,\!\kb x)\\
&=\! x^\rho\,
\overline{\psi}(\ka x)U(\ka
x,\tau x)\Gamma^i_\alpha 
F^j_{\beta\rho}(\tau x)U(\tau x,\kb x)\psi(\kb
x)\,,
\nonumber
\end{align}
where again some possible flavour structure has been suppressed. These
operators are related to the following generalizations of the so-called 
Shuryak-Vainshtein operators \cite{Shu82,BB88},
\begin{eqnarray}
S^\pm_\beta(\ka x,\tau x,\kb x)
=
x^\alpha \big(
{}^{I=1}\! \OO_{\alpha\beta}^{11}(\ka x,\tau x,\kb x) \pm \ii\, 
{}^{I=1}\! \OO_{\alpha\beta}^{22}(\ka x,\tau x,\kb x) 
\big)x^\alpha , \nonumber
\end{eqnarray}
which, in the flavour singlet case, mix with the 
following three--gluon operators:
\begin{align}
&
{}^{I=1}\! \OO^{i}_{\alpha\beta}(\ka x,\tau x,\kb x)
=
x^\rho
x^\sigma\,
{}^{I=1}\! \OO^{i}_{\alpha\beta\rho\sigma}(\ka x,\tau x,\kb x)\\
&\qquad\qquad
=\!x^\rho x^\sigma\,
F^a_{\alpha\nu}(\ka x)U^{ab}(\ka x,\tau x) 
F^{i\,bc}_{\beta\rho}(\tau
x)U^{cd}(\tau x,\kb x)F^{d\nu}_\sigma(\kb x)\,,
\nonumber
\end{align}
where $F^{ab}_{\alpha\beta}(x)\equiv f^{acb}F^c_{\alpha\beta}(x)$
and the phase factors are taken in the adjoint representation.
 
Because of their construction these special
trilocal operators have the property
\begin{eqnarray}
\label{prop0}
x^\beta\, {}^{I}\! \OO^{ij}_{\alpha\beta}(\ka x,\tau
x,\kb x) \equiv 0\,.
\end{eqnarray} 
They contain twist--3 up to twist--6 contributions which, because of 
the antisymmetry of the gluon field strength, have to be determined by
means of the Young tableaux (ii) -- (iv). %---  
Because of (\ref{prop0}) the scalar operators
$x^\alpha x^\beta\, {}^{I}\! \OO^{ij}_{\alpha\beta}(\ka x,\tau x,\kb x)$ 
vanish identically.

The Taylor expansion of the operator 
${}^{I=0}\! \OO^{11}_{\alpha\beta}(\ka x,\tau x,\kb x)$ reads
\begin{equation}
\label{Oij}
{}^{I=0}\! \OO^{11}_{\alpha\beta}(\ka x,\tau x,\kb  x)
=
\sum_{N=0}^\infty \frac{1}{N!} x^{\mu_1}\ldots x^{\mu_N} x^\rho \,
{}^{I=0}\! \OO^{11}_{\alpha\beta\rho\mu_1\ldots\mu_N}(\ka,\tau,\kb)\,,
\end{equation}
 with
\begin{equation}
\label{O_local3}
{}^{I=0}\!\OO^{11}_{\alpha\beta\rho\mu_1\ldots\mu_N}(\ka,\tau,\kb)
\equiv 
\overline\psi(0)
\gamma_\alpha
{\bf D}_{[\beta\rho]\mu_1\ldots\mu_N}(\ka,\tau,\kb) \psi(0)\,,
%\nonumber
\end{equation}
where the field strength--dependent derivative already has been given
by (\ref{D_F}). The Taylor expansion of the related three--gluon 
operator obtains as
follows:
\begin{align}
{}^{I=1}\! \OO^{11}_{\alpha\beta}(\ka x,\tau x,\kb x)
=&
\sum_{N=0}^\infty \frac{1}{N!} x^{\mu_1}\ldots x^{\mu_N} x^\rho x^\sigma
\;{}^{I=1}\!\OO^A_{\alpha\beta\rho\sigma\mu_1\ldots\mu_N}(\ka,\tau,\kb)
\end{align}
where
\begin{align}
\label{O_local4}
{}^{I=1}\!
\OO^{11}_{\alpha\beta\rho\sigma\mu_1\ldots\mu_N}(\ka,\tau,\kb)
\equiv
&
F^a_{\alpha\nu}(0)
{\bf D}^{ab}_{[\beta\rho]\mu_1\ldots\mu_N}(\ka,\tau,\kb)
F_{\sigma}^{b\nu}(0)\,,
%\nonumber
\end{align}
with the same field strength dependent generalized covariant derivative
(\ref{D_F}), but now in the adjoint representation. The generalization
arbitrary values $i$ and $j$ is obvious. 

As it became obvious by the above considerations all the nonlocal 
three--particle operators ${}^{I}\!\OO^{ij}_{\alpha\beta}$ have the same 
general local structure and, therefore, decompose according to the same symmetry 
patterns. They only differ in the rank of the local tensor operators
which is due to the external powers of $x$. Again, this 
must be taken into account by a change of the integration 
measure according to\footnote{
Observe, that because of notational simplicity in the explicit 
expressions below we shifted the variable $I \rightarrow I-1$.} 
\begin{equation*}
\d\lambda \longrightarrow
\d\lambda\,\lambda^{I+1}\,.
\end{equation*}
From now on we omit the indices $i$ and $j$.

In order to determine the twist--3 light--cone operators 
${}^{I}\! \OO_{\alpha\beta}(\ka\xx,\tau\xx,\kb\xx)$ for $I=0$ and $I=1$,
let us consider the following Young tableaux of symmetry pattern (iiB)
\noindent
\unitlength0.40cm
\begin{picture}(8.5,2)
\linethickness{0.05mm}
\put(1,0){\framebox(1,1){$\SC\beta$}}
\put(1,1){\framebox(1,1){$\SC\rho$}}
\put(2,1){\framebox(1,1){$\SC\mu_1$}}
\put(3,1){\framebox(3,1){$\SC\ldots$}}
\put(6,1){\framebox(1,1){$\SC\mu_N$}}
\put(7,1){\framebox(1,1){$\SC\alpha$}}
\end{picture}
and
\unitlength0.4cm
\begin{picture}(8.5,2)
\linethickness{0.05mm}
\put(1,0){\framebox(1,1){$\SC\beta$}}
\put(1,1){\framebox(1,1){$\SC\rho$}}
\put(2,1){\framebox(1,1){$\SC\sigma$}}
\put(3,1){\framebox(1,1){$\SC\mu_1$}}
\put(4,1){\framebox(3,1){$\SC\ldots$}}
\put(7,1){\framebox(1,1){$\SC\mu_N$}}
\put(8,1){\framebox(1,1){$\SC\alpha$}}
\end{picture}
\hspace{.2cm}
, respectively:
\begin{align}
%\hspace{-1cm}
\label{O_tw3_1}
{}^{I}\!\OO^{\mathrm{tw}3}_{\alpha\beta}(\ka x,\tau x,\kb x)
&=\hbox{\large$\frac{1}{2}$}\!\!Π\int_{0}^{1}\!\!\!
\d\lambda\,\lambda^I
\Big[\big(1\!+\!\lambda^2\big)\delta_\beta^\mu\pd_\alpha
\!+\!\big(1\!-\!\lambda^2\big)\delta_\alpha^\mu\pd_\beta\Big]
{}^{I}\!\tl \OO_{\mu}(\ka\lambda x,\!\tau\lambda x,\!\kb\lambda x)
\nonumber\\
%\hspace{-1cm}
&={}^{I}\!\OO^{\mathrm{tw}3}_{[\alpha\beta]} (\ka x,\tau x,\kb x)
+{}^{I}\!\OO^{\mathrm{tw}3}_{(\alpha\beta)} (\ka x,\tau x,\kb x)\,;
\end{align}
here, and in the following, we write 
$\OO_\mu \equiv x^\rho\OO_{\rho\mu}$.
Let us remark that, first, because of property (\ref{prop0}), only the
above mentioned Young tableaux contribute and, second, the difference
between the expressions (\ref{O_tw3_1}) and (\ref{M_tw2_nl}) result 
also from that property, namely the truncation by $x^\rho$ in the second 
term of the integrand, $-x^\mu \pd_\alpha\pd_\beta
{}^{I}\!\tl \OO_{\mu}(\ka x,\tau x,\kb x) = 
2\delta_{(\alpha}^\mu\pd_{\beta)} {}^{I}\!\tl\OO_{\mu}(\ka x,\tau x,\kb x)$,
after partial integrations leads to the above expression (\ref{O_tw3_1}).
In the case $I=0$ this operator is in agreement with the expression 
given by Balitsky and Braun (cf.~eq.~(5.14) of \cite{BB88}). However,
their operator lacks to be really of twist--3 since it is not traceless; 
therefore it contains also twist--4, twist--5 as well as twist--6 operators 
resulting from the trace terms. 

Now, let us to project onto the light--cone. As in the case of the
gluon tensor only the first terms, $k=0,1,2$, in the expansion
of ${}^{I}\!\tl \OO_{\mu}(\ka\lambda x,\!\tau\lambda x,\!\kb\lambda x)$ 
are to be taken into account.
For the antisymmetric part of the twist--3 operator one gets
\begin{equation}
\label{OO_tw3_iiasy}
{}^{I}\!\OO^{\mathrm{tw}3}_{[\alpha\beta]}(\ka\lcx,\tau\lcx,\kb\lcx)
=
\int_0^1\d\lambda\,
\lambda^{2+I}\pd_{[\alpha}
{}^{I}\!\OO_{\beta]}(\ka\lambda\lcx,\tau\lambda\lcx,\kb\lambda\lcx)
-{}^{I}\!\OO^{\mathrm{>(ii)}}_{[\alpha\beta]}(\ka\lcx,\tau\lcx,\kb\lcx)
\end{equation}
where
\begin{align}
\label{}{}^{I}\!\OO^{\mathrm{>(ii)}}_{[\alpha\beta]}
(\ka\lcx,\tau\lcx,\kb\lcx)
=\hbox{\large$\frac{1}{2}$}
&\int_{0}^{1}\d\lambda\,\lambda^{1+I}
(1-\lambda)
\big(2x_{[\alpha}\pd_{\beta]}\pd^\mu-x_{[\alpha}\delta_{\beta]}^\mu\square
\big)\nonumber\\
&\times{}^{I}\! \OO_{\mu}(\ka\lambda x,\tau\lambda x,\kb\lambda x)
\big|_{x=\xx}.
\end{align}
The decomposition of
${}^{I}\!\OO^{\mathrm{>(ii)}}_{[\alpha\beta]}
(\ka\lcx,\tau\lcx,\kb\lcx)$
arises as
\begin{align}
%\label{M_tw2_b}
{}^{I}\!\OO^{\mathrm{>(ii)}}_{[\alpha\beta]}
(\kappa_1\tilde{x},\tau\lcx,\kappa_2\tilde{x})
=
{}^{I}\!\OO^{\mathrm{tw4(ii)}}_{[\alpha\beta]}
(\kappa_1\tilde{x},\tau\lcx,\kappa_2\tilde{x})+
{}^{I}\!\OO^{\mathrm{tw5(ii)}}_{[\alpha\beta]}
(\kappa_1\tilde{x},\tau\lcx,\kappa_2\tilde{x})\,,
\end{align}
with
\begin{eqnarray}
\hspace{-.5cm}
\label{OO_tw4_iiasy}
{}^{I}\!\OO^{\mathrm{tw4(ii)}}_{[\alpha\beta]}
(\kappa_1\tilde{x},\tau\lcx,\kappa_2\tilde{x})
\!\!\!&=&\!\!\!
\hbox{\large$\frac{1}{2}$}x_{[\alpha}\pd_{\beta]}\pd^{\mu}
%x^\nu
\int_{0}^{1}\d\lambda\,\lambda^I
(1-\lambda^2)\,
{}^{I}\!\OO_{\mu}(\kappa_1\lambda x,\tau\lambda x,\kappa_2\lambda x)
\big|_{x=\xx}\,,
\nonumber\\
\hspace{-.5cm}
{}^{I}\!\OO^{\mathrm{tw5(ii)}}_{[\alpha\beta]}
(\kappa_1\tilde{x},\tau\lcx,\kappa_2\tilde{x})
\!\!\!&=&\!\!\!
\label{OO_tw5_iiasy}
-
\hbox{\large$\frac{1}{4}$}
x_{[\alpha}\big(\delta_{\beta]}^{\mu}(x\pd)
-x^{\mu}\pd_{\beta]}\big)\square\nonumber\\
&&\qquad\qquad\times
\int_{0}^{1}\!\!\!\d\lambda\,\lambda^I
(1\!-\!\lambda)^2\,
{}^{I}\!\OO_{\mu}(\kappa_1\lambda x,\tau\lambda x,\kappa_2\lambda x)
\big|_{x=\xx}\,.
\nonumber
\end{eqnarray}
The symmetric part reads
\begin{equation}
\label{OO_tw3_iisy}
{}^{I}\!\OO^{\mathrm{tw}3}_{(\alpha\beta)}
(\ka\lcx,\tau\lcx,\kb\lcx)
=
\int_0^1\d\lambda\,\lambda^{I}\pd_{(\alpha}
{}^{I}\!\OO_{\beta)}(\ka\lambda\lcx,\tau\lambda\lcx,\kb\lambda\lcx)
-{}^{I}\!\OO^{\mathrm{>(ii)}}_{(\alpha\beta)}(\ka\lcx,\tau\lcx,\kb\lcx)
\end{equation}
where
\begin{eqnarray}
\label{OO_high_(ii)}
\lefteqn{{}^{I}\!\OO^{\mathrm{>(ii)}}_{(\alpha\beta)}
(\ka\lcx,\tau\lcx,\kb\lcx)
=
%\hbox{\large$\frac{1}{2}$}
\int_{0}^{1}\!\d\lambda\,\lambda
^I\Big\{
\hbox{\large$\frac{1}{2}$}(1-\lambda^2)
g_{\alpha\beta}\pd^\mu
+\hbox{\large$\frac{1}{2}$}(1-\lambda)\delta^\mu_{(\alpha}x_{\beta)}\square}\qquad\\
&&+\lambda(1-\lambda)x_{(\alpha}\pd_{\beta)}\pd^\mu
%%\nonumber\\
-\hbox{\large$\frac{1}{4}$}(1-\lambda)^ 2x_\alpha x_\beta\pd^\mu\square
\Big\}
{}^{I}\!\OO_{\mu}(\ka\lambda x,\tau\lambda x,\kb \lambda x)
\big|_{x=\xx}.\nonumber
\end{eqnarray}
Let us remark that the conditions of tracelessness
for the light-cone operators are:
\begin{eqnarray}
g^{\alpha\beta}\,{}^{I}\!\OO^{\mathrm{tw}3}_{(\alpha\beta)}
(\ka\xx,\tau\xx,\kb\xx)
=
0\,,\qquad\d^\alpha\,
{}^{I}\!\OO^{\mathrm{tw}3}_{(\alpha\beta)}(\ka\xx,\tau\xx,\kb\xx)
&=&0
\,,
%\qquad
\nonumber\\
\d^\alpha\,
{}^{I}\!\OO^{\mathrm{tw}3}_{[\alpha\beta]}(\ka\xx,\tau\xx,\kb\xx)
&=&0\,.
\nonumber
\end{eqnarray}

Now we can calculate the
twist--3 vector operator from the twist--3 tensor 
operator.
The twist--3 light--cone vector
operator
reads
\begin{align}
\label{OO_tw3_vec}
&{}^{I}\!\OO^{\mathrm{tw3}}_\alpha(\ka\lcx,\tau\lcx,\kb\lcx)
=\int_0^1\d\lambda\,
\lambda^{2+I}
\Big[\delta_\alpha^\mu(x\pd)-x^\mu\pd_\alpha -x_\alpha\pd^\mu\Big]
{}^{I}\!\OO_\mu(\ka\lambda x,\tau\lambda x,\kb\lambda x)\big|_{x=\tilde{x}}
\nonumber\\
&\qquad={}^{I}\!\OO_\alpha(\ka\lcx,\tau\lcx,\kb\lcx)
-\lcx_\alpha\int_0^1\d\lambda\, \lambda^{2+I}\pd^\mu
\,{}^{I}\!\OO_\mu(\ka\lambda x,\tau\lambda x,\kb\lambda x)
\big|_{x=\tilde{x}},
\end{align}
and the twist--4 vector operator
which is contained in the trace terms
of the twist--3 vector operator
\begin{equation}
\label{OO_tw4_ivec}
{}^{I}\!\OO^{\mathrm{tw4(ii)}}_\alpha(\ka\lcx,\tau\lcx,\kb\lcx)=
\lcx_\alpha\int_0^1\d\lambda\,\lambda^{2+I}\pd^\mu\,
{}^{I}\!\OO_\mu(\ka\lambda x,\tau\lambda x,\kb\lambda x)
\big|_{x=\tilde{x}}\, .
\end{equation}

The determination of the antisymmetric twist--4 operator having symmetry 
type (iii) obtains from Young tableaux 
\unitlength0.4cm
\begin{picture}(8,3)
\linethickness{0.05mm}
\put(1,0){\framebox(1,1){$\SC\alpha$}}
\put(1,1){\framebox(1,1){$\SC\beta$}}
\put(1,2){\framebox(1,1){$\SC\rho$}}
\put(2,2){\framebox(1,1){$\SC\mu_1$}}
\put(3,2){\framebox(3,1){$\SC\ldots$}}
\put(6,2){\framebox(1,1){$\SC\mu_N$}}
\end{picture}
and
\unitlength0.4cm
\begin{picture}(9,3)
\linethickness{0.05mm}
\put(1,0){\framebox(1,1){$\SC\alpha$}}
\put(1,1){\framebox(1,1){$\SC\beta$}}
\put(1,2){\framebox(1,1){$\SC\rho$}}
\put(2,2){\framebox(1,1){$\SC\sigma$}}
\put(3,2){\framebox(1,1){$\SC\mu_1$}}
\put(4,2){\framebox(3,1){$\SC\ldots$}}
\put(7,2){\framebox(1,1){$\SC\mu_N$}}
\end{picture},
respectively.
Again, making use of property (\ref{prop0}), the result is
\begin{align}
\label{OO_tw4_iiiasy}
{}^{I}\!\OO^{\rm
tw4(iii)}_{[\alpha\beta]}(\ka\lcx,\tau\lcx,\kb\lcx)
=&\!
\int_{0}^{1}\!\!\!\d\lambda\lambda^{2+I}
\delta^\mu_{[\alpha}\!\left( (x\pd)\delta_{\beta]}^\nu
\!-\! 2x^\nu\pd_{\beta]}\right)\!\!
{}^{I}\!\OO_{[\mu\nu]}(\ka\lambda x,\!\tau\lambda x,\!\kb\lambda x)
\big|_{x=\xx}\nonumber
\\
& 
- {}^{I}\!\OO^{\rm tw5(iii)}_{[\alpha\beta]}(\ka\lcx,\tau\lcx,\kb\lcx),
%%\nonumber
\end{align}
%where the
twist--5 part is determined by the trace namely with
\begin{align}
%\label{M_tw4}
%\lefteqn{
%\hspace{-.5cm}
{}^{I}\!\OO^
{\rm tw5(iii)}_{[\alpha\beta]}(\ka\lcx,\tau\lcx,\kappa\lcx)
= 
&
-
\int_{0}^{1}\d\lambda\,\lambda^I
(1-\lambda^2)\Big\{x_{[\alpha}\big(\delta_{\beta]}^{[\mu}(x\pd)
-x^{[\mu}\pd_{\beta]}\big)\pd^{\nu]}
\nonumber\\
&\ -
x_{[\alpha}\delta_{\beta]}^{[\mu}x^{\nu]}\square\Big\}
{}^{I}\!\OO_{[\mu\nu]}(\ka\lambda x,\tau\lambda x,\kb\lambda x)
\big|_{x=\xx}.% \qquad
%%\nonumber
\end{align}
The antisymmetric twist--5 tensor operator is given by:
\begin{align}
\label{OO_tw5_ii}
\lefteqn{
\hspace{-1cm}
{}^{I}\!\OO^{\rm tw5}_{[\alpha\beta]}(\ka\lcx,\tau\lcx,\kappa\lcx)
= 
{}^{I}\!\OO^{\rm tw5(ii)}_{[\alpha\beta]}(\ka\lcx,\tau\lcx,\kappa\lcx)+
{}^{I}\!\OO^{\rm tw5(iii)}_{[\alpha\beta]}(\ka\lcx,\tau\lcx,\kappa\lcx)
}\nonumber\\
&%\quad
=-\hbox{\large$\frac{1}{4}$}
x_{[\alpha}\big(\delta_{\beta]}^{\mu}(x\pd)
-x^{\mu}\pd_{\beta]}\big)\square\!\!
%x^\nu
\int_{0}^{1}\!\!\!\d\lambda\,\lambda^I
(1\!-\!\lambda)^2\,
{}^{I}\!\OO_{\mu}(\kappa_1\lambda x,\tau\lambda x,\kappa_2\lambda x)
\big|_{x=\xx}\nonumber\\
&%\quad
-\int_{0}^{1}\d\lambda\,\lambda^I
(1-\lambda^2)\Big\{
x_{[\alpha}\big(\delta_{\beta]}^{[\mu}(x\pd)
-x^{[\mu}\pd_{\beta]}\big)\pd^{\nu]}
\nonumber\\
&\qquad\qquad\qquad\qquad\qquad
-x_{[\alpha}\delta_{\beta]}^{[\mu}x^{\nu]}\square\Big\}
{}^{I}\!\OO_{[\mu\nu]}(\ka\lambda
x,\tau\lambda x,\kb\lambda x)
\big|_{x=\xx}.% \qquad
%%\nonumber
\end{align}
Finally, we obtain the complete decomposition of the antisymmetric 
tensor operator:
\begin{align}{}^{I}\!\OO_{[\alpha\beta]}(\ka\lcx,\tau\lcx,\kb\lcx)&=
{}^{I}\!\OO^{\rm tw3(ii)}_{[\alpha\beta]}(\ka\lcx,\tau\lcx,\kb\lcx)+
{}^{I}\!\OO^{\rm tw4(iii)}_{[\alpha\beta]}(\ka\lcx,\tau\lcx,\kb\lcx)
\nonumber\\
&+{}^{I}\!\OO^{\rm tw4(ii)}_{[\alpha\beta]}(\ka\lcx,\tau\lcx,\kb\lcx)+
{}^{I}\!\OO^{\rm tw5}_{[\alpha\beta]}(\ka\lcx,\tau\lcx,\kb\lcx).
\end{align}

The determination of the symmetric
twist--4 operator having symmetry 
type (iv) obtains from Young
tableaux
\unitlength0.4cm
\begin{picture}(9,2)
\linethickness{0.05mm}
\put(1,0){\framebox(1,1){$\SC\beta$}}
\put(2,0){\framebox(1,1){$\SC\alpha$}}
\put(1,1){\framebox(1,1){$\SC\rho$}}
\put(2,1){\framebox(1,1){$\SC\mu_1$}}
\put(3,1){\framebox(1,1){$\SC\mu_2$}}
\put(4,1){\framebox(3,1){$\SC\ldots$}}
\put(7,1){\framebox(1,1){$\SC\mu_N$}}
\end{picture}
and
\unitlength0.4cm
\begin{picture}(9,2)
\linethickness{0.05mm}
\put(1,0){\framebox(1,1){$\SC\beta$}}
\put(2,0){\framebox(1,1){$\SC\alpha$}}
\put(1,1){\framebox(1,1){$\SC\rho$}}
\put(2,1){\framebox(1,1){$\SC\sigma$}}
\put(3,1){\framebox(1,1){$\SC\mu_1$}}
\put(4,1){\framebox(3,1){$\SC\ldots$}}
\put(7,1){\framebox(1,1){$\SC\mu_N$}}
\end{picture},
respectively.
Making use of property (\ref{prop0}) the result is %[compare(\ref{})]
\begin{eqnarray}
\label{OO_tw4_ivsy}
\lefteqn{{}^{I}\!\OO^{\mathrm{tw4(iv)}
}_{(\alpha\beta)} 
(\ka\lcx,\tau\lcx,\kb\lcx)
=}\nonumber\\
&&\int_0^1\!\!\d\lambda\,\lambda^{I}(1-\lambda)
\Big(\delta_\alpha^\mu\delta_\beta^\nu x^\rho (x\pd)\pd_\rho
-2x^\mu(x\pd)\delta_{(\beta}^\nu\pd_{\alpha)}
\Big) {}^{I}\!\OO_{(\mu\nu)}(\ka\lambda x,\tau\lambda x,\kb\lambda
x)
\big|_{x=\xx}\quad\nonumber\\
&&-{}^{I}\!\OO^{\mathrm{>(iv)}}_{(\alpha\beta)}(\ka\lcx,\tau\lcx,\kb\lcx)\,,
\end{eqnarray}
where
\begin{align}
%%\lefteqn{
\label{OO_high_(iv)_sy}
&{}^{I}\!\OO^{\mathrm{>(iv)}}_{(\alpha\beta)}(\ka\lcx,\tau\lcx,\kb\lcx)
=\\
%\hbox{\large$\frac{1}{2}$}
&
\int_{0}^{1}\!\!\d\lambda\Big\{\Big(2\lambda^{2+I}
x_{(\alpha}\delta_{\beta)}^\nu\pd^\mu
-\lambda^{1+I}(1-\lambda)x_\alpha
x_\beta\pd^\mu\pd^\nu\Big)
{}^{I}\!\OO_{(\mu\nu)}(\ka\lambda x,\tau\lambda x,\kb\lambda x)
\nonumber\\
&-
\hbox{\large$\frac{1}{2}$}
\Big( \lambda^{I} g_{\alpha\beta}\pd^\mu
+\lambda^{I}(1-\lambda^2)x_{(\alpha}\delta^\mu_{\beta)}\square
-\hbox{\large$\frac{1}{2}$}\lambda^{I}(1-\lambda)^2 
x_\alpha x_\beta\square\pd^\mu \Big)
{}^{I}\!\OO_{\mu}(\ka\lambda x,\tau\lambda x,\kb\lambda x) \nonumber\\
&
+
%%\hbox{\large$\frac{1}{2}$}
\Big(
\hbox{\large$\frac{1}{2}$}\lambda^I g_{\alpha\beta}(x\pd)
-\lambda^{2+I}
x_{(\alpha}\pd_{\beta)}
-\hbox{\large$\frac{1}{2}$}\lambda^{2+I} (\ln\lambda)
x_\alpha x_\beta\square\Big) {}^{I}\!\OO^\rho_{\ \rho}
(\ka\lambda x,\tau\lambda x,\kb\lambda x)
\Big\}\Big|_{x=\tilde{x}}\!.\nonumber
\end{align}

Now, we use the property (\ref{prop0}) and sum up the symmetric 
{\em higher twist} operators, which appear in the trace terms of the 
symmetric twist operators with Young symmetry (ii)
and (iv), to a complete twist--4, twist--5 and twist--6 operator. The
complete twist--4 operators are  constructed by means of the Young tableau 
(i). The `scalar part' of the twist--4 tensor operator is given by
\begin{align}
\label{OO_tw4_sc}
{}^{I}\!\OO_{(\alpha\beta)}^{\mathrm{tw4,s}}(\ka\lcx,\tau\lcx,\kb\lcx)
=\hbox{\large$\frac{1}{2}$}
g_{\alpha\beta}\bigg\{
&(x\pd)\int_0^1\d\lambda\,\lambda^I\,
{}^{I}\!\OO^\rho_{\ \rho}(\ka\lambda x,\tau\lambda x,\kb\lambda x)\\
&-\pd^\mu\int_0^1\d\lambda\,\lambda^{2+I}\,
{}^{I}\!\OO_{\mu}(\ka\lambda
x,\tau\lambda x,\kb\lambda x)
\bigg\}\bigg|_{x=\tilde{x}}\nonumber
\end{align}
and the twist--4 `vector part' reads
\begin{align}
\label{OO_tw4_vec}
%%\lefteqn{
&{}^{I}\!\OO_{(\alpha\beta)}^{\mathrm{tw4,v}}(\ka\lcx,\tau\lcx,\kb\lcx)
=-\lcx_{(\alpha}
%%%\times\nonumber}\\
%%%&\!\!\!
\bigg\{\int_0^1\!\!\d\lambda\,\lambda^I\,
\Big\{\pd_{\beta)}+\hbox{\large$\frac{1}{2}$}x_{\beta)}
(\ln\lambda)\square\Big\}
{}^{I}\!\OO^\rho_{\ \rho}(\ka\lambda x,\tau\lambda x,\kb\lambda x)\nonumber\\
&-\int_0^1\!\!\d\lambda\,\lambda^I\,
\Big\{\big(1-\lambda^2\big)\pd_{\beta)
}+\hbox{\large$\frac{1}{4}$}x_{\beta)}
\big(1-\lambda^2+2\ln\lambda\big)\square\Big\}\pd^\mu\,
{}^{I}\!\OO_{\mu}(\ka\lambda x,\tau\lambda x,\kb\lambda x)
\bigg\}\bigg|_{x=\tilde{x}}\!\!.
%%\nonumber
\end{align}
Furthermore, the complete twist--5 vector operator is again constructed by means of the 
Young tableau (ii). The `vector part' of the twist--5 tensor operator is 
given by
\begin{align}
\label{OO_tw5_vec}
\lefteqn{{}^{I}\!\OO_{(\alpha\beta)}^{\mathrm{tw5,v}}
(\ka\xx,\tau\xx,\kb\xx)=
-\lcx_{(\alpha}\times}\\
&\bigg\{\hbox{\large$\frac{1}{4}$}
\int_0^1\d\lambda\,\lambda^I\, \big(1-\lambda\big)^2
\big(\delta^\nu_{\beta)}(x\pd)-x^\nu\pd_{\beta)}-x_{\beta)}\pd^\nu\big)
\square
\,{}^{I}\!\OO_\nu(\ka\lambda x,\tau\lambda x,\kb\lambda x)\nonumber\\
&-\!\!\int_0^1\!\!\d\lambda\,\lambda^I\Big\{
\big(1-\lambda^2\big)
\big(\delta^\nu_{\beta)}(x\pd)-x^\nu\pd_{\beta)}-x_{\beta)}\pd^\nu\big)\pd^\mu 
-\Big(\hbox{\large$\frac{1}{2}$}\big(1-\lambda^2\big)
+\ln\lambda\Big)
x_{\beta)}\square\pd^\mu x^\nu\Big\}\nonumber\\
&\hspace{8cm}\times
\,{}^{I}\!\OO_{(\mu\nu)}(\ka\lambda x,\tau\lambda x,\kb\lambda x)
\bigg\}\bigg|_{x=\tilde{x}}.\nonumber
\end{align}
The full twist--6 scalar operator has Young symmetry (i).  
The `scalar part' of the twist--6 tensor operator is given by
\begin{align}
\label{OO_tw6_sc}
&{}^{I}\!\OO_{(\alpha\beta)}^{\mathrm{tw6,s}}(\ka\lcx,\tau\lcx,\kb\lcx)
=\lcx_\alpha\lcx_\beta\bigg\{\hbox{\large$\frac{1}{2}$}\square
\int_0^1\d\lambda\,\lambda^I
\big(1-\lambda^2\big)\ln(\lambda)\,
{}^{I}\!\OO^\rho_{\ \rho}(\ka\lambda x,\tau\lambda x,\kb\lambda x)\nonumber\\
&\qquad\qquad
-\hbox{\large$\frac{1}{4}$}\square\pd^\mu
\int_0^1\d\lambda\,\lambda^I
\big(1-\lambda\big)^2\,
{}^{I}\!\OO_\mu(\ka\lambda x,\tau\lambda x,\kb\lambda x)\nonumber\\
&\qquad\qquad
+\pd^\mu\pd^\nu\int_0^1\d\lambda\,\lambda^{I}(1-\lambda)\,
{}^{I}\!\OO_{(\mu\nu)}(\ka\lambda x,\tau\lambda x,\kb\lambda x)
\bigg\}\bigg|_{x=\tilde{x}}.
\end{align}

Thus, together with the twist--3 part, eq.~(\ref{OO_tw3_iisy}), and the 
twist--4 part, eq.~(\ref{OO_tw4_ivsy}), 
we finally obtain the complete decomposition of the
symmetric tensor operator:
\begin{eqnarray}
\lefteqn{{}^{I}\!\OO_{(\alpha\beta)}(\ka\lcx,\tau\lcx,\kb\lcx)=
{}^{I}\!\OO^{\rm tw3(ii)}_{(\alpha\beta)}(\ka\lcx,\tau\lcx,\kb\lcx)+
{}^{I}\!\OO^{\rm tw4(iv)}_{(\alpha\beta)}(\ka\lcx,\tau\lcx,\kb\lcx)}
\nonumber\\
&&+{}^{I}\!\OO^{\rm tw4,s}_{(\alpha\beta)}(\ka\lcx,\tau\lcx,\kb\lcx)+
{}^{I}\!\OO^{\rm tw4,v}_{(\alpha\beta)}(\ka\lcx,\tau\lcx,\kb\lcx)+
{}^{I}\!\OO^{\rm tw5,v}_{(\alpha\beta)}(\ka\lcx,\tau\lcx,\kb\lcx)\nonumber\\
&&+{}^{I}\!\OO^{\rm tw6,s}_{(\alpha\beta)}(\ka\lcx,\tau\lcx,\kb\lcx).
\end{eqnarray}

\section{Conclusion}

Let us shortly summarize our results. Making use of the tensorial harmonic 
polynomials up to rank 2 we determined for the first time  and in a 
systematic way the various contributions of definite twist 
%resulting from the (only relevant) symmetry types 
for a generic bilocal 2nd rank tensor operator 
$G_{\alpha\beta}(\ka x, \kb x)$ as well as its symmetric %(traceless) 
and antisymmetric parts
$G_{(\alpha\beta)}(\ka x, \kb x)$ and 
$G_{[\alpha\beta]}(\ka x, \kb x)$, respectively. In
addition, we 
determined the related vector and scalar operators which occur by 
truncating with $x^\beta$ and $x^\alpha x^\beta$ or
$g^{\alpha\beta}$, 
respectively. All these operators are harmonic tensor functions.
By projection onto the light--cone 
%$x\rightarrow \xx, \xx^2 = 0$, 
we obtained the decomposition of a generic
bilocal light--ray tensor operator of 2nd rank -- and its reductions to 
(anti)symmetric tensors as
well as the related vector and scalar operators -- into operators of definite 
twist. In order to make the main results quite obvious we summarize them in 
Table 1 and 2. There, we indicate the different twist contributions 
to $G_{\alpha\beta}(\ka \xx, \kb \xx) =
G_{(\alpha\beta)}(\ka \xx, \kb \xx)+G_{[\alpha\beta]}(\ka \xx, \kb \xx)$
by their symmetry type (i) up to (iv) and the number of the
corresponding equations of Chapt.~3; in the second half of the tables 
the contributions from the trace terms are shown. 

These expressions also apply to generic trilocal light--ray 
operators. The only difference is that the respective values of twist 
increase by a definite amount which is equal for any given
operator and possibly, depending on the number $I$ of external 
$x$--factors, by a change in the integration measure, 
$\d\lambda \rightarrow \d\lambda\ \lambda^I$. 
The applicability of our procedure which has been described 
in Chapt.~2 for bilocal operators rests upon the fact that the general form
of the Taylor expansion (around $y = 0$) for any of the nonlocal operators 
is the same: It is given by an infinite sum of terms of $N$th 
order being multiplied by $x^{\mu_1} x^{\mu_2} \ldots x^{\mu_N}$ where each 
term consists of a finite sum of local operators having exactly the same 
tensor structure. In principle, this works
also for multi-local operators of higher order.

However, there are special trilocal operators which do not immediately 
fit into the general scheme of Chapt.~3. An example of such operators 
has been considered in Chapt.~4 on the same line of reasoning as for 
the generic case. Its twist decomposition is obtained by restricting 
the expressions of the generic case to the  special conditions under 
consideration. The results are classified in Table 3 and 4, again by 
indicating their symmetry type and the number of the corresponding 
equation. In the same manner one should proceed for other kinds of 
special multilocal operators like that of eq.~(\ref{OO'}).

\begin{table}[h]
\caption{Twist decomposition of general tensor operators
$G_{\alpha\beta} = G_{(\alpha\beta)} + G_{[\alpha\beta]}$}
  \begin{center}
%%    \leavevmode
\begin{tabular}{|c|c|c|c|}
\hline
Eq.& YT &   
$G_{(\alpha\beta)}(\kappa_1\lcx,\kappa_2\lcx)$
& 
$G_{[\alpha\beta]}(\kappa_1\lcx,\kappa_2\lcx)$\\
%%%& 
%%%${}^{I}\!\QQ_{\alpha\beta}(.,.,.)$
%%$\widetilde{D}_{\alpha\beta}(\ka\lcx,\tau\lcx,\kb\lcx) $ 
%%%&
%%%$W_{[\alpha\beta]}(.,.,.)$\\ 
%%$\widetilde{W}_{[\alpha\beta]}(\ka\lcx,\tau\lcx,\kb\lcx)$\\
%%& & & $\widetilde{V}_{\alpha\beta}(\ka\lcx,\tau\lcx,\kb\lcx) $ 
%%& 
%%$\widetilde{D}_{\alpha\beta}(\ka\lcx,\tau\lcx,\kb\lcx) $ 
%%&
%%$\widetilde{W}_{[\alpha\beta]}(\ka\lcx,\tau\lcx,\kb\lcx) $ \\
\hline
\hline
\ref{Gtw2_gir} & (i)    &  $\tau=2$ & --  \\
\ref{G3_symm}  & (ii)   & $\tau=3$ & -- \\
\ref{G3_anti}  & (ii)   & -- & $\tau=3$  \\
\ref{G_tw4_iii} & (iii) & -- & $\tau=4$  \\
\ref{G_tw4_iv} & (iv)   & $\tau=4$  &  -- \\
\ref{G_tw4sc_i} & (i)   & $\tau=4$ &  -- \\
\ref{G_tw4ve_i} & (i)   &  $\tau=4$ &  -- \\
\ref{G_tw4_as} & (i)    & -- & $\tau=4$ \\
\ref{G_tw5_ii_sy} & (ii) & $\tau=5$ &  -- \\
\ref{G_tw5_as} & (ii)   & -- & $\tau=5$  \\
\ref{G_tw6_i}  & (i)    & $\tau=6$ &  -- \\
\hline
\end{tabular}
\end{center}
\end{table}
\begin{table}[h]
\caption{Twist decomposition of general vector operators
$G_\alpha = G_{(\alpha\bullet)} + G_{[\alpha\bullet]}$}
\begin{center}
%%    \leavevmode
\begin{tabular}{|c|c|c|c|}
\hline
Eq.& YT &  
$G_{(\alpha\bullet)}(\kappa_1\lcx,\kappa_2\lcx)$
& 
$G_{[\alpha\bullet]}(\kappa_1\lcx,\kappa_2\lcx)$\\
%%%& 
%%%${}^{I}\!\QQ_\alpha(.,.,.)$
%%$\widetilde{D}_{\alpha\beta}(\ka\lcx,\tau\lcx,\kb\lcx) $ 
%%%&
%%%$W_{\alpha}(.,.,.)$\\ 
%%$\widetilde{W}_{[\alpha\beta]}(\ka\lcx,\tau\lcx,\kb\lcx)$\\
%% & & $\widetilde{V}_{\alpha}(\ka\lcx,\tau\lcx,\kb\lcx) $ 
%%& 
%%$\widetilde{D}_{\alpha}(\ka\lcx,\tau\lcx,\kb\lcx) $ 
%%&
%%$\widetilde{W}_{\alpha}(\ka\lcx,\tau\lcx,\kb\lcx) $ \\
\hline
\hline
\ref{G_tw2_vec} & (i)  & $\tau=2$   &  -- \\
\ref{G_tw3_vec_sy} & (ii) & $\tau=3$  & -- \\
\ref{G_tw3_vec_asy} & (ii) & -- & $\tau=3$  \\
\ref{G_tw4_vec_sy} & (i)  & $\tau=4$   &  -- \\
\ref{G_tw4_vec_asy} & (i) & -- & $\tau=4$  \\
\hline
\end{tabular}
\end{center}
\end{table}

\begin{table}[h]
\caption{Twist decomposition of special tensor operators 
${}^{I}\! \OO_{\alpha\beta}$
 with the property $x^\beta\, {}^{I}\! \OO_{\alpha\beta} \equiv 0$}
  \begin{center}
%%    \leavevmode
\begin{tabular}{|c|c|c|c|}
\hline
Eq.& YT &  type & 
${}^{I}\!\OO_{\alpha\beta}(\ka\lcx,\tau\lcx,\kb\lcx)$
\\
\hline
\hline
\ref{OO_tw3_iisy} & (ii) & sym  & $\tau=3$  \\
\ref{OO_tw3_iiasy} & (ii) & asym & $\tau=3$  \\
\ref{OO_tw4_iiiasy} & (iii) & asym & $\tau=4$  \\
\ref{OO_tw4_ivsy} & (iv) & sym  & $\tau=4$  \\
\ref{OO_tw4_sc} & (i)   & sym & $\tau=4$  \\
\ref{OO_tw4_vec} & (i)   & sym &  $\tau=4$  \\
\ref{OO_tw4_iiasy} & (i)   & asym & $\tau=4$  \\
\ref{OO_tw5_vec} & (ii) & sym  & $\tau=5$  \\
\ref{OO_tw5_ii} & (ii) & asym & $\tau=5$  \\
\ref{OO_tw6_sc} & (i)   & sym   & $\tau=6$  \\
\hline
\end{tabular}
\end{center}
\end{table}

\begin{table}[h]
\caption{Twist decomposition of special vector operators 
${}^{I}\! \OO_{\alpha}$  with the property
$x^\alpha\, {}^{I}\! \OO_{\alpha} \equiv 0$}
\begin{center}
%%    \leavevmode
\begin{tabular}{|c|c|c|}
\hline
Eq.& YT &  
${}^{I}\!\OO_{\alpha}(\ka\lcx,\tau\lcx,\kb\lcx)$
\\
\hline
\hline
\ref{OO_tw3_vec} & (ii) & $\tau=3$  \\
\ref{OO_tw4_ivec} & (i) & $\tau=4$   \\
\hline
\end{tabular}
\end{center}
\end{table}

The present study made also obvious how difficult an exact treatment of
higher twist contributions will be. There is a complicate interplay
between the higher twists resulting from different symmetry types
and the corresponding trace terms. The results also suffer from 
the equations of motion (EOM) which are frequently used in the 
applications
because they do not contribute between physical states \cite{P}. 
The twist decomposition of these EOM operators
\begin{align}
{}^{\text{EOM}\!}\OO_\Gamma(\ka\lcx,\kb\lcx)&=
\overline{\psi}(\ka\lcx) \Gamma
U(\ka\lcx,\kb\lcx)\Big[\ii\RD D\Dslas-m\Big]\psi(\kb\lcx)\nonumber\\
&-\overline{\psi}(\ka\lcx) \Big[\ii\LD D\Dslas-m\Big]\Gamma
U(\ka\lcx,\kb\lcx)\psi(\kb\lcx),\quad
\Gamma=\{1,\gamma_\alpha,\sigma_{\alpha\beta}\}
\nonumber
\end{align}
is analogous to the corresponding quark operators
$\OO_\Gamma(\ka\lcx,\kb\lcx)$ which is given in~\cite{GLR990}.
The only difference is that for the equation of motion operators
the canonical dimension and the twist raises by one unit.

The various tensor operators which we considered here are nonlocal 
generalizations of local operators already considered earlier in the 
literature for twist--3, cf. \cite{twist_3}, and twist--4, 
cf.~\cite{twist_4}. There, as far as possible an explicit twist 
decomposition has been circumvented by truncating {\em any} of the 
(constant) totally symmetrized tensor indices of the local operators by 
some light--like vector $n^\mu$, i.e., by reducing to the scalar 
case. To the best of our knowledge the only work where the twist 
decomposition of (scalar) bilocal operators has been considered is that 
of Balitsky and Braun \cite{BB88}. However, their work is based on the
external field formalism and does not have an obvious
group theoretical systematics. 

In principle, our procedure may be extended also to tensors of
arbitrary high rank. Despite being defined by an obvious algorithm, its 
application to the next step, the twist decomposition of a generic 
3rd rank tensor operator, will be very cumbersome. First of all, it would
be necessary to determine the projection operators onto all traceless 
3rd rank tensorial harmonic polynomials and, secondly, also additional 
Young patterns had to be taken into account. Fortunately, such kind 
of nonlocal operators  -- at least in the near future --  may not be of 
physical relevance. Therefore, any further study in that direction seems 
to be reasonable only from a group theoretical point of view.

However, another observation deserves mentioning. This is the appearance 
of the inner derivation on the cone and its relation to the conformal 
group. The inner derivative not only may be used in defining the property of 
tracelessness of the nonlocal light--ray operators. In some cases it is 
possible to construct the nontrivial tensor operators by applying 
(products of) the inner derivative on the scalar light--ray operators. 
Therefore, it seems to be of immediate value for a more direct 
determination of the complicated expressions obtained here and
for a simplification of our procedure. 
%%This will  be studied in the near future.

\bigskip
\bigskip
%%%\newpage
\noindent
{\large\bf Acknowledgement}\\
\noindent 
The authors are very much indebted
to D. Robaschik and S. Neumeier for stimulating discussion.
M.L. acknowledges financial support by Graduate College
``Quantum field theory'' at Center for Theoretical Sciences of
Leipzig University.
\bigskip

%%%\newpage
\begin{appendix}
\section{Harmonic tensor polynomials }
\renewcommand{\theequation}{\thesection.\arabic{equation}}
\setcounter{equation}{0}
\label{trace}

In this Appendix we 
derive the operators projecting onto the traceless part of\\
-~~ completely symmetric tensors of rank $n$, 
$T_{(\mu_1 \ldots \mu_n)}$,\\ 
-~~ tensors $T_{\alpha(\mu_1 \ldots \mu_n)}$
of rank $n+1$ being symmetric in $n$ of its indices,\\
-~~ tensors $T_{[\alpha\beta](\mu_1 \ldots \mu_n)}$
of rank $n+2$ being symmetric in $n$ and antisymmetric in \\
\phantom{-~~} two indices, 
and \\
-~~ tensors $T_{(\alpha\beta)(\mu_1 \ldots \mu_n)}$ of rank $n+2$ 
being symmetric in $n$ and, independently, \\
\phantom{-~~} symmetric in two indices. \\
In order to construct vector and tensor polynomials corresponding
to $T_{\alpha(\mu_1 \ldots \mu_n)}$, $T_{[\alpha\beta](\mu_1 \ldots \mu_n)}$,
and $T_{(\alpha\beta)(\mu_1 \ldots \mu_n)}$
we generalize the homogeneous polynomial technique which is well-known
in constructing irreducible representations of the orthogonal groups
and which has been used for two-point functions by Todorov et. al.~\cite{Tod69}
and in the scalar case by Nachtmann \cite{N}. 
To achieve this we use the fact that, after contracting the indices 
of the symmetric part with some vector $x$, 
the resulting scalar, vector and (anti)symmetric tensor polynomials
of order $n$ obey the following conditions of tracelessness:
\begin{eqnarray}
\label{H0}
\square \tl T_n(x) \!\!\!&=&\!\!\! 0,
\\
\label{H1}
\square \tl T_{\alpha n}(x) \!\!\!&=&\!\!\! 0,
\quad~~\,
\pd^\alpha \tl T_{\alpha n}(x) = 0,
\\
\label{H2}
\square \tl T_{[\alpha \beta] n}(x) \!\!\!&=&\!\!\! 0,
\quad
\pd^\alpha \tl T_{[\alpha \beta] n}(x) = 0,
%\quad
%\pd^\beta  \tl T_{[\alpha \beta] n}(x) = 0,
\\
\label{H3}
\square \tl T_{(\alpha \beta) n}(x) \!\!\!&=&\!\!\! 0,
\quad
\pd^\alpha \tl T_{(\alpha \beta) n}(x) = 0,
%\quad
%\pd^\beta  \tl T_{(\alpha \beta) n}(x) = 0,
\quad
g^{\alpha\beta}\tl T_{(\alpha \beta) n}(x) = 0.%\qquad
\end{eqnarray}
A solution of the first three sets of equations already have 
been given in Part I. Below, for the readers convenience we 
list them for arbitrary dimensions $D=2h$. In Chapt. 3 they 
will be used for $D=4$. In addition, we also solve the last 
set of equations; these symmetric harmonic tensor polynomials 
are wanted in the consideration of symmetry type (iv) of Chapter 3.4.

\noindent
(i)\hspace{.5cm}
The solutions of eq.~(\ref{H0}) are the {\em (scalar) 
harmonic polynomials} of order $n$ corresponding to symmetric 
traceless tensors of rank $n$. They are given by
(see e.g. \cite{Vil}, Chapter IX) %and \cite{Bargmann})
\begin{eqnarray}
\label{T_harm_d}
\tl T_n(x) 
&=&
H^{(2h)}_n\!\left(x^2|\square\right)\!T_n(x)
%\nonumber
\end{eqnarray}
with the harmonic projection operator
\begin{eqnarray}
\label{Harm4}
H^{(2h)}_n\!\left(x^2|\square\right)
&=&
\sum_{k=0}^{[\frac{n}{2}]}
\frac{(-1)^k(n+h-k-2)!}{4^k k! (n+h-2)!}\,
x^{2k}\,\square^{k} \,.
\end{eqnarray}

\noindent
(ii)\hspace{.5cm}
The solutions of eqs.~(\ref{H1}) are the {\em harmonic vector 
polynomials} $T_{\alpha n}(x)$ of order $n$ corresponding to 
the traceless tensors $T_{\alpha(\mu_1 \ldots \mu_n)}$, 
i.e.,~$D$ harmonic polynomials transforming as a vector under 
$SO(D)$. They are given by
\begin{align}
\label{Proj}
\tl T_{\alpha n}(x)
=&
\left\{\delta_{\alpha}^{\beta}
-\hbox{\large$\frac{1}{(h+n-1)(2h+n-3)}$}
(h-2+x\pd)\left[x_\alpha\pd^\beta
-\hbox{\large$\frac{1}{2}$} x^2\pd_\alpha\pd^\beta\right]
\right\}
\nonumber\\
& \times
H^{(2h)}_n\!\left(x^2|\square\right) T_{\beta n}(x)\, .
\end{align}
After contraction with $x^\alpha$ the scalar harmonic polynomial 
$\tl T_{n+1}(x)$ obtains.
\\

\noindent
(iii)\hspace{.5cm}
The solutions of eqs.~(\ref{H2}) are the {\em antisymmetric 
harmonic tensor polynomials} $\tl T_{[\alpha\beta] n}(x)$ of 
order $n$. They are given by
\begin{align}
\label{Proj5}
\hspace{-.3cm}
%\lefteqn{
\tl T_{[\alpha\beta] n}(x)
=&
\Big\{\delta_{[\alpha}^\mu\delta_{\beta]}^\nu
+\hbox{\large$\frac{2}{(h+n-1)(2h+n-3)}$}
(h-2+x\pd)\left( x_{[\alpha}\delta_{\beta]}^{[\mu}\pd^{\nu]} 
-\hbox{\large$\frac{1}{2}$} x^2\pd_{[\alpha}\delta_{\beta]}^{[\mu}\pd^{\nu]}\right)
\nonumber
\\
\hspace{-.3cm}
&-\hbox{\large$\frac{2}{(h+n-1)(2h+n-4)(2h+n-2)}$}
 x_{[\alpha}\pd_{\beta]}x^{[\mu}\pd^{\nu]}\Big\}
H^{(4)}_n\!\left(x^2|\square\right)T_{[\mu\nu]n}(x)\,.
%\nonumber
\end{align}

\noindent
(iv)\hspace{.5cm} Now we want to determine the {\em symmetric harmonic
tensor polynomials} $\tl T_{(\alpha\beta) n}(x)$. 
Its general structure is as follows:
\begin{align}
\label{Proj6}
\tl T_{(\alpha\beta) n}
=&\;
\Big\{\delta_\alpha^\mu\delta_\beta^\nu
+a_n g_{\alpha\beta} x^{(\nu}\pd^{\mu)}
+b_n x_{(\alpha}\delta_{\beta)}^{(\nu}\pd^{\mu)}
+c_n x_{(\alpha}x^{(\nu}\pd^{\mu)}\pd_{\beta)}
\nonumber\\
&+d_n x^2\delta_{(\alpha}^{(\nu}\pd_{\beta)}\pd^{\mu)}
%%\nonumber\\
+e_n x^2 x^{(\nu}\pd^{\mu)} \pd_{\alpha}\pd_{\beta}
+f_n x^2 x_{(\alpha}\pd_{\beta)} \pd^{\mu}\pd^{\nu}
\nonumber\\
&+g_n x_{\alpha}x_{\beta} \pd^{\mu}\pd^{\nu}
+k_n x^2 g_{\alpha\beta} \pd^{\mu}\pd^{\nu}
+l_n x^4 \pd_{\alpha}\pd_{\beta} \pd^{\mu}\pd^{\nu}
\Big\}
\breve{T}_{(\mu\nu) n}(x)
\end{align}
with the partially traceless polynomials
\begin{equation}
\breve{T}_{(\mu\nu) n}(x)=\Big(\delta^\rho_\mu\delta^\sigma_\nu
-\hbox{\large$\frac{1}{2h}$}\;g_{\mu\nu}g^{\rho\sigma}\Big)
H^{(2h)}_n\!\left(x^2|\square\right)
T_{(\rho\sigma) n}(x).
\end{equation}
It holds by construction 
\begin{equation}
\square \breve{T}_{(\mu\nu) n}(x)=0,\qquad 
g^{\mu\nu} \breve{T}_{(\mu\nu) n}(x)=0.
\end{equation}
From eqs.~(\ref{H3}) the following system of linear equations
for the nine unknown coefficients results:
\begin{eqnarray*}
0 &=& (n-1)c_n+b_n+2h a_n\\
0 &=& d_n+g_n+(n-2) f_n+2h k_n\\
0 &=& 2(h+n-2)d_n+c_n+b_n\\
0 &=& 2(h+n-2)e_n+c_n\\
0 &=& 2(h+n-2)f_n+c_n+2g_n\\
0 &=& 2(h+n-2)k_n+a_n+g_n\\
0 &=& 4(h+n-3)l_n+f_n+e_n\\
0 &=& (2h+n)b_n+2(n-1)d_n +2a_n+2\\
0 &=& (2h+n)c_n+2d_n +2a_n+4(n-2)e_n\\
0 &=& (2h+n)f_n+2k_n+d_n+8(n-3)l_n+2e_n\\
0 &=& 2(2h+n-1)g_n+4k_n+c_n+2(n-2)f_n+b_n.
\end{eqnarray*}
Its unique solution is
\begin{eqnarray*}
a_n&=& \frac{h+n-2}{(h-1)(h+n)(2h+n-2)},  \\
b_n&=& - 2\,\frac{h(h+n-1)-n+1}{(h-1)(h+n)(h+n-1)(2h+n-2)},   \\
c_n&=& - 2\,\frac{h+n-2}{(h-1)(h+n)(h+n-1)(2h+n-2)},  \\
d_n&=& \frac {h(h+n-1)-n+2}{(h-1)(h+n)(h+n-1)(2h+n-2)},  \\
e_n&=&\frac{1}{(h-1)(h+n)(h+n-1)(2h+n-2)},   \\
f_n&=& -\frac {h(h+n-4)-2(n-3)}{(h-1)(h+n)(h+n-1)(2h+n-2)(2h+n-3)},  \\
g_n&=&\frac{\big(h(h+n-1)-n+3\big)(h+n-2)}{(h-1)(h+n)(h+n-1)(2h+n-2)(2h+n-3)},\\ 
k_n&=&-\frac{h(h+n-3)-2n+3+\hbox{\large$\frac{1}{2}$}(h+n)(h+n-1)}{(h-1)(h+n)(h+n-1)(
2h+n-2)(2h+n-3)}, \\
l_n&=&\frac{h-3}{(h-1)(h+n)(h+n-1)(2h+n-2)(2h+n-3)}.
\end{eqnarray*}
In the case $D=4$ these coefficients simplify; they are given as follows:
\begin{align*}
a_n&=\frac {n}{(n +2)^2},
&&b_n= - 2\, \frac{n(n + 3)}{(n + 1)(n + 2)^2},  \\
c_n&= - 2\, \frac {n}{(n + 1)(n + 2)^2}, 
&&d_n=\frac {4 + n}{(n + 1)(n + 2)^2}, \\
e_n&=\frac {1}{(n + 1)({n} + 2)^2}, 
&&f_n=-2\,\frac{1}{(n + 1)^2(n + 2)^2},\\ 
g_n&=\frac {n(n + 3)}{(n+ 1)^{2}(n + 2)^2},  
&&k_n=-\frac{1}{2}\frac{n^2+3n+4}{(n+1)^2(n +2)^2}, \\
l_n&=-\frac{1}{4}\frac {1}{({n} + 1)^2({n} + 2)^2}\,.  
\end{align*}
 
This finishes the determination of the harmonic tensor polynomials 
being necessary for the present study. In principle, the extension 
of the procedure to arbitrary tensors of higher order is obvious. 
However, the explicit computation,  without any additional information 
about their general properties, will be quite complicated. Let us 
remark that, to the best of our knowledge, such quantities have not 
been considered in the mathematical literature.

In contrast to the harmonic scalar polynomials which carry irreducible 
representations of $SO(2h)$, the harmonic vector and tensor polynomials 
are not irreducible; they are only traceless. In order to obtain 
irreducible harmonic vector and tensor polynomials one has to 
(anti)symmetrize
according to the possible Young patterns as we have done to construct 
operators with well-defined twist.

Finally, we remark that another method exists for the construction of 
scalar harmonic polynomials which seems to be well adapted to the 
problem of twist decomposition of light-cone operators. This method 
uses the space of homogeneous polynomials of degree $n$ on the complex 
cone as the carrier space for symmetric traceless tensors and its 
harmonic extensions~\cite{Bargmann}. However, no general theoretical
frame exists for the vector and tensor case.
%There, extensive use has been made by  the interior derivative. 
%It has been used for the construction of various
%conformal operators~\cite{Dobrev76,Dobrev82,Dobrev82b,Dobrev85}
\end{appendix}


\begin{thebibliography}{99}
\bibitem{exp}
BCDMS Collab., A.C. Benvenuti et al., Phys. Lett. {\bf B 237} (1990) 592;\\
L.W.Whitlow et al. Phys. Lett {\bf B 282} (1992) 475;\\
CCFR-NuTeV Collab., W.G. Seligman et al., Phys. Rev. Lett., {\bf 79} (1997)
1213;\\
E140X Collab., L.H. Tao et al., Z. Phys. {\bf C70} (1996) 387;\\
NM Collab., M. Arneodo et al., Nucl. Phys. {\bf B 483} (1997) 3;\\
E143 Collab., K. Abe et al., Phys. Rev. Lett. (hep-ph 9808028).

\bibitem{HTEXP}
M. Virchaux and A. Milsztajn, Phys. Lett. {\bf  274B} (1992) 211;\\
A.V. Sidorov and M.V. Tokarev, Nuovo Cim. {\bf  A110} (1997) 1401;\\
A.L. Kataev, A.V. Kotikov, G. Parente, and A.V. Sidorov, 
Phys. Lett. {\bf  417B} (1998) 374;\\
A.V. Kotikov, V.G. Krivokhizhin, hep-ph/9805353;\\
A.L. Kataev, G. Parente, and A.V. Sidorov, hep-ph/ 9809500, 
hep-ph/9905310;\\
G. Ricco and S. Simula, hep-ph/9809264;\\
U.K. Yang and A. Bodek, Phys. Rev. Lett. {\bf  82} (1999) 2467;\\
A.I. Alekhin and A.L. Kataev, Phys. Lett. {\bf  452B} (1999) 402;\\
A.I. Alekhin, Phys. Rev. {\bf  D59} (1999) 114016; hep-ph/9902241; hep-ph/9907350.

\bibitem{Gro71}
D.J.~Gross and S.B.~Treiman, Phys.~Rev.~{\bf D4} (1971) 1059.

\bibitem{KS70}
K.~Kogut and D.E.~Soper, Phys. Rev. {\bf D1} (1970) 2901.

\bibitem{Jaf92}
R.L~Jaffe and X.~Ji, Nucl. Phys. {\bf B375} (1992) 527;\\
X.~Ji, Nucl. Phys. {\bf B402} (1993) 217.

\bibitem{GLR990}
B.~Geyer, M.~Lazar, and D.~Robaschik,
Nucl.~Phys.~{\bf B559} (1999) 339; \\
%%%{\tt hep-th$/$9901090}
%%%\bibitem{GLR99b}
B.~Geyer, M.~Lazar, and D.~Robaschik,
Nucl.~Phys.~B (Proc.~Suppl.) {\bf 79} (1999) 560.

\bibitem{BGR99}
J.~Bl\"umlein, B.~Geyer, and D.~Robaschik, Nucl. Phys. {\bf B560} (1999) 283.
%%%{\tt hep-ph/9903520}

\bibitem{AS78}
S.A. Anikin and O.I. Zavialov, Ann. Phys. (NY) {\bf 116} (1978) 135;\\
O.I.~Zavialov, {\sf Renormalized Feynman Diagrams},
(Nauka, Moscow, 1979), in Russian;  
{\sf Renormalized Quantum Field Theory} (Kluwer Academic Press,
Dordrecht, 1990), extended English translation.

\bibitem{SLAC}
M.~Bordag and D.~Robaschik, Nucl. Phys. {\bf B169} (1980) 445.
%%M.~Bordag, B.~Geyer, J.~Ho\v{r}ej\v{s}i, and D.~Robaschik,
%%Zs. Phys. {\bf C26} (1985) 591;\\
%%B.~Geyer, D.~M\"uller, and D.~Robaschik, preprint SLAC-Pub 1993;

\bibitem{LEIP}
D.~M\"uller, D.~Robaschik, B.~Geyer, F.-M.~Dittes, and J.~Ho\v{r}ej\v{s}i: 
{\it Wave functions, evolution equations
and evolution kernels from light-ray operators of QCD}, 
Fortsch.~Phys.~{\bf 42} (1994) 101;
reprinted as {\tt hep-ph/9812448}.

\bibitem{BR}
A.O.~Barut and R.~Raczka,
{\sf Theory of Group Representations and Applications}
PWN -- Polish Scientific Publishers, Warszawa, 1977.

\bibitem{Vil}
%%N.Ya.~Vilenkin, 
%%{\sf Special functions and the theory of group representations}
%%Nauka, Moscow 1965 (in Russian),\\
N.Ya.~Vilenkin and A.U.~Klimyk,
{\sf Representations of Lie groups and Special
Functions}, Vol. 2 (Kluwer Academic Publishers, Dordrecht 1993).

\bibitem{BB88}%
%%%I.I.~Balitsky and V.M.~Braun, {\sf Proc. XXV LNPI Winter 
%%%School on Physics}, Leningrad (1990) p. 105
I.I.~Balitsky and V.M.~Braun, Nucl. Phys. {\bf B311} (1988/89) 541;\\
see also: I.I.~Balitsky, Phys. Lett. {\bf 124B} (1983) 230.

\bibitem{Bargmann}
V.~Bargmann and I.T.~Todorov, J. Math. Phys.~{\bf 18} (1977) 1141.

\bibitem{Dobrev77}
V.K.~Dobrev, G.~Mack, V.B.~Petkova, S.G.~Petrova, and I.T.~Todorov,
{\sf Harmonic Analysis of the n-Dimensional Lorentz Group and its 
Applications to Conformal Quantum Field Theory},
Lecture Notes in Physics, No.~{\bf 63} (Springer, 1977).

\bibitem{Dobrev76}
V.K.~Dobrev, V.B.~Petkova, S.G.~Petrova, and I.T.~Todorov, 
Phys. Rev.~D~{\bf13} (1976) 887;\\
N.S.~Craigie, V.K.~Dobrev, and I.T.~Todorov,
Ann. Phys.~(N.Y.)~{\bf159} (1985) 411.

\bibitem{Dobrev82b}
V.K.~Dobrev, A.Ch.~Ganchev, and O.I.~Yordanov,
Phys. Lett. {\bf 119B} (1982) 372;\\
V.K.~Dobrev and A.Ch.~Ganchev,
{\it Conformal operators from spinor fields:
Antisymmetric tensor case}, Dubna preprint E2-82-881 (1982).

\bibitem{GL99}
B.~Geyer and M.~Lazar, {\it Twist decomposition of nonlocal light-ray 
operators and harmonic tensor functions},
{\tt hep-th$/$9911024}; to appear in: Proceedings Int. Conf. 
``Quantum Theory and Symmetries'', Goslar, July 1999.

\bibitem{N}
O. Nachtmann, Nucl. Phys., {\bf  B63} (1973) 237.

\bibitem{G}
S. Gottlieb, Nucl. Phys. {\bf  B139} (1978) 125.

\bibitem{P}
H.D. Politzer, Nucl. Phys. {\bf  B172} (1980) 349.

\bibitem{EFP}
R.K. Ellis, W. Furmanski, and R. Petronzio, 
Nucl. Phys. {\bf  B207} (1982) 1, {\bf  B212} (1983) 29;\\
%%[Feynman graph method, generalized quark-gluon correlation functions]
E.V. Shuryak and A.I. Vainshtein, Phys. Lett. {\bf  B105} (1981) 65, 
Nucl. Phys. {\bf  B199} (1982) 951; Nucl. Phys. {\bf  B201} (1982) 141;\\
R.L. Jaffe, Nucl. Phys. {\bf  B229} (1983) 205;\\
%%\bibitem{BFKL}
A.P. Bukhvostov,  G.V. Frolov, E.A. Kuraev, and L.N. Lipatov,
Nucl. Phys. {\bf  B258} (1985) 601;\\
%%[introduction of quasi-partonic operators]
A.P. Bukhvostov and G.V. Frolov, Yad. Fiz. {\bf  45} (1987) 1136 
[Sov. J. Nucl. Phys. {\bf  45} (1987) 704].

\bibitem{twist_4}
%%A. DeRujula, H. Georgi, and H.D. Politzer, 
%%Ann. Phys. (N.Y.) {\bf  103} (1977) 355;
S. Wada, Progr. Theor. Phys. {\bf 62} (1979) 475, Nucl. Phys. {\bf B202} (1983) 201;\\
M. Okawa, Nucl. Phys. {\bf  B172} (1980) 481, {\bf  B187} (1981) 71;\\
S.P. Luttrell, S. Wada, and B.R. Webber, Nucl. Phys. {\bf  B188} (1981) 219;\\
S.P. Luttrell and S. Wada, Nucl. Phys. {\bf  B197} (1982) 290;\\
%%%\bibitem{JS82}
R.L.~Jaffe and M. Soldate, Phys. Lett. {\bf  B105} (1981) 467,
Phys.~Rev.{\bf D26} (1982) 49;\\
J.W. Qiu, Phys. Rev. {\bf  D42} (1990) 30.

\bibitem{twist_3}
M.A. Ahmed and G.G. Ross, Phys. Lett. {\bf 56B} (1975) 385,
Nucl. Phys. {\bf B111} (1976) 441;\\
A.P. Bukhvostov,  E.A. Kuraev, and L.N. Lipatov,  
ZhETF  {\bf  87} (1984) 37;\\
P.G. Ratcliffe, Nucl. Phys. {\bf  B264} (1986) 493.
%%%[quasi-partonic twist-4 operators]

\bibitem{Shu82}
E.V.~Shuryak and A.I.~Vainshtein, Nucl.~Phys.~B~{\bf 201} (1982) 141.

\bibitem{Gey96b}
B.~Geyer, D.~M\"uller, and D.~Robaschik,
Nucl.~Phys.~B (Proc.~Suppl.) {\bf 51~C} (1996) 106;\\
%%%{\tt hep-ph$/$9611452}
B.~Geyer, D.~M\"uller, and D.~Robaschik,
{\it The evolution of non-singlet twist-3 parton distribution
functions}, DESY 96-239, Proc. of Third Meeting on Prospects in 
Nucleon-Nucleon Spin Physics at HERA, Dubna, 1996,
hep-ph/9611452.

\bibitem{BBK89}
I.I.~Balitsky, V.M.~Braun, and A.V.Kolesnichenko, 
Nucl. Phys. {\bf B312} (1989) 509;\\
I.I.~Balitsky and V.M.~Braun, 
Nucl. Phys. {\bf B361} (1991) 93.

%%\bibitem{BBKT98}
%%P.~Ball, V.M.~Braun, Y.~Koike, and K.~Tanaka,
%%Nucl. Phys. {\bf B529} (1998) 323
\bibitem{BBKT99}
P.~Ball, V.M.~Braun, Y. Koike, and K. Tanaka, Nucl. Phys. {\bf B529} (1998) 509;\\
P.~Ball and V.M.~Braun, Nucl. Phys. {\bf B543} (1999) 201.

\bibitem{JS82}
R.L.~Jaffe and M. Soldate, Phys.~Rev.{\bf D26} (1982) 49;\\
%%%\bibitem{BFKL}
A.P. Bukhvostov,  G.V. Frolov, E.A. Kuraev, and L.N. Lipatov,
Nucl. Phys. {\bf  B258} (1985) 601.

\bibitem{BBS99}
J. Bartels, C. Bontus, and H. Spiesberger,
{\it Factorization of Twist-Four Gluon Operator Contributions},
hep-ph/9908411.

\bibitem{Tod69}
I.T.~Todorov and R.P.~Zaikov, J. Math. Phys.~{\bf 10} (1969) 2014;\\
A.I.~Oksak and I.T.~Todorov, Comm. Math. Phys.~{\bf 14} (1969) 271.

\end{thebibliography}
\end{document}